\documentclass[preprint2]{aastex}
\usepackage{float,lscape}

\usepackage{graphicx}
\usepackage{pdflscape}
\usepackage{subfigure}
\newcommand{\se}{{\tt SExtractor\ }}
\newcommand{\psf}{{\tt PSFEx\ }}
\newcommand{\dos}{{\tt SExtractor+PSFEx\ }}

\shorttitle{Searching for Extragalactic Sources in the VVV Survey}
\shortauthors{Baravalle et al.}

\begin{document}


%
\title{Searching for Extragalactic Sources in \\ the VISTA Variables in 
the V\'ia L\'actea Survey} 

\author{Laura D. Baravalle\altaffilmark{1}}
\author{M. Victoria Alonso\altaffilmark{1,2}}
\author{Jos\'e Luis Nilo Castell\'on\altaffilmark{3,4}}
\author{Juan Carlos Beam\'in\altaffilmark{5,6}} 
\author{Dante Minniti\altaffilmark{6,7,8}}


\altaffiltext{1}{Instituto de Astronom\'ia Te\'orica y Experimental, (IATE-CONICET), Laprida 922, C\'ordoba, Argentina.}
\altaffiltext{2}{Observatorio Astron\'omico de C\'ordoba, Observatorio Astron\'omico de C\'ordoba, Laprida 854, C\'ordoba, Argentina.}
\altaffiltext{3}{Departamento de F\'isica y Astronom\'ia, Facultad de Ciencias, Universidad de La Serena,  Av. Juan Cisternas 1200 Norte, La Serena, Chile.}
\altaffiltext{4}{Direcci\'on de Investigaci\'on y Desarrollo, Universidad de La 
Serena, Av. Ra\'ul Bitr\'an Nachary 1305, La Serena, Chile.}
\altaffiltext{5}{Instituto de F\'isica y Astronom\'ia, Facultad de Ciencias, Universidad de Valpara\'iso, Ave. Gran Breta\~na 1111, Playa Ancha, Valpara\'iso, Chile.}
\altaffiltext{6}{Millennium Institute of Astrophysics, Chile.} 
\altaffiltext{7}{Departamento de Ciencias F\'{\i}sicas, Universidad Andr\'es Bello, Rep\'ublica 220, Santiago, Chile.}
\altaffiltext{8}{Vatican Observatory, Vatican City State V-00120, Italy.}


\begin{abstract}

We search for extragalactic sources in the 
VISTA Variables in the V\'ia L\'actea survey that are hidden by the Galaxy.

Herein, we describe our photometric procedure to find and characterize
extragalactic objects using a combination of \se 
and {\tt PSFEx}.  It was applied
 in two tiles of the survey: d010 and d115, without previous extragalactic IR
 detections, in order to obtain photometric parameters of the detected sources.
The adopted criteria to define extragalactic candidates
include  ${\tt CLASS\_STAR} < 0.3$; 1.0 $<$ R$_{1/2} < 5.0~arcsec$;
2.1 $<$ C $<$ 5; 
and $\Phi > 0.002$ and the colors:
0.5 $<$ (J - K$_s$) $<$ 2.0 mag; 0.0 $<$ (J - H) $<$ 1.0 mag; 
0.0 $<$ (H - K$_s$) $<$ 2.0 mag and (J - H) + 0.9 (H - K$_s$) $>$ 0.44 mag.

We detected 345 and 185 extragalactic candidates in the d010 and d115 tiles, respectively. 
 All of them were visually inspected and  confirmed to be galaxies. 
  In general, they are small and more circular objects, due to the near-IR
  sensitivity to select more compact objects with higher surface brightness.
 
 The procedure will be used to identify 
extragalactic objects in other tiles of the VVV disk, which
will allow us to study the distribution of galaxies
and filaments hidden by the Milky Way.

\end{abstract}
 

\section{Introduction}

The  Milky Way (MW) complicates the study of extragalactic 
objects that are 
behind our Galaxy due to the presence of dust, gas and stars that absorb,
obscure and 
reduce their brightness.
However, some efforts were made in the past to find these objects with
one of the pioneers being Renee Kraan-Kortewerg, who led in the 
1990´s several studies in the plane of the Galaxy.  Kraan-Korteweg (1999)
discussed the importance of deep surveys in the optical, near-infrared
(near-IR) and blind HI wavelength regimes, as well as associated systematic
issues.
Galactic dust obscuration attenuates the light coming from 
extragalactic sources in the optical region 
  and  it is transparent to the 21-cm line radiation of neutral hydrogen, favoring the detection
  of HI-rich galaxies.
In the optical wavelengths, these authors defined a limiting diameter to discriminate
between stars and galaxies, with objects of diameters greater than 
0.2 arcmin being considered to be galaxy candidates.  
Using this procedure, Kraan-Korteweg (2000) observed objects in the Antlia 
and Hydra galaxy clusters, in the direction of the Cosmic Microwave 
Background radiation dipole and found more than 8000 new galaxy candidates.
In addition, Woudt \& Kraan-Korteweg (2001) studied the 
neighborhoods of the Crux (289$^{\circ} < l < $ 318$^{\circ}$ and 
$|b| < $ 10$^{\circ}$)
and the Great Attractor (289$^{\circ} < l <$ 338$^{\circ}$ and 
$|b| < $ 10$^{\circ}$) regions.  

 Recently, McIntyre et al. (2015) presented a blind HI survey named the Arecibo 
L-band Feed 
Array Zone of Avoidance (ALFA ZOA) deep Survey, using the Arecibo Radio 
Telescope.  On analyzing an area of 15 square degrees centered on 
l = 192$^{\circ}$ and b = −2$^{\circ}$, they encountered 61 galaxies  and
  concluded that bright galaxies are likely to be identified
in  infrared surveys,  with less probability at higher extinctions.  
Ramatsoku et al. (2016) focused their 
21cm HI-line imaging survey on the 
Perseus-Pisces 
Supercluster filament crossing the ZOA and detected 211 galaxies in an
area of 9.6$^{\circ}$, with 62\% of these having near-IR counterparts in the
 UKIRT (UK InfraRed Telescope) Infrared
 Deep Sky Survey (UKIDSS; Lawrence et al. 2007).  They also noted that near-IR
 selection favored
early-type galaxies, in contrast with the late-types typically detected in HI 
samples. Moreover, Staveley--Smith et al. 
(2016) presented a HI survey of the 
extragalactic sources in the
southern regions of the Galaxy (l=212$^{\circ}$ to l=36$^{\circ}$ 
and $|b|<5^{\circ}$) and discovered complex 
structures  with three new galaxy concentrations in the Great Attractor region.

 The older stellar population in galaxies emits most of the light in the near-infrared wavelength range. 
One of the most important near-IR surveys that has been carried out is 
the 2 Micron All Sky Survey (2MASS, Skrutskie et al. 2006),
covering the whole sky in the J, H, and K passbands  including 
the Point Source Catalog, and also the Extended Source Catalog (2MASSX).
Jarrett et al. (2000a) described the basic algorithms utilized in this survey
for the 
object detections and their characterizations, and Jarrett et al. (2000b)
reported extended sources beyond the 
plane of our Galaxy.
A complete detection of galaxies brighter than K$_s \sim 13.5$, 
H $\sim$ 14.3 and
J $\sim$ 15.0 mag was presented, over a wide range of surface brightnesses.  
In their star-galaxy separation, in order to resolve confusion among real
galaxies, 
Galactic nebulae, double stars
and other artifacts that could be bright stars or meteor streaks, they also
used the (J - H) vs (H - K$_s$) color plane and a color separation of 
(J - K$_s$) $ \sim 1.0$ between stars and galaxies. In this way,
Jarrett et al. (2000b) confirmed 14 of the 2MASS galaxy candidates using 
follow-up observations in the ZOA using 
HI 21cm and optical spectroscopy.  Deeper near-IR surveys than 2MASS are
UKIDSS (Lawrence et al. 2007) and
VISTA  (Visible and Infrared Survey Telescope
for Astronomy) Kilo-degree INfrared Galaxy survey
(VIKING; Arnaboldi et al. 2007) probing other areas of the
sky.    Moreover, Williams, Kraan-Korteweg \& Woudt (2014)
and Said et al. (2016a, 2016b) 
obtained 
near-IR photometric properties of spiral galaxies using the Parkes 
deep HI survey of the ZOA (HIZOA, Henning, Kraan-Korteweg \& Staveley-Smith
2005). They  were interested in 
distance determinations and peculiar velocities using the Tully-Fischer
relationship (Tully \& Fischer, 1977).

The VISTA Variables in the 
V\'ia L\'actea (VVV, Minniti et al. 2010) is 
a public European Southern Observatory (ESO) near-IR variability survey covering the MW Bulge 
 (-10$^{\circ}$ $<$ l $<$ +10$^{\circ}$ and -10$^{\circ}$ $<$ b $<$ +5$^{\circ}$) and an 
adjacent section of the mid-plane (-65$^{\circ}$ $<$ l $<$ -10$^{\circ}$ and 
-2$^{\circ}$ $<$ b $<$ +2$^{\circ}$).  This large survey offers an excellent
opportunity to study extragalactic sources 
behind the MW, such as background galaxies and quasars.  
Using the VVV data, Am{\^o}res et al. (2012) visually identified 
204 galaxy candidates in the d003 tile
(l = 298.356$^{\circ}$, b = -1.650$^{\circ}$) by analyzing their size and colors.
Later, Coldwell et al. (2014) obtained the photometric properties 
for the galaxy cluster Suzaku J1759-3450 at z=0.13 in the b261 tile
(l = 356.597$^{\circ}$, b = -5.321$^{\circ}$).

In order to take advantage of the
VVV data, the main goal of this work was to find and 
characterize extended objects behind our Galaxy.  
This is the first paper that describes in detail a methodology
to search, detect and analyze the photometric properties of
these objects in two tiles of the VVV survey.  
This paper is organized as follows: 
in  $\S$2 we briefly describe the VVV survey and 
in  $\S$3 the photometric pipeline of the \dos 
combination is presented.  In $\S$4, our procedure is compared with those
obtained by 
the Cambridge 
Astronomical Survey Unit.  We present in $\S$5, the criteria to
select the extragalactic sources, and in $\S$6, 
their main photometric characteristics.  
Finally, $\S$7 summarizes the main results of this study and suggests the
future work.

\section{Observational database: The Vista Variables in the V\'ia L\'actea}

The 4m VISTA telescope 
at ESO  is equipped with a VIRCAM camera (Emerson, McPherson \& 
Sutherland 2006; Dalton et al. 2006), which is an array of 
16$\times$2048$\times$2048 Raytheon VIRGO IR detectors with a scale of
0.339 arcsec/pix with five broad passbands: Z, Y, J, H and K$_s$ and three narrow 
passbands at 0.98, 0.99, and 1.18 $\mu$m.  The  VVV  is 
one of the six largest ESO public surveys conducted using this telescope without
any overlapping regions. The VVV images (Tiles) are produced by six 
single pointing observations with a 
total field of view of
1.64 square degrees. The survey area is fully covered by 348 tiles: 
196 in the bulge and 152 in the disk of the MW.
These data were reduced using the pipeline developed by the Cambridge 
Astronomical Survey Unit (hereafter CASU) within the VISTA data flow system (Lewis et al. 2006, 
Emerson, McPherson \& Sutherland 2006).
Due to the large amount of data, processing was performed on a 
night-by-night basis and consisted mainly of:  mean dark current
subtraction; a linear correction applied to the detectors; a flat-field 
correction made by dividing the mean twilight flat-field; and a sky background 
correction that removed the large-scale spatial background emission.
The photometric data provided by CASU  (Saito et al. 2012) mainly included the 
equatorial coordinates, the fluxes at different
apertures,  and a simple classification as stellar (flag=-1) or non-stellar 
(flag=+1) objects.

\begin{figure*}
\begin{centering}
  	\includegraphics[width=150mm,height=90mm]{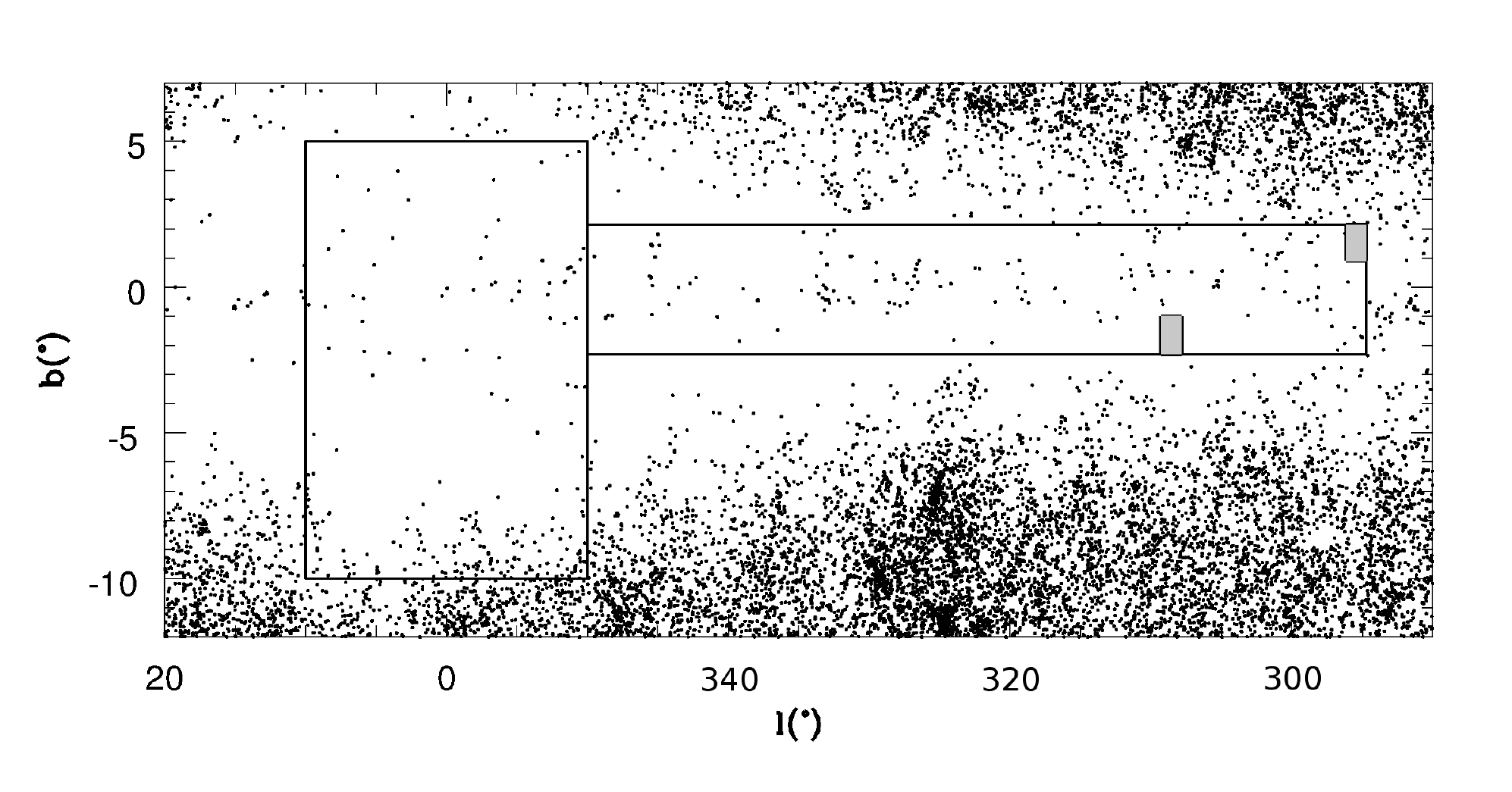}
	\caption{Distribution of the 2MASSX objects in galactic coordinates.
The area of the VVV survey is drawn with the studied tiles represented by the 
rectangles at positive
(d115) and negative (d010) galactic latitudes.}
	\label{lb2mass}
\end{centering}
\end{figure*}

The main goal of the VVV project is to identify variable stars in the MW.
However, there are additional scientific objectives such as searching for 
new star clusters, background quasars and extended objects. In this sense,
our goal was to find extended objects in these regions by searching for
extragalactic sources 
and to obtain and characterize 
their photometric 
properties. In this work, we analyzed two VVV regions in the
disk with opposite galactic latitudes, corresponding to the tiles: d010 centered at 
J2000 $RA=13h43m7.27s$, 
$Dec=-63^{\circ}57^{\prime}15.84^{\prime\prime}$ (l=308.569$^{\circ}$, 
b=-1.650$^{\circ}$) and d115 at J2000 
$RA=11h50m18.72s$, $Dec=-60^{\circ}21^{\prime}9.00^{\prime\prime}$ 
(l=295.438$^{\circ}$,
b=1.627$^{\circ}$), with observing time of 20s for Z and Y; 
10s for J and H; and 4s for K$_s$ passbands.  The two tiles were selected because there are no
near-IR extended sources previously
detected at these latitudes.  In fact,  in the d010 tile, the galaxy
HIPASS J1341-64 was
reported by Kilborn
et al. (2002)  with a radial velocity of
2473 km~s$^{-1}$.  Recently, the HIZOA-S survey (Staveley-Smith et al.
2016) observed two galaxies in the region: HI J1341-64 and HI J1347-64. 
Figure~\ref{lb2mass} shows the distribution in
galactic coordinates of the extended sources from 2MASSX at lower
galactic regions, and includes a sketch of the area covered
by the VVV. The studied tiles are highlighted in gray, which are located
at the outermost parts of the VVV disk in the less
crowded regions with smaller interstellar extinctions. 
  Figure~\ref{extinction}
  shows the A$_{K_s}$ distributions and the extinction in the K$_s$ passband for
  the two tiles
obtained from the
Galactic dust extinction maps of Schlafly \& Finkbeiner (2011). 
The median A$_{K_s}$  values are 0.86 $\pm$ 0.32 and 0.42 $\pm$ 0.08
for the d010 and d115 tiles, respectively, with the d010 tile having
a much more 
widespread 
distribution in A$_{K_s}$ and showing regions of higher absorptions.

\section{The photometric pipeline}

Different photometric procedures which can separate stellar and extended
objects and perform photometry have been utilized to determine sources
in astronomical images. For example, \se (Bertin \& Arnouts 1996) is a program
that analyzes
 images with the aim of producing a 
large 
amount of photometric data, and \psf (PSF Extractor, Bertin 2011) 
generates the Point Spread Function (PSF)
from the images processed by {\tt SExtractor}. The resulting PSF models can
then be used for model-fitting
photometry and morphological analyses\footnote{http://www.astromatic.net/software/sextractor and http://www.astromatic.net/software/psfex}. 
Annunziatella et al. (2012) compared the extraction performances
of using the combination of \se  with \psf against DAOPHOT (Stetson 1987) and 
ALLSTAR (Stetson 1994), which are the most traditional applications. The main
conclusion was 
that DAOPHOT and ALLSTAR both provide optimal solutions for point-source 
photometry in stellar fields, whereas \se and PSF photometry gave more 
accurate photometry for galaxies. 
Mauro et al. (2013) also created an automatic PSF-fitting 
photometric pipeline (the VVV-SkZ pipeline) based on DAOPHOT and 
produced a deeper stellar photometry with the VVV data.  
Here, based on the Annunziatella et al. (2012)  results, we decided to use the 
\se $v2.19.5$ and \psf $v3.17.1$ combination to detect and analyze the 
extended objects in the VVV survey, and in this section we describe our adopted procedure.

\subsection{\bf{SExtractor}}

\se has the ability to detect astronomical sources by identifying regions in images with 
different properties or features, such as brightness, color and texture, and
has been used  in different extragalactic 
studies.  Varela et al. (2009) obtained the B and V photometry of 
the WIde Field Nearby Galaxy-cluster Survey (WINGS), while Durret et al. 
(2011) counted clusters of galaxies as a function of mass and redshift 
in the Canada France Hawaii Telescope Legacy Survey (CFHTLS).
 In addition, Nilo Castell\'on et al. (2014) obtained the galaxy morphological 
parameters of galaxy clusters with low-X ray emissions.  

\se uses various steps to separate objects. In the first part of the process,
\se identifies and 
separates an object from the background noise. Then, the object is defined as 
the 
sum of adjacent pixels, with signals coming from the contribution of the object 
itself, and also the background.
A background map is created by defining a grid over the image, which divides
the original frame into several boxes 
where the local background is calculated. A combination of 
$\kappa-\sigma$ clipping and mode estimation is applied over the grid,
with a median filter used to suppress possible 
local background overestimation, and a bi-cubic-spline interpolation utilized  
to smooth out the background map.  At this stage, the filtering is necessary
to separate low surface 
brightness objects from spurious detections, such as optical defects, 
inhomogeneities in the images, cosmic rays and bright 
spikes from saturated objects. \se applies convolution filters over 
the image, with the filtering selection depending on the 
image characteristics, 
atmospheric conditions, studied regions and source morphologies.
 In our case, we decided to choose the {\tt Gaussian 
filter}, because it works well on the detection 
of faint and extended objects, and Gaussian 
functions were defined, with 5$\times$5 pixels with convolution mask of a full width half maximum 
(FWHM)  of 3 pixels.  Finally, thresholding and deblending were performed after applying the 
convolution filter.  For the detection and separation of objects from the 
background noise,  {\tt SExtractor} uses a minimum number of adjacent pixels 
defined with a flux in the  K$_s$ passband of $1.0 \sigma$ above the local background, where 
$\sigma$ is its dispersion. 
The {\it deblending} facility is one of the most important {\tt SExtractor}
features,
which can analyze overlapping  objects.  This is carried out 
using a multi-thresholding algorithm, that employs a multiple isophotal analysis 
technique (Beard, McGillivray \& Thanish 1990).

\subsection{\bf{PSFEx}}

\begin{figure*}
\begin{centering}
	\includegraphics[width=100mm]{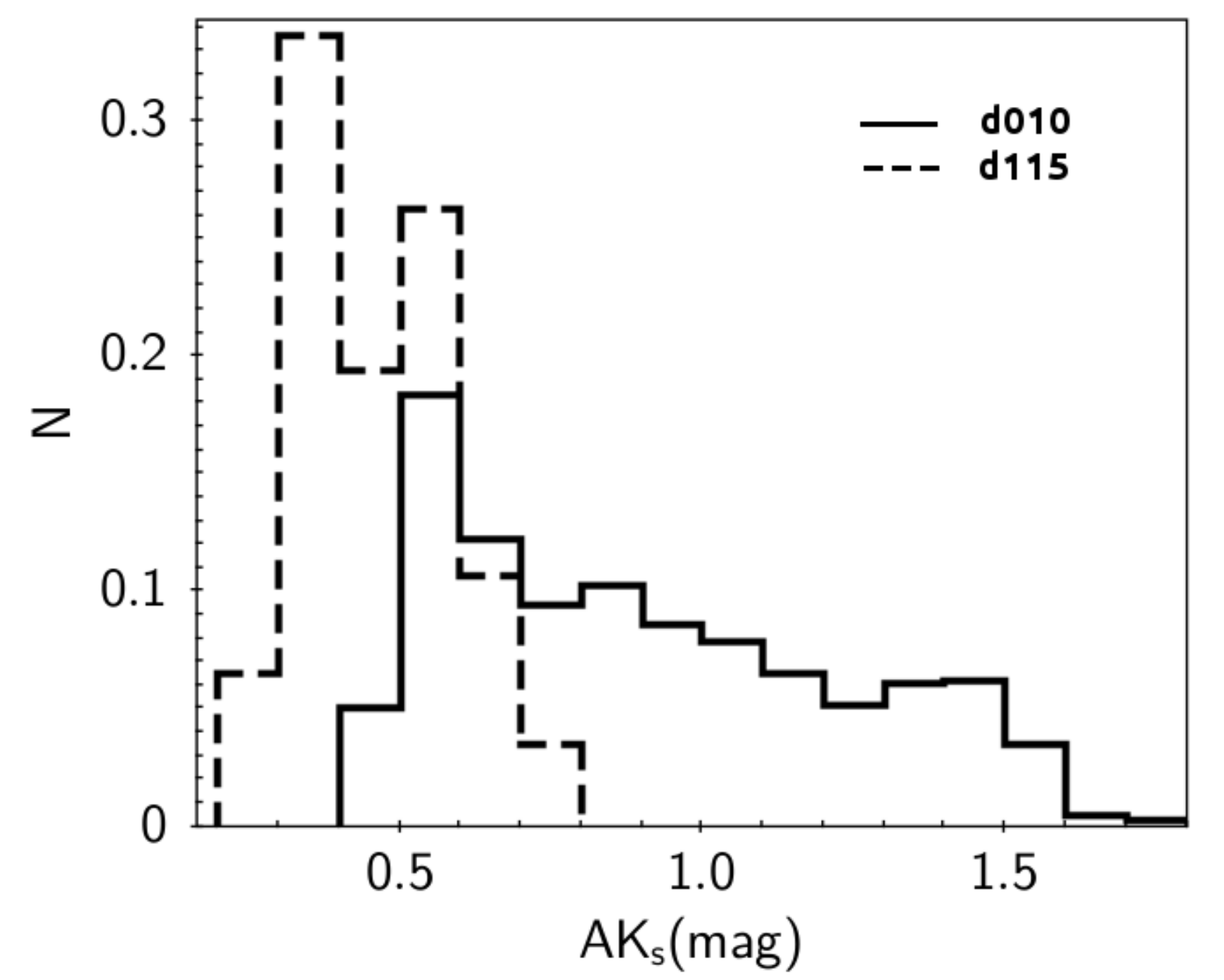}
	\caption{Normalized distributions of Galactic extinction in the K$_s$ 
passband of the regions of the two studied tiles d010 (solid histogram) and
d115 (dashed histogram).}
	\label{extinction}
\end{centering}
\end{figure*}

\begin{figure*}
  \includegraphics[width=170mm]{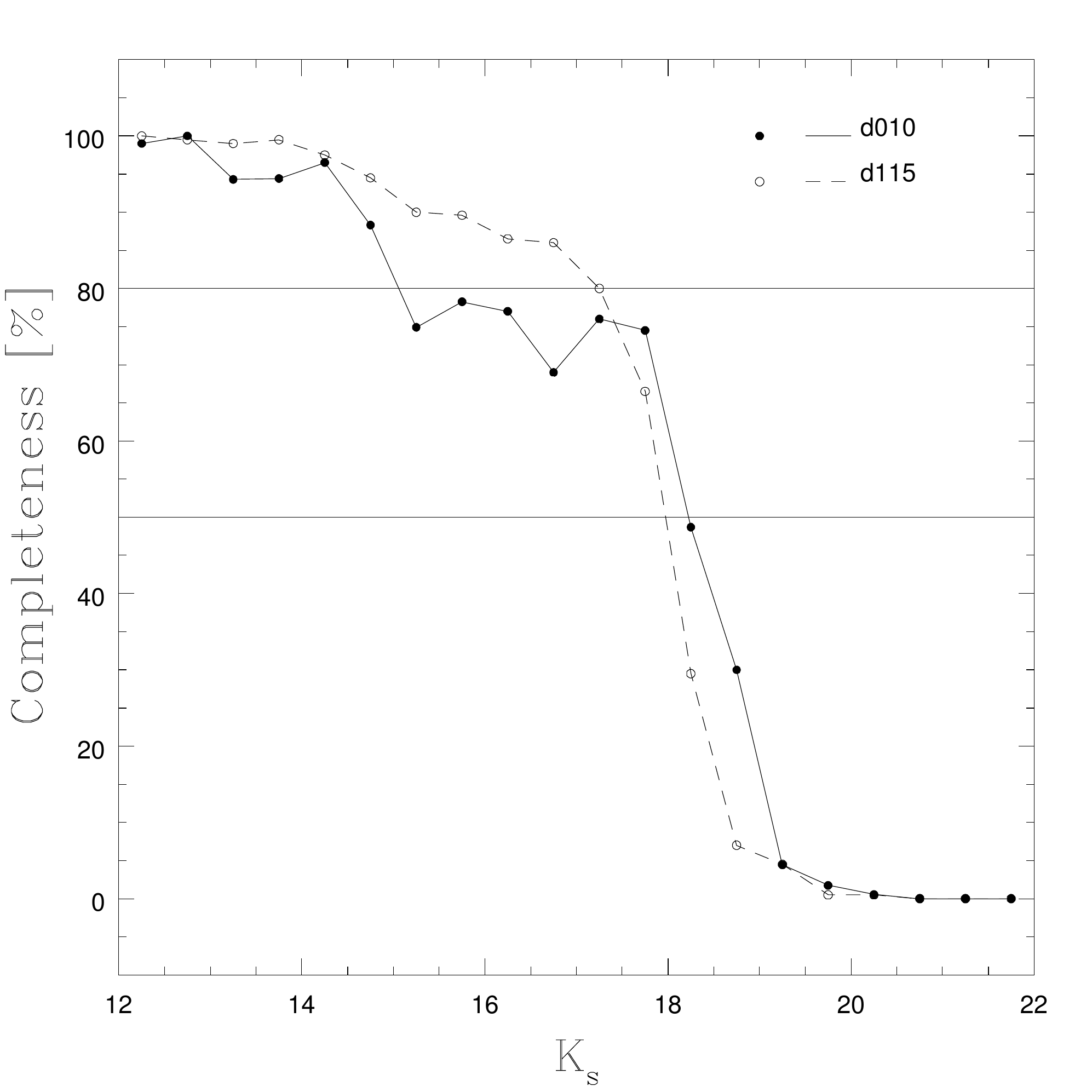}
  \caption{Completeness in percentage for input K$_s$ magnitudes of
    simulated point and galaxy detections.  Black dots and solid line correspond to
    the d010 tile and empty dots and dashed line to the d115 tile. It is also
    shown the 80 and 50\% completeness levels.}
	\label{completeness}
\end{figure*}

\begin{figure*}
\begin{centering}
  \includegraphics[width=80mm]{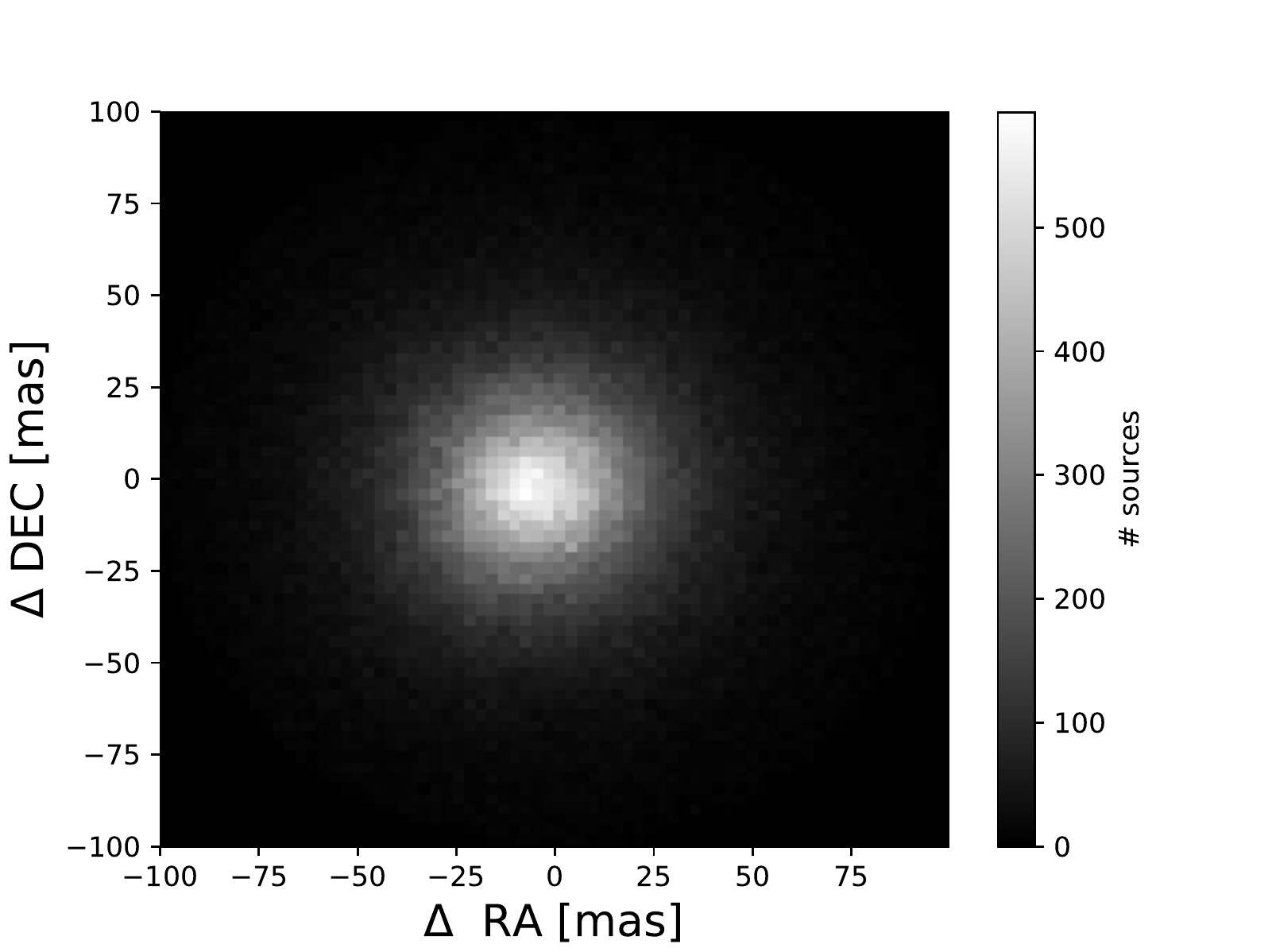}
  \includegraphics[width=80mm]{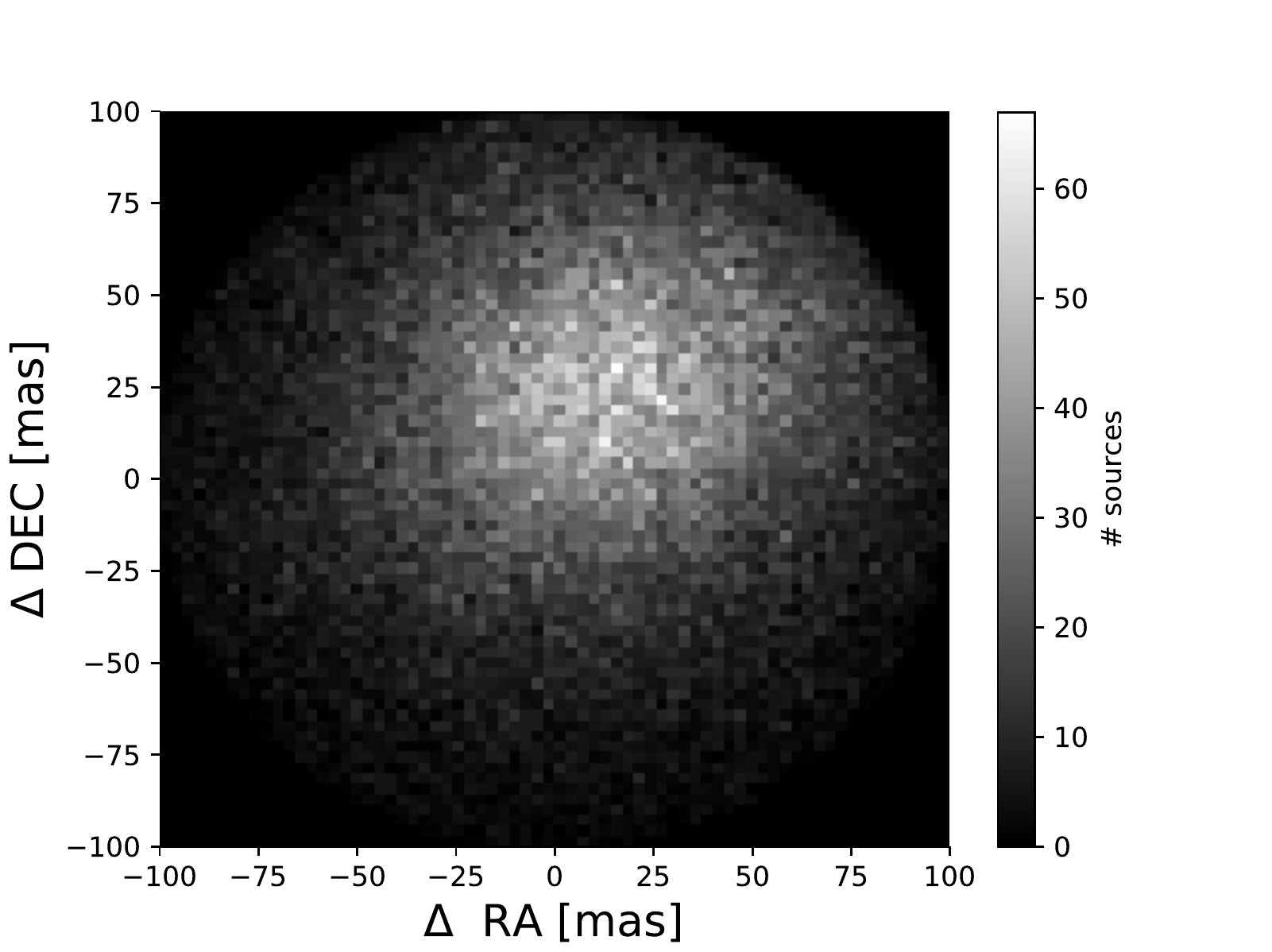}
  \caption{Density plots of the astrometric difference in gray color scale
    of the point sources detected using our 
          procedure and those of CASU.  The panels show the differences
          $\Delta$ = \dos - CASU in RA and 
Dec coordinates for the d010 (left) and d115 (right) tiles. }
	\label{detection}
\end{centering}
\end{figure*}

\begin{figure*}
 \includegraphics[width=54mm,height=52mm]{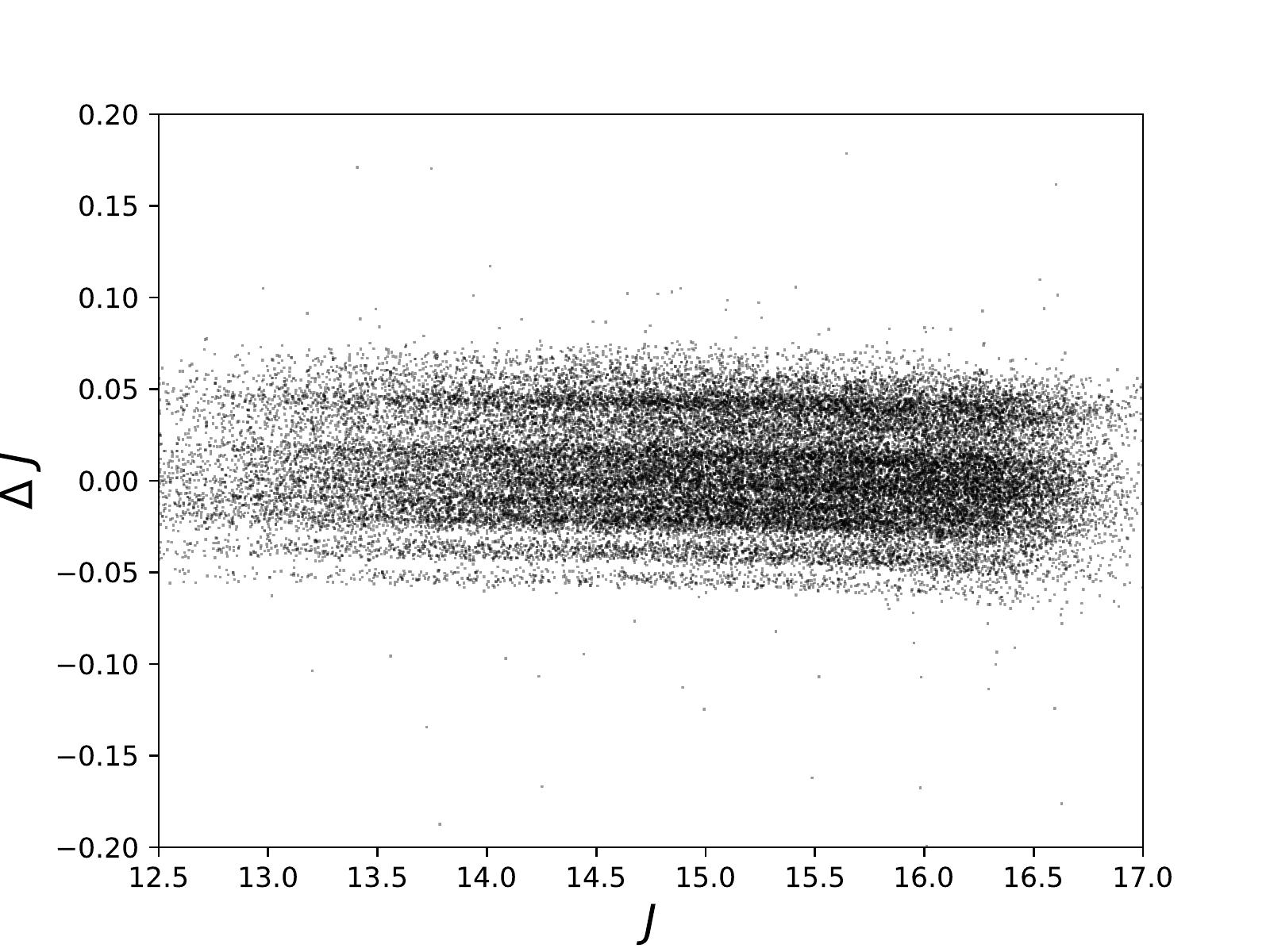}
\includegraphics[width=54mm,height=52mm]{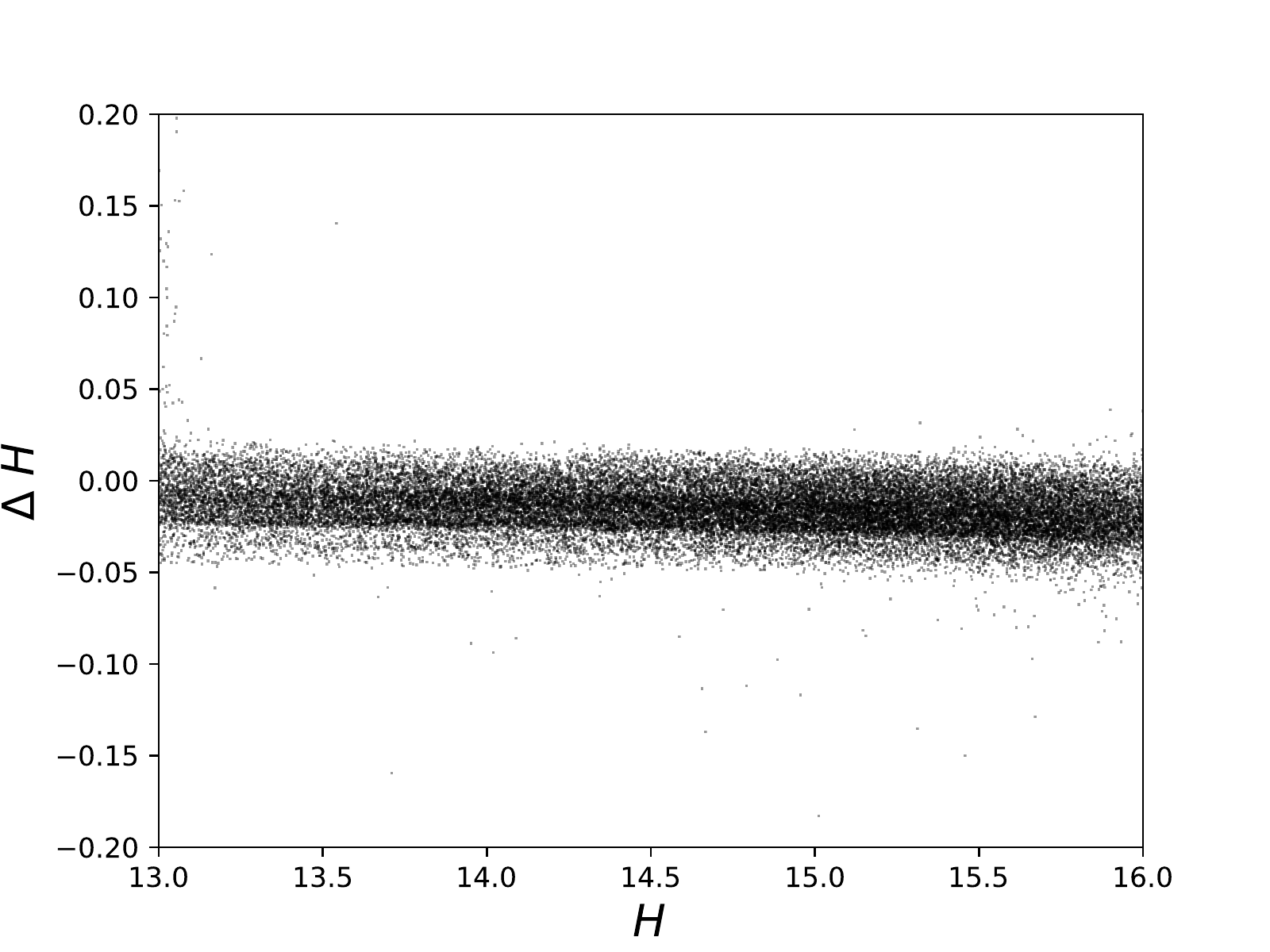}
\includegraphics[width=54mm,height=52mm]{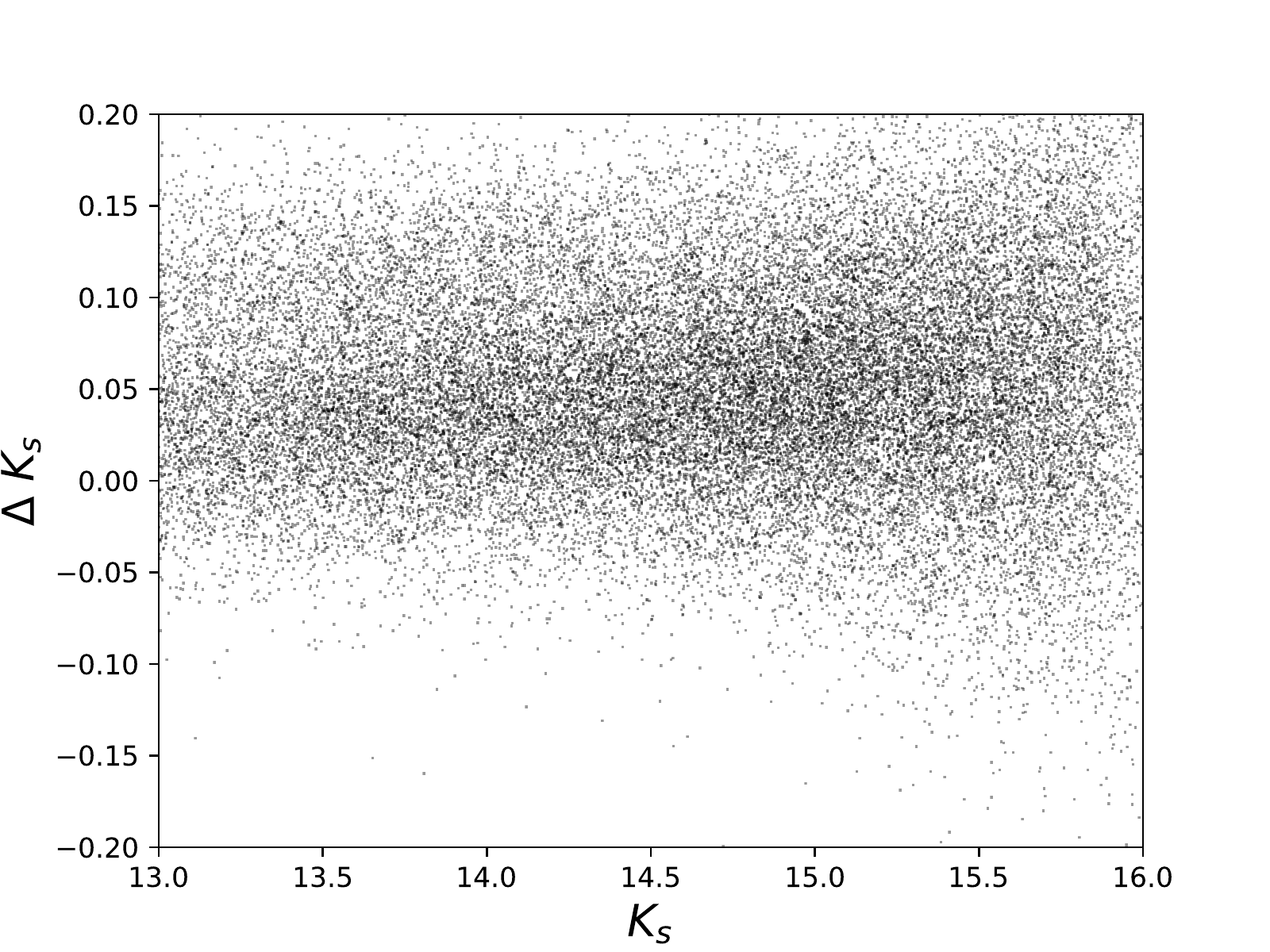}
\caption{Magnitude comparisons for point sources.  The panels show
  the differences in
magnitudes between \dos and CASU for $\Delta$J, $\Delta$H 
and $\Delta$K$_s$ (left to right) as a function of our magnitudes.}
	\label{mag}
\end{figure*}

Based on small images previously 
processed 
by {{\tt SExtractor}, \psf automatically selects non-saturated stars to compute
  different PSF models.
In our study we adopted a 20$\times$20 pixel kernel, and followed variations 
to the 3$^{rd}$ order.  The PSF is modelled as a linear combination of basis
functions as the Gaussian, or normal distributions derived 
using point 
sources with an S/N $>$ 20; elongations 
higher than 0.98; and a
half-light Radius (R$_{1/2}$, the radius that 
encloses 50\% of the object total flux) in the range of 0.5 to 0.9 arcsec.  
The best PSF model was a 
two-dimensional modeling which minimized the $\chi^2$ goodness between the 
observed flux distribution and the model.  It was applied to the images
to perform PSF photometry, with the magnitudes being estimated by integrating
the sources over the model.

\subsection{\bf{SExtractor and PSFEx} }

Briefly, the photometric pipeline  consists of three  steps: 

	\begin{enumerate}
		\item {\tt SExtractor} creates catalogs,
		 which include the position of the objects, the morphology, 
some flags and small images associated with each detection.
		\item {\tt PSFEx} takes the  \se catalogs and creates 
the best PSF model, by looking for unsaturated, circular,
		isolated and well defined objects, in order to model the PSF of each point source.  
	      \item Finally, {\tt SExtractor} applies the PSF model to
                each source to obtain the astrometric, photometric
                and morphological properties. 
	\end{enumerate}

        \dos was used in the d010 and d115 tiles of the VVV survey, and we  
detected 752,233 and 310,283  sources, respectively. 
 The results include astrometric and photometric data 
 such as coordinates and PSF magnitudes and  circular aperture magnitudes
 within a diameter of 
2 arcsec  that allow us to have a lower stellar contamination.  The PSF magnitudes are total magnitude estimates
obtained as an improvement of \se MAG\_AUTO magnitudes taking into account
the flux wings (Annunziatella et al. 2013).   With the total flux \dos also
give the R$_{1/2}$, the half-light radius and the ellipticity.  Also, for the
spheroid Sersic index (n, Sersic 1968), \dos compute the model for a single
Sersic component convolved with the PSF model.  The concentration index (C,
Conselice, Bershady \& Jangren 2000) is calculated as the ratio of two circular
radii that contain 80 and 20 percent of the total Petrosian flux.
The colors are obtained using the circular aperture magnitudes.
  All these parameters are used 
to separate stellar and extended objects, and also
to characterize the photometric
and morphological properties of the extended sources.

\section{Comparison with CASU}

\subsection{Completeness and detection efficiency}

  We determined confidence levels and limiting magnitudes of our photometric
  procedure using all the detections obtained with {\tt SExtractor+PSFEx}.
  We have
used the Completeness Estimator of source extraction (ComEst)
by Chiu, Desai \& Liu (2016) to obtain the completeness of our detections.  
This program
derives the detection rate of synthetic point sources
and galaxies on optical and near-IR images.  The observed sources detected
by \se were first removed from the images creating a
source-free image.  Then, artificial sources were added to this image.
Simulated galaxies were created using
GalSim (Rowe et al. 2015), which considered a 
mixture of bulge and disk components convolved with PSF models
varying the half-light radius,
the major-to-minor axis ratio, and the position angle in a range of K$_s$
magnitudes between 14 and 20 mag.  The sources are randomly 
placed on the source-free images assuming a number density of 15 galaxies
per arcmin square. Figure~\ref{completeness} shows the K$_s$ passband
completeness of the two studied tiles. We find that
the source detection efficiency reaches 50\% for point and extended sources
at about K$_s$ = 18.0 mag.  This is in agreement with
the CASU completeness results of Saito et al.
(2012).   The different behavior for the two
tiles are related to the extinctions shown in Figure~\ref{extinction}.

\subsection{Astrometry and photometry comparison}

 In order to check our photometric procedure, in this section,  we compare 
the positions and magnitudes of the objects detected using the \dos photometry 
with those of CASU.   Point sources were selected from the CASU catalogs\footnote{
    http://horus.roe.ac.uk/vsa/}
  (flag=-1) and
  \dos catalogs (CLASS\_STAR higher than 0.9).  We cross-matched the sources
  in both catalogs
  within a distance radius of 0.1 arcsec and we used circular aperture
  magnitudes within a diameter of 2
  arcsec considering only  errors in the photometry smaller than 0.1 mag.  
  Both magnitude systems are aperture corrected.

 Figure~\ref{detection} shows the comparisons in positions of detected point
sources as density plots for the two tiles using both the \dos and CASU procedures.
There are small offsets with 
median differences of $\Delta$RA cos(Dec) =  -4.38 $\pm$ 0.04 $mas$ and
10.12 $\pm$
0.11 $mas$, and $\Delta$Dec = -2.30$\pm$ 0.04 $mas$ and 19.19 $\pm$ 0.11 $mas$,  for the
d010 and d115 tiles, respectively. The astrometric accuracy is quite good, and the position uncertainties
are lower than 175 $mas$ for K$_s$=18.0 mag reported by Saito et al. (2012) for the VVV survey.

Figure~\ref{mag} shows the differences for point source detections
in the d115 tile between our  circular aperture J, H, and K$_s$ 
magnitudes and those of CASU  as $\Delta$J, $\Delta$H and $\Delta$K$_s$,
respectively vs our magnitudes.
  The median differences are $\Delta$J = 0.004$\pm$0.001 mag,  
$\Delta$H = -0.014$\pm$0.001 mag and 
$\Delta$K$_s$ = 0.046$\pm$0.001 mag. Similar results have been
obtained for the d010 tile.
 The comparisons show some separate offsets between
  the magnitude systems, specially for the J passband.  This effect is related
  to the CASU tiling process (Gonz\'alez-Fern\'andez et al. 2017).  The PSF
  variations along the tiles are shown as systematic spatial offsets in the
  photometry.  The CASU photometry takes this into account using a variable
  aperture correction.  The differences with our procedure are lower than
  0.1 mag, that is under our uncertainties.   
Figure~\ref{colour} shows the same differences in J, H and  K$_s$ aperture
magnitudes compared with our colors.   The tiling effect is also present in
  these comparisons, which are better than 0.2 mag with
no clear color dependence.

 From the comparisons between CASU and \dos photometry, 
 we are confident in using
  the \dos procedure 
  for the search and analysis of extended objects in the VVV
  survey.

\section{Searching for extragalactic objects}

In this section, we present
the results of our search of extended objects detected using 
\dos on the two studied VVV tiles.  After automatic
identification of these candidates using their morphological and 
photometric properties, we performed a visual inspection 
to check the detection quality.

\subsection{Object classification}

 We classified the detected objects and we divided them in point and
  extended sources using a combination of 
four morphological parameters provided by {\tt SExtractor+PSFEx}:  
the {\tt CLASS\_STAR} index; R$_{1/2}$; the concentration index, C; and the {\tt SPREAD\_MODEL ($\Phi$)} parameter.

The {\tt CLASS\_STAR} is a stellarity index associated with the light 
distribution of the source, which depends on the FWHM and the pixel scale of
the 
image and ​​ranges from 1 (for perfectly 
circular objects such as stars) to 0 (for extended objects, for example
galaxies). 
On the other hand, $\Phi$ is a normalized linear 
simplified discriminant included in the new 
\se versions as another star-galaxy classifier. 
The use of neural networks and the {\tt SPREAD\_MODEL} parameter for object 
classification (Annunziatella et al. 2012) has produced deep catalogs,
with a good star-galaxy separation as in the case of Desai et al. (2012).  
The parameters R$_{1/2}$; and C have been defined previously in $\S$ 3.2.

Taking into account the morphological properties of the objects
detected simultaneously in the
J, H and  K$_s$ passbands, our
adopted criteria to define extended objects were: ${\tt CLASS\_STAR} < 0.3$;
1.0 $<$ R$_{1/2} < 5.0~arcsec$; 2.1 $<$ C $<$ 5; 
and $\Phi > 0.002$. In this way, we obtained 17,889 and 7,055 
extended objects for d010 and d115 tiles, respectively, which
represents about 2.4\% of the total detected objects.
Figure~\ref{criterios} shows for all detections (from left to right), 
the parameters: R$_{1/2}$; concentration index; {\tt SPREAD\_MODEL} and
{\tt CLASS\_STAR} as a function of the K$_s$ magnitudes without any extinction
correction.
Upper (bottom) panels display the
results for the d010 (d115) tile.   Gray points represent all detections and
black points are
the extended sources that satisfy our adopted criteria.
The morphological parameters are better defined for sources in the d115 tile, related to 
lower extinction values with a better defined distribution found for this tile and shown in
Figure~\ref{extinction}.

\subsection{The color criterion}

 The {\tt SExtractor+PSFEx} combination extracts sources above  a defined threshold} in surface
  brightness.  Some extended sources are detected in the near-IR passbands at
  longer
wavelengths (J, H and K$_s$ passbands), and they are very faint or not even
visible at shorter wavelengths 
(Z and Y passbands). In contrast, the stellar objects were clearly visible in the 
five passbands of the survey.
Figure~\ref{gxs_star} shows the Z, Y, J, H and K$_s$ images of two
examples of extended sources with detections in three
(upper panels) and five (bottom panels) passbands.

 \dos may confuse some faint objects, 
with  
clumps of stars or star associations in the MW and they might be interpreted
as single objects. In order to  reduce the number of
false detections and to better characterize the extended objects, we added
the color as
an additional criterion.  The magnitudes  and colors were first
corrected by extinction along the line-of-sight using
the maps of Schlafly \& Finkbeiner (2011) and the
relative extinctions  of Catelan et al. (2011) for the VVV IR passbands.

At lower latitudes, the contamination by foreground stars is important.
 Williams, Kraan-Korteweg \& Woudt (2014)
        used the KILLALL routine
        (Buta \& McCall 1999) for star subtraction.  We performed a visual inspection as a
cross-check of the automatic 
identification and classification of extended sources
obtained by {\tt SExtractor+PSFEx}.  About 10\% of these objects have
the presence of nearby stars that affect the magnitudes.
We implemented a similar process
  based on DAOPHOT routines in IRAF\footnote{IRAF: the Image Reduction and Analysis Facility is distributed by the National Optical Astronomy Observatories, which is operated by the Association of Universities for Research in Astronomy, Inc. (AURA) under cooperative agreement with the National Science Foundation (NSF).} (Tody 1993): we detected stars above 4.0$\sigma$ of the
  background; we performed the 
  photometry and PSF determinations based on an appropriate star selection;
  and finally, the stars were subtracted from the images.  Then, we re-run
  {\dos} on
  these star-subtracted images to obtain the corrected photometric parameters.

Am{\^o}res et al. (2012) and Jarrett et al. (2000a and 2000b), have previously
used color cuts to select and separate their sources, with 
Am{\^o}res et al. (2012) reporting that the galaxy
candidates had colors 0.5 $<$ (J - H) $<$ 1.8 mag;
0.3 $<$ (H - K$_s$) $<$ 1.3 mag; 0.5 $<$ (Y - J) $<$ 1.5 mag and
  0.5 $<$ (Z - Y) $<$ 1.3 mag (their Figure 4).
Jarrett et al. (2000a) noted that in the IR the light of the galaxies was
dominated by older and 
redder stellar populations, and defined a color score to separate 
extragalactic sources from stars, with the former having redder colors and a
more 
extended appearance than other objects.  Jarrett et al. (2000b) confirmed by
spectroscopy the extragalactic nature of their extended
sources, which had intrinsically red colors, (J - K$_s$) $>$ 1.0 mag.

Figure~\ref{diagrama1} shows the Color-Magnitude and Color-Color 
Diagrams for the
extended objects detected in three 
J, H and K$_s$ passbands in the d010 (upper panels) and d115
(bottom panels) tiles, which are
represented by small dots.   We visually found the presence of false detections,
mainly double stars, triple stars, and more complex stellar associations.
We used the following color cuts: 0.5 $<$ (J - K$_s$) $<$ 2.0 mag;
0.0 $<$ (J - H) $<$ 1.0 mag and 0.0 $<$ (H - K$_s$) $<$ 2.0 mag to
eliminate these false detections based on previous analyses
of Am{\^o}res et al. (2012) and Jarrett et al. (2000a and 2000b) and our
visual inspection.  These color cuts are represented with solid lines in
the Figure~\ref{diagrama1}.  We also defined a straight line (dashed line
in the figure) and the extragalactic candidates have ''distances'' to this
line defined as (J - H) + 0.9 (H - K$_s$) $>$ 0.44 mag.  These objects 
 are represented by larger circles in
the figure.  This additional
constraint is similar to the color score defined in Jarrett et al. (2000a).
There are some objects with reddest colors and they have the
strongest stellar contamination.  Their magnitudes were corrected but they
should be taken with caution.

  In total, we have 345 and 185 extragalactic candidates
detected in the d010 and d115 tiles, respectively.  
Of them, 193 (in d010 tile) and 43 (in d115 tile) were detected only
in  J, H and K$_s$ passbands.  The others,  152 (in d010 tile) and 142
(in d115 tile) were detected in the 5 passbands.  These objects should also
satisfy -0.3 $<$ (Y - J) $<$ 1.0 mag and 
-0.3 $<$ (Z - Y) $<$ 1.0 mag. Figure~\ref{color2} shows
the  Y - J vs J - H and Z - Y vs Y - J Color-Color Diagrams for these
detections in the d010 (d115) tile
  represented by open (filled) circles.

  Figure~\ref{flowchart} shows a flowchart
representing our complete selection algorithm.   
    All the extragalactic candidates were visually checked and confirm the robustness of our
    adopted selection criteria. 
    They  are, in general,  extended and widespread in the
  images and have the reddest colors.  We
may consider our visual inspection as a lower limit on the reliability of our
photometric procedure to detect extragalactic sources. However, it is important to note that the
final confirmation of the
extragalactic nature  is always given by spectroscopic data.  

The distribution of the  extragalactic candidates  in galactic
  coordinates for the two tiles is shown in Figure~\ref{lbvvv}.  We identified
  with different symbols those detections in three (filled circles) and five
  (crosses) passbands.    The 2MASSX objects in the neighborhood
of the studied regions are also included, which reveals the
lack of previous studied sources.  The two galaxies reported by Staveley--Smith et al. 
(2016) in the HIZOA-S survey are not included here since they do not have
IR counterparts.  The photometric catalog of these  extragalactic candidates
is published in its entirety in machine-readable format.  The first 10
sources are shown in Table~\ref{cat} for guidance regarding form and
content.  The table gives
the identification (column 1), the J2000 coordinates (columns 2 and 3),
the  PSF Z, Y, J, H and K$_s$ magnitudes and aperture magnitudes within
a diameter of 2 arcsec (columns 4 to 13), and the morphological parameters: R$_{1/2}$, C,
ellipticity and spheroid Sersic index (columns 14 to 17).  Column 18
  includes comments about the object morphology and, if
  present, the contamination by nearby stars. Some examples of these
detections are shown in color composed 
images in Figure~\ref{rgbd115}.

\section{Photometric properties of the extragalactic candidates  }

In this section, we analyze general properties of the extragalactic candidates
found in the
two tiles of the VVV survey.  Figure~\ref{magdist}
shows the normalized extinction corrected magnitude and
(H - K$_s$) color distributions for these extragalactic candidates.  The
  distributions for the detections in the five passbands (Z, Y, J, H and K$_s$) are
  represented with solid lines and those detections in only three passbands (J, H and K$_s$) with dashed lines.   Figure~\ref{all1} shows the
  normalized distributions of some of the
structural properties:
R$_{1/2}$, C,
ellipticity and spheroid n Sersic index (on a logarithmic scale). The histograms are represented as in the previous figure.   In general, the two distributions
  are similar.  In the K$_s$ distribution, the detections in the three passbands
are slightly shifted towards brighter K$_s$ magnitudes.  In the C
distribution, there are more detections in the three 
passbands for C $>$ 3.5 and more detections in the five 
passbands for C $<$ 3.5.

In the studied regions, there is no extragalactic sources with IR data coming
from other surveys and a direct comparison cannot be
made.  However, we have IR properties of extended objects behind the MW plane
from the studies of HI galaxies as Williams, Kraan-Korteweg \& Woudt (2014)
and Said et al. (2016b).  Their main goal was to perform surface
photometry on the star-subtracted images to produce deep near-IR catalogs,
2 magnitudes deeper than 2MASS from HIZOA galaxies.  They obtained
ellipticities, isophotal magnitudes and extrapolated total magnitudes
of 578 galaxies with recession velocities out to 6000 km~s$^{-1}$ (Williams,
Kraan-Korteweg \& Woudt 2014) and 674 galaxies with confirmed counterparts in the HIZOA catalogs
(Said
et al. 2016b).  
Comparing their results with our distributions,  our sample of  extragalactic sources  contains fainter objects
than that of late-type galaxies in the HIZOA survey.
 For ellipticities smaller than 0.3, there is a slight increase of
extragalactic sources  with five passband
  detections and, on the contrary, the other distribution has an increase
  for higher values. 
 Our results are different from Said et al. (2016b), who found similar number of objects with
ellipticies between 0.2 to 0.6 
  but their selection
 favored late-type galaxies.  In general, our sample
 had smaller and more circular objects,  mainly due to the near-IR
   sensitivity to select more compact objects with higher surface brightness.

Andrews et al. (2014) analyzed three near-IR surveys:
2MASS, UKIDSS and VIKING and made a comparison of some
structural measurements, such as the Sersic index. 
For optical wavelengths, the distribution of the Sersic indices is bimodal 
(Kelvin et al. 2012), with peaks centered at 1
and 3.5 that correspond to late and early-type
galaxies, respectively. The near-IR is more sensitive to
the older stellar population in galaxies and the two peaks are less
defined as shown in Andrews et al. (2014). The Sersic index distributions
found in this study are similar to those of previous analyses.  As a conclusion,
the extragalactic sources 
 are found, in general, smaller; with R$_{1/2}$
values having a peak at about 1.3 arcsec; with
concentration indices
between 2.5 to 3.5; ellipticities ranging from 0.1 to  0.6;  and the Sersic
index varying from 1 to 7, with a peak around 4.
This latter index was mainly associated with bulges
with a peak of around 4.  Lower n values corresponded to late-types whereas
higher values indicated more concentrated objects.  

Table~\ref{medianas} summarizes the median 
values of PSF and aperture near-IR magnitudes,  colors, and some structural 
parameters 
obtained for the extragalactic sources detected in three and five
passbands. In general, these results
are similar between the two sets of detections. 

\section{Comments and Future Plans}

The VVV  
is an ESO near-IR variability survey of the MW, whose goal is to
study the stellar objects in the Galaxy, mainly variable stars.  This has
provided us with the opportunity to search
for extragalactic sources that are hidden by the presence of our Galaxy.

Herein, we described our procedure to use the VVV high quality images 
to search for extended objects and we utilized 
photometric parameters based on Point-Spread 
Function fitting photometry.   
The photometric pipeline is the combination of 
\se and {\tt PSFEx},  and was applied for the first time to the
tiles d010 and d115 of the survey, thereby obtaining astrometric, photometric and 
morphological parameters of the detected sources.  

Our results for stellar objects were compared with those provided by 
CASU.  On applying the  \dos combination for the two
tiles, we found differences of $\Delta$RA $\sim$ 3 $mas$ and 
$\Delta$Dec $\sim$ 8 $mas$ in the source positions.  
 The magnitude comparison has 
 median differences  of $\Delta$J = 0.004$\pm$0.001 mag,  
$\Delta$H = -0.014$\pm$0.001 mag and 
$\Delta$K$_s$ = 0.046$\pm$0.001 mag without any dependence on color.  

Using {\tt SExtractor+PSFEx}, we obtained photometric data of the d010 and
d115 tiles in 
the five passbands of the survey: Z, Y, J, 
H and K$_s$ passbands, which include equatorial coordinates, magnitudes, 
ellipticities, the half-light radius, fitting model and morphological
parameters.  In order to define  extragalactic candidates, the adopted
criteria include the structural parameters 
${\tt CLASS\_STAR} < 0.3$, 1.0 $<$ R$_{1/2} < 5.0~arcsec$, 2.1 $<$ C $<$ 5, 
and $\Phi > 0.002$.  The following color criteria was also added to
eliminate false detections:
0.5 $<$ (J - K$_s$) $<$ 2.0 mag, 0.0 $<$ (J - H) $<$ 1.0 mag,  
0.0 $<$ (H - K$_s$) $<$ 2.0 mag, and (J - H) + 0.9 (H - K$_s$) $>$ 0.44 mag.

 Summarizing,  345 extragalactic candidates were found in the 
 d010 tile and 185 in the d115 tile, making a total of 530 sources detected
 in the J, H and K$_s$ passbands.  Some of these sources: 152 (in the d010 tile)
  and 142 (in the d115 tile) had also detections in Z and Y
  passbands, or a  total of 294 sources. They also satisfied
  -0.3 $<$ (Y - J) $<$ 1.0 mag and
-0.3 $<$ (Z - Y) $<$ 1.0 mag.  All of these extragalactic
  candidates were visually inspected and they are confirmed to be galaxies.
  In general, it is observed a slight increase of
  the number of  extragalactic candidates with five passband
  detections for ellipticities smaller than 0.3.  The opposite is observed
  for ellipticities higher than 0.3.
The R$_{1/2}$
distribution has a peak at a median value of about 1.3 arcsec and the Sersic
index varying from 1 to 7, with a peak around 4.  The  extragalactic sources
found in the two tiles of the VVV survey are, in general, small;
more circular and red, mainly due to the near-IR sensitivity to
   select more compact objects with higher surface brightness.

This is the first paper of a series that defines the methodology to
search  for extragalactic sources in the VVV survey.  We also plan
to apply this analysis to  other regions of the disk.  
In this sense, we are analyzing the d015 tile as a galaxy cluster
candidate has been identified (Baravalle et al. 2017) and we have
obtained spectroscopic data 
using Flamingos-2 at the Gemini 
telescope (GS-2016A-FT-18) to establish cluster membership.
It would be very useful to be able to confirm the extragalactic 
nature of other sources through spectroscopic measurements.
The new VVV eXtended Survey (VVVX) will cover from
−130$^{\circ}$ $<$ l $<$ +20$^{\circ}$ and contain new northern and southern
bulge extensions together with
a new northern disk  and a large southern disk extensions.   The whole VVVX
area will be observed using the J, H and K$_s$ passbands, and this will allow us
to search for interesting connections in the filaments linked with the
GA in order to provide information about the distribution of galaxies in
galaxy clusters and 
filaments
in this hidden part of the sky.

\vskip 0.5cm

{\bf ACKNOWLEDGEMENTS}
\vskip 0.5cm
 
We would like to thank the anonymous referee for the suggestions to improve
the manuscript. We would also thank I-Non Chiu for assisting us with the
use of the
ComEst package and Dar\'io Gra\~na for his technical support. 
This work was partially supported by grants from the Secretar\'ia
de Ciencia y T\'ecnica (Secyt) of Universidad Nacional de C\'ordoba (UNC)
 and Consejo de Investigaciones Cient\'ificas y 
T\'ecnicas (CONICET, PIP No. 112-20110101014).  LDB acknowledges the financial 
support for her PhD thesis 
from Secyt (UNC) and CONICET.  JLNC is also grateful for financial support
received from the 
Programa de Incentivo a la 
Investigaci\'on Acad\'emica de la Direcci\'on de Investigaci\'on de la 
Universidad de La Serena (PIA-DIULS), Programa DIULS de Iniciaci\'on 
Cient\'ifica No. PI15142. JLNC also acknowledges the 
financial support from the 
GRANT PROGRAM No. FA9550-15-1-0167 of the Southern Office of Aerospace 
Research and Development (SOARD), a branch of the Air Force Office of the
Scientific Research's International Office of the United States (AFOSR/IO).
JCB acknowledges support from programa ESO-C\'omite mixto, Gobierno de Chile.

We gratefully acknowledge data from the ESO Public Survey program ID 179.B-2002 taken with the VISTA telescope, and products from the
Cambridge Astronomical Survey Unit (CASU). 
DM is supported by the BASAL Center for Astrophysics and Associated Technologies (CATA) through grant PFB-06, by the Ministry for the
Economy, Development and Tourism, Programa Iniciativa Cient\'ifica Milenio grant IC120009, awarded to the Millennium Institute of
Astrophysics (MAS), and by FONDECYT No. 1170121.

\clearpage



\begin{figure*}
\includegraphics[width=54mm,height=52mm]{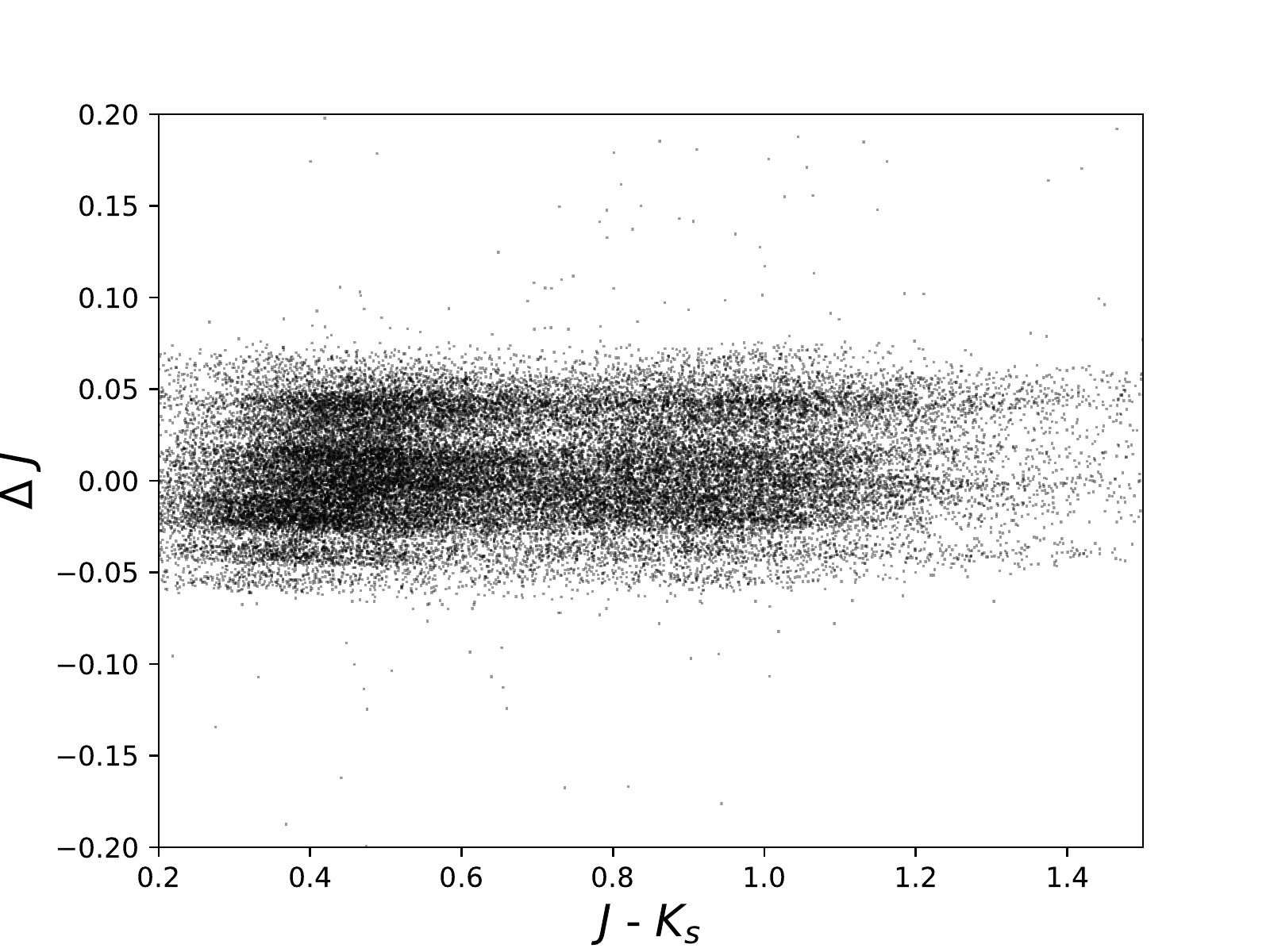}
\includegraphics[width=54mm,height=52mm]{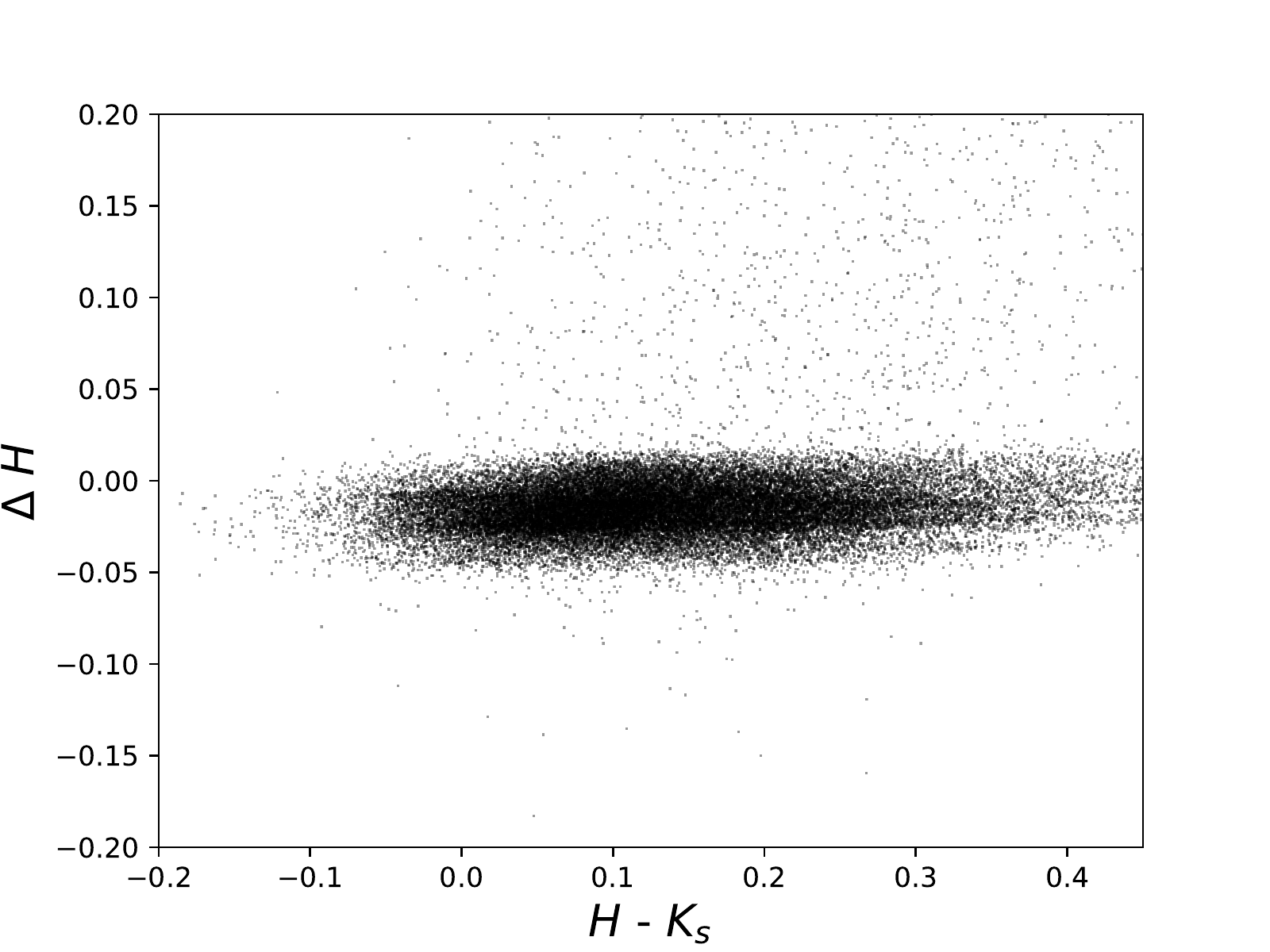}
\includegraphics[width=54mm,height=52mm]{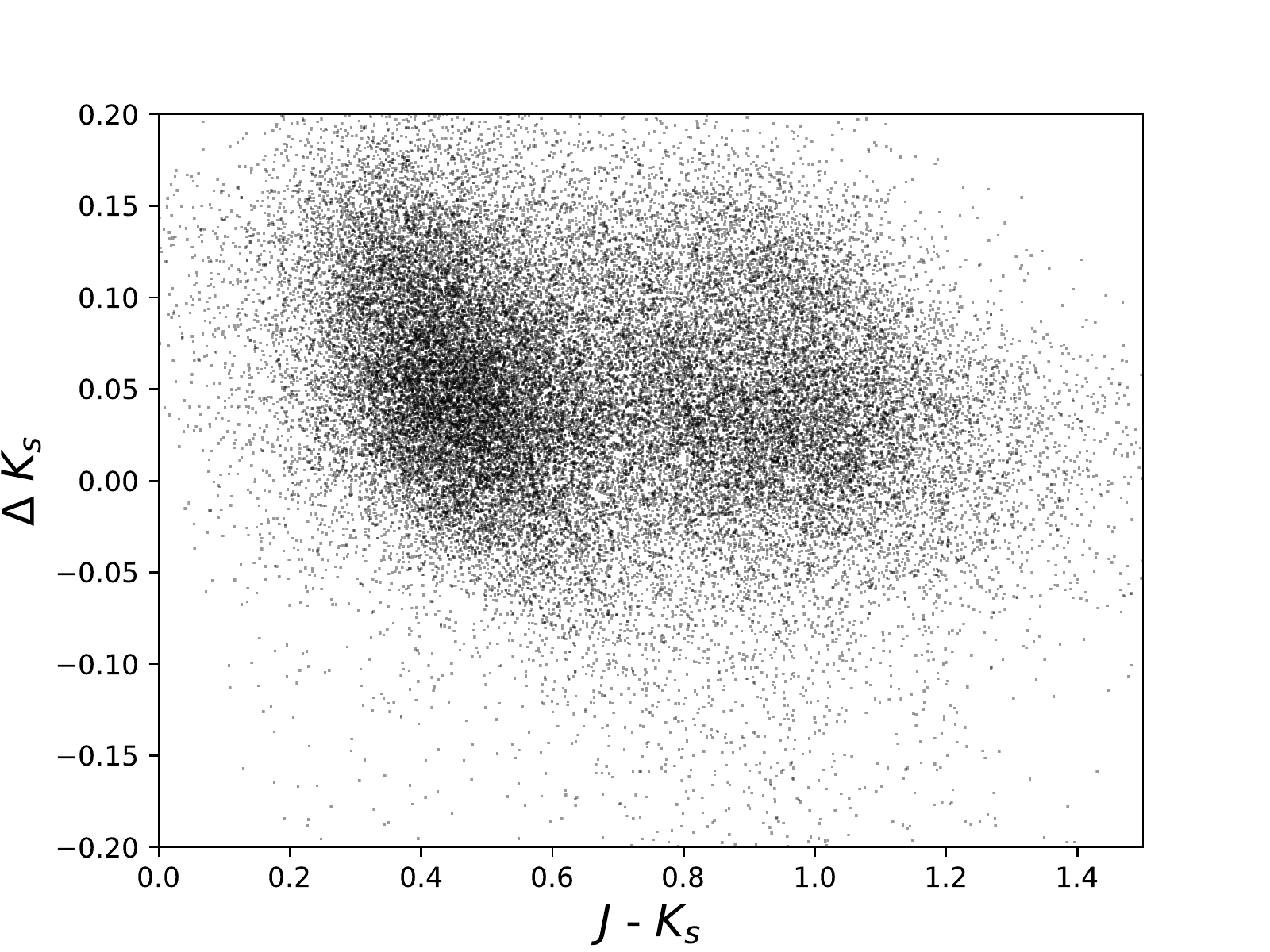}
	\caption{Color dependence of point sources.  The panels show $\Delta$J, $\Delta$H 
and $\Delta$K$_s$ (left to right) as a function of our colors. }
	\label{colour}
\end{figure*}

\begin{figure*}
\begin{centering}
            \includegraphics[width=40mm,height=40mm]{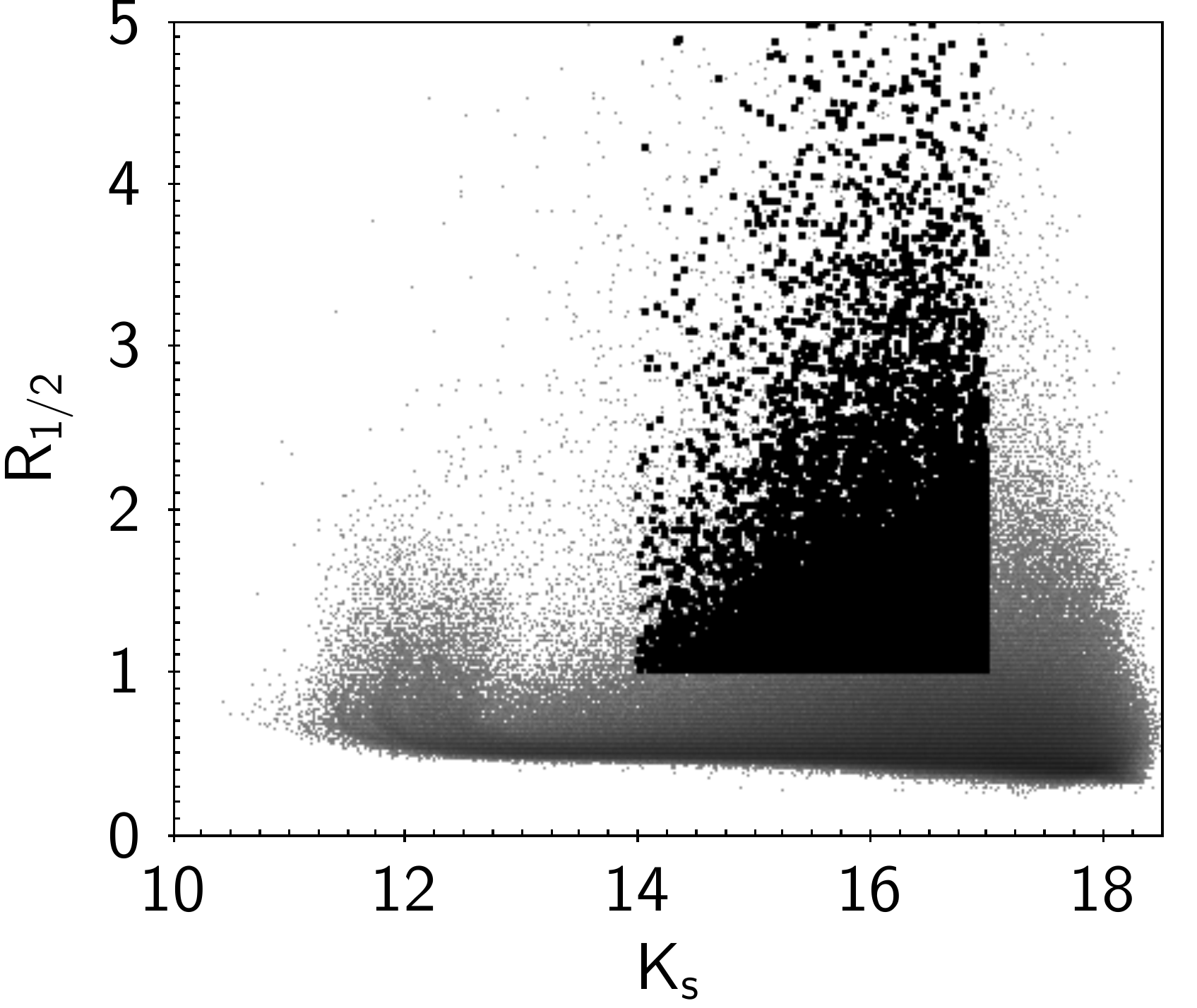}
           \includegraphics[width=40mm,height=40mm]{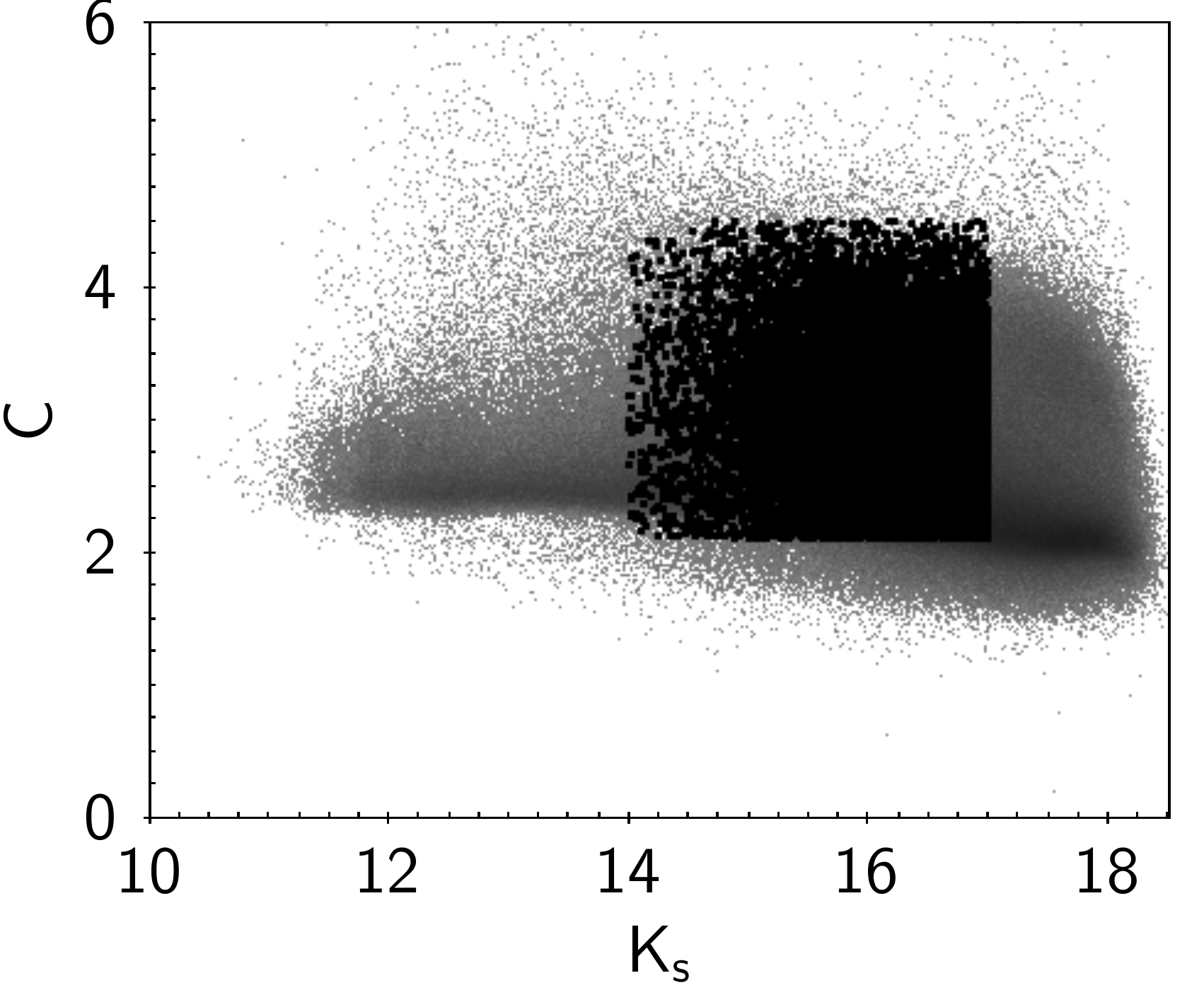}
           \includegraphics[width=40mm,height=40mm]{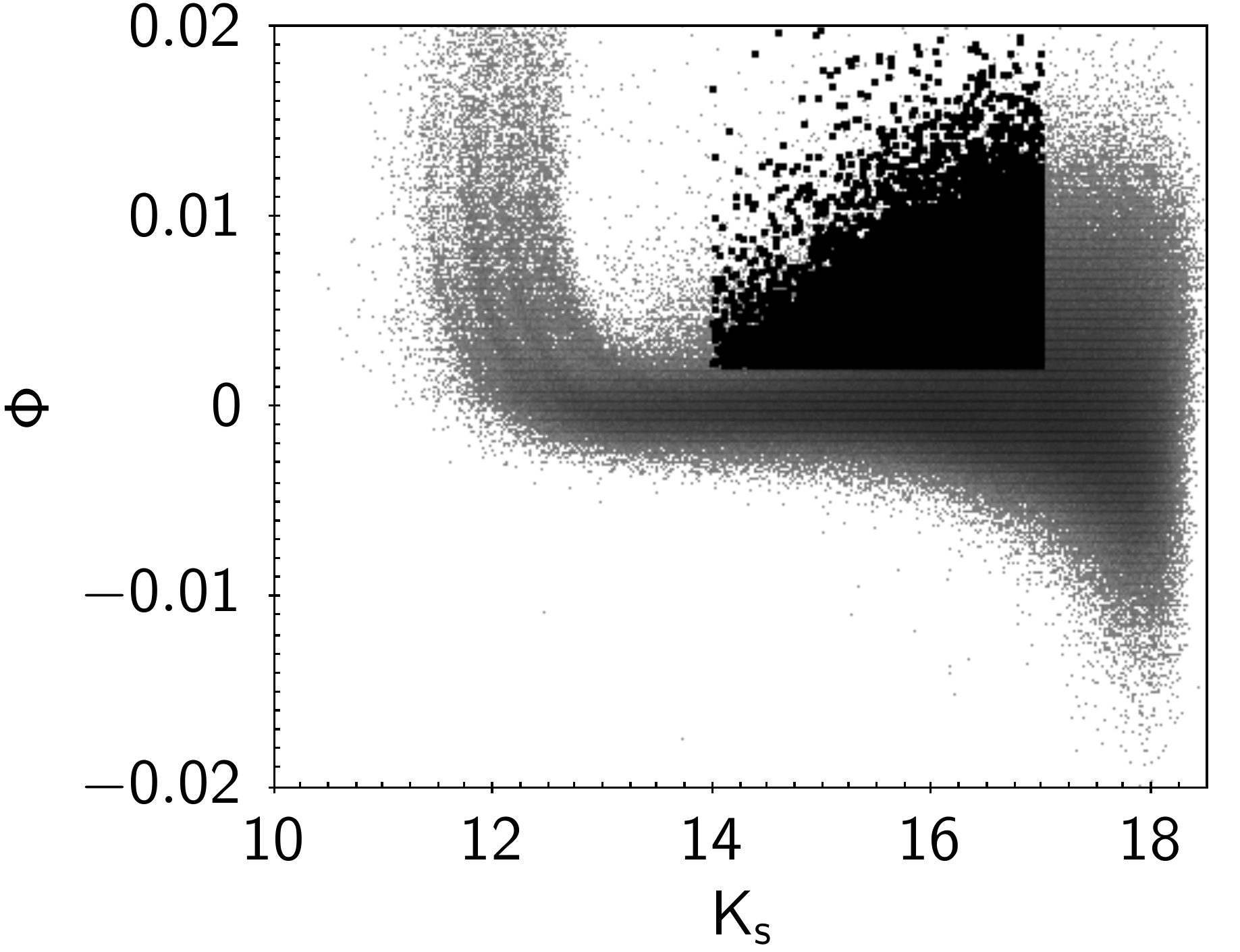}
           \includegraphics[width=40mm,height=40mm]{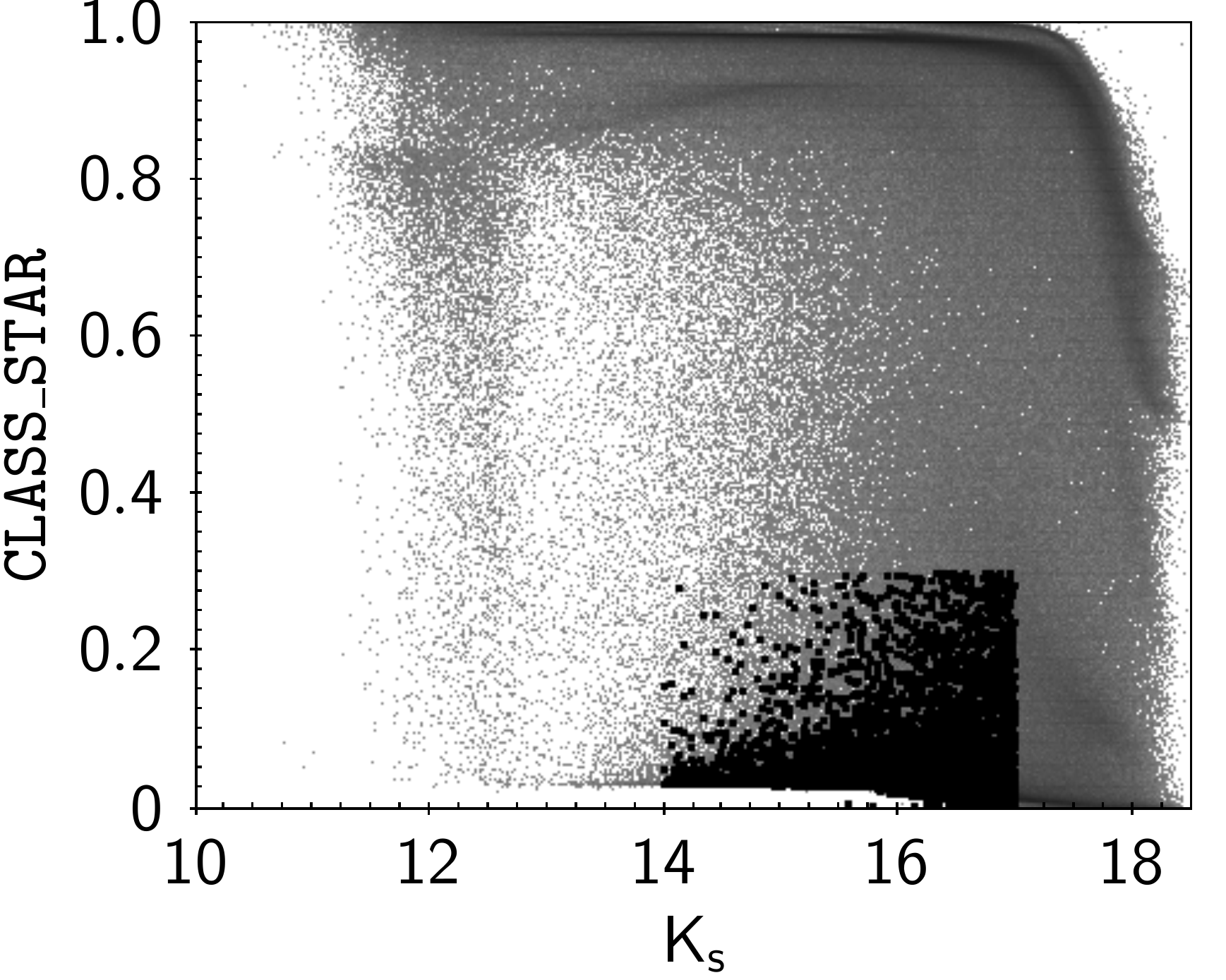}
          \includegraphics[width=40mm,height=40mm]{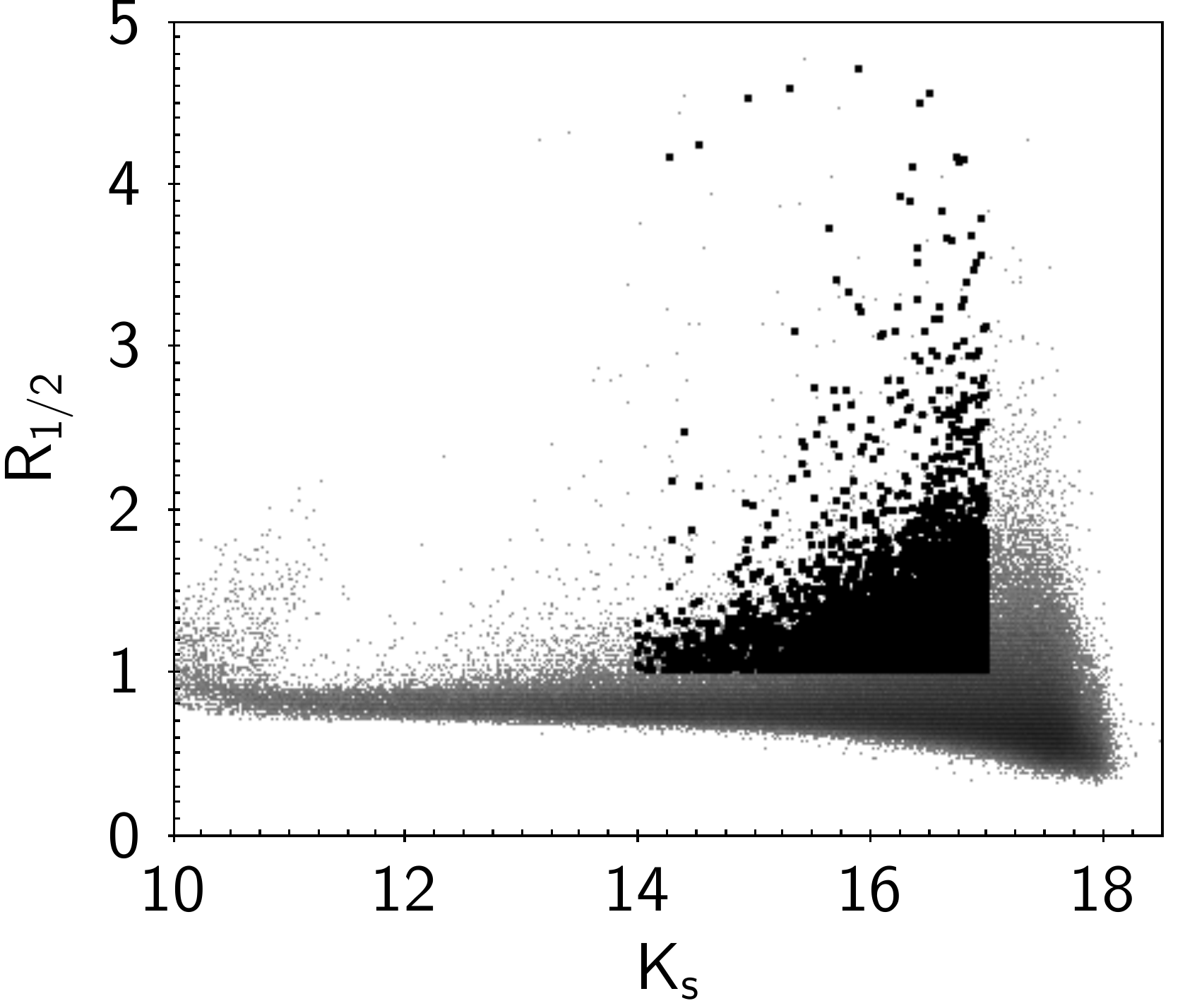}
\includegraphics[width=40mm,height=40mm]{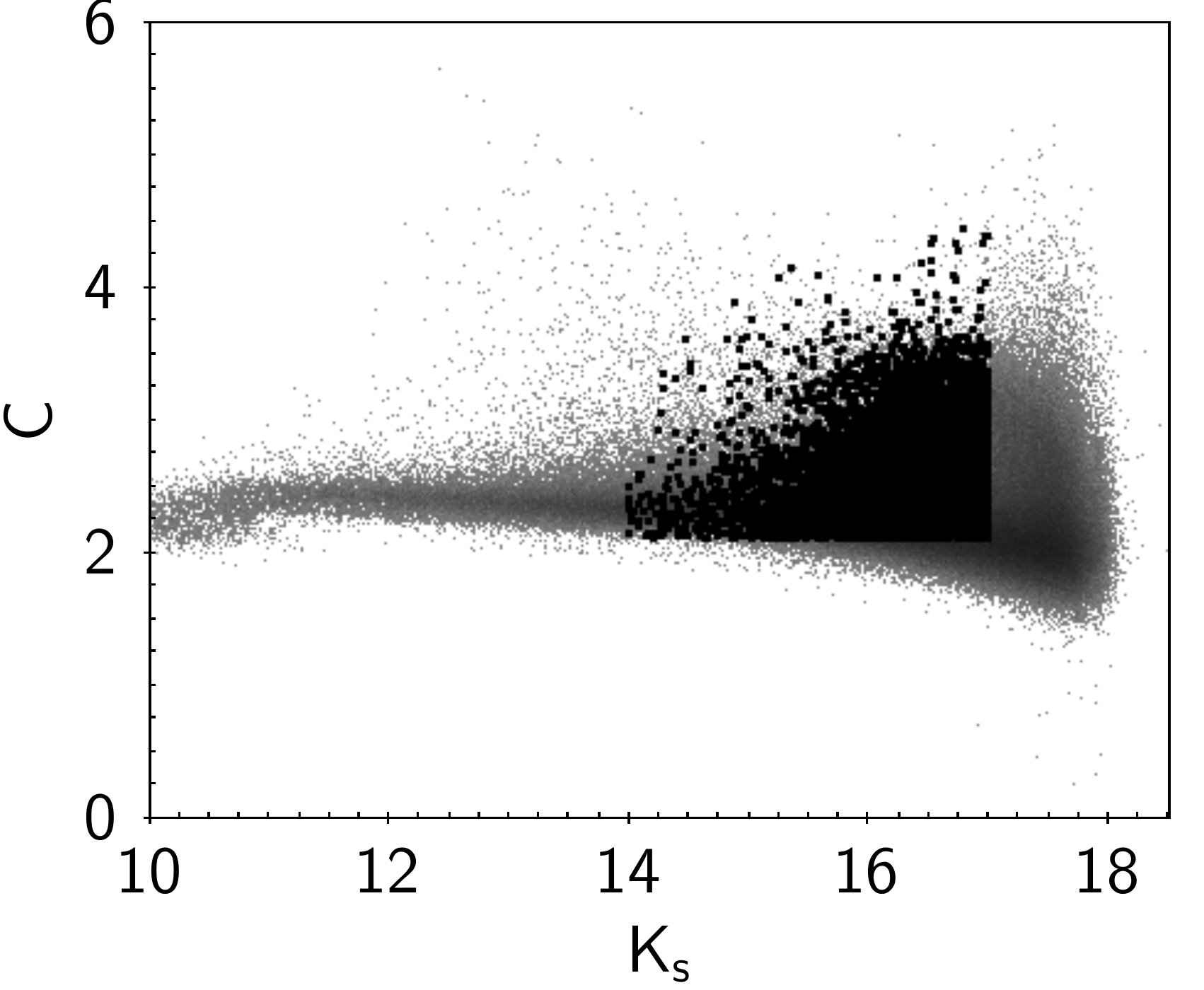}
\includegraphics[width=40mm,height=40mm]{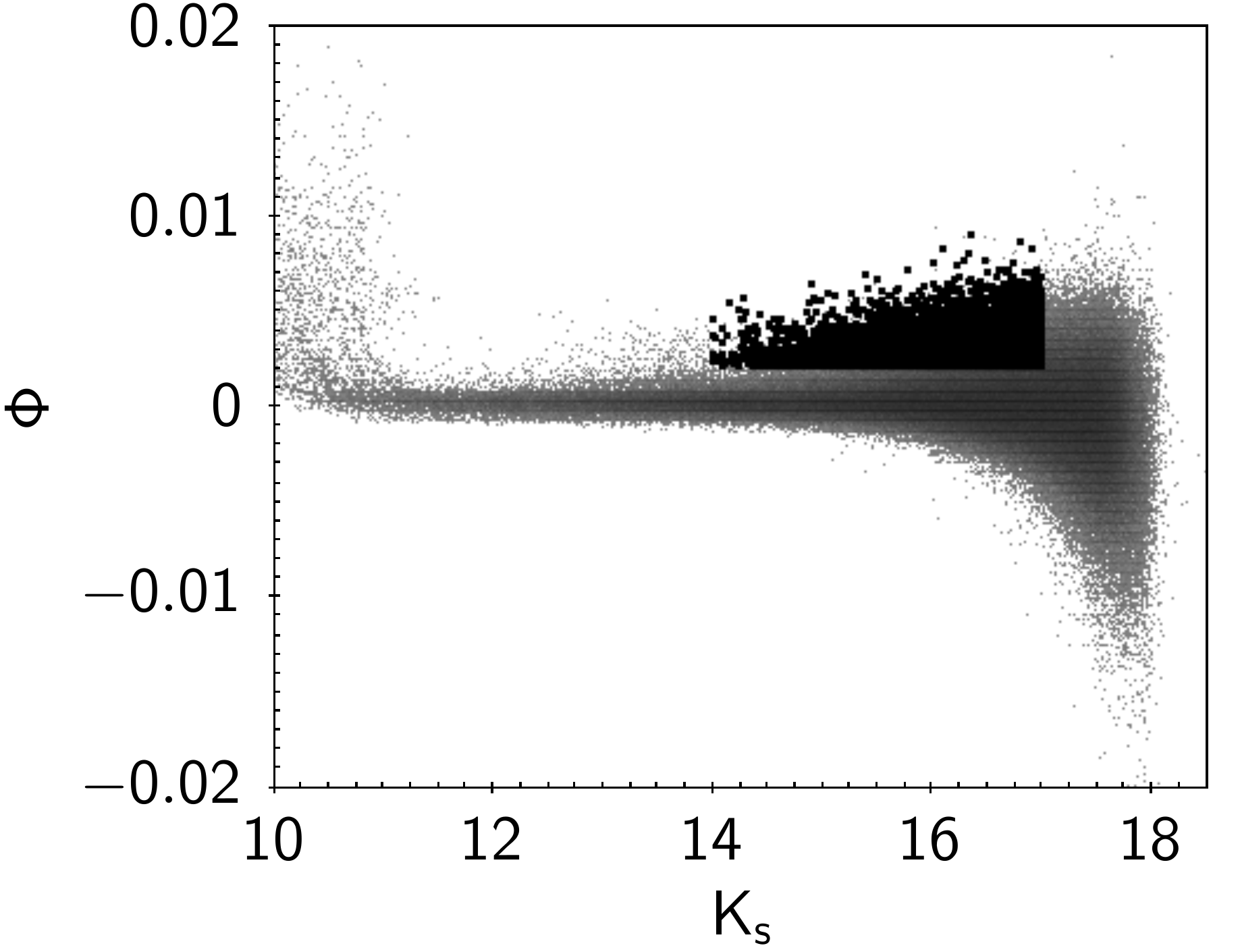}
\includegraphics[width=40mm,height=40mm]{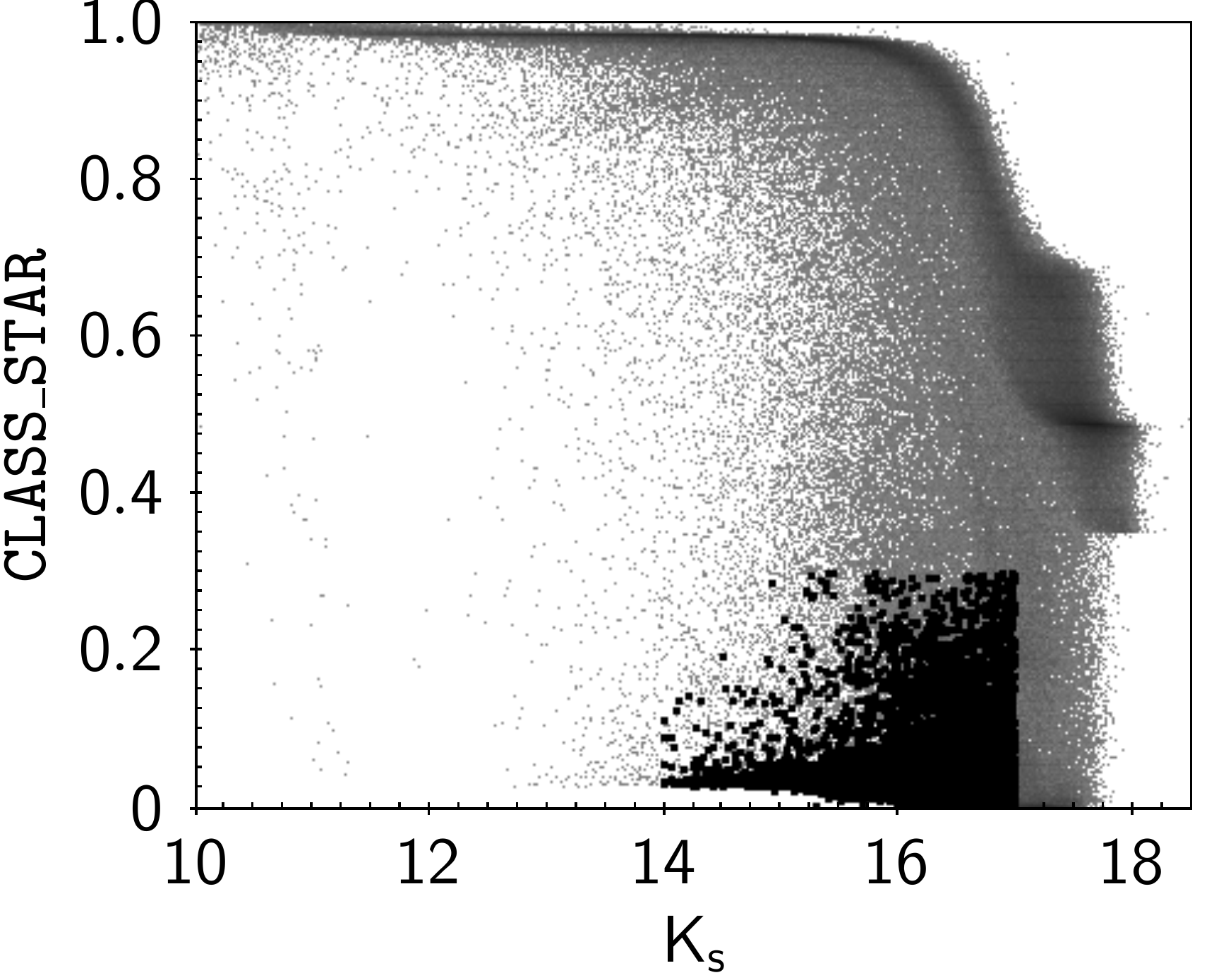}
\caption{Different parameters used to classify the objects detected
  using \dos photometry for the d010 tile (upper panels) and the d115 tile
(bottom panels). 
  Panels show from left to right: R$_{1/2} $; C;  $\Phi$; and  {\tt CLASS\_STAR} vs PSF K$_s$ magnitudes.  Black (gray) points represent the extended  (stellar)
  objects.
}
            \label{criterios}
\end{centering}
\end{figure*}

\begin{figure*}
\begin{centering}
  \includegraphics[width=180mm] {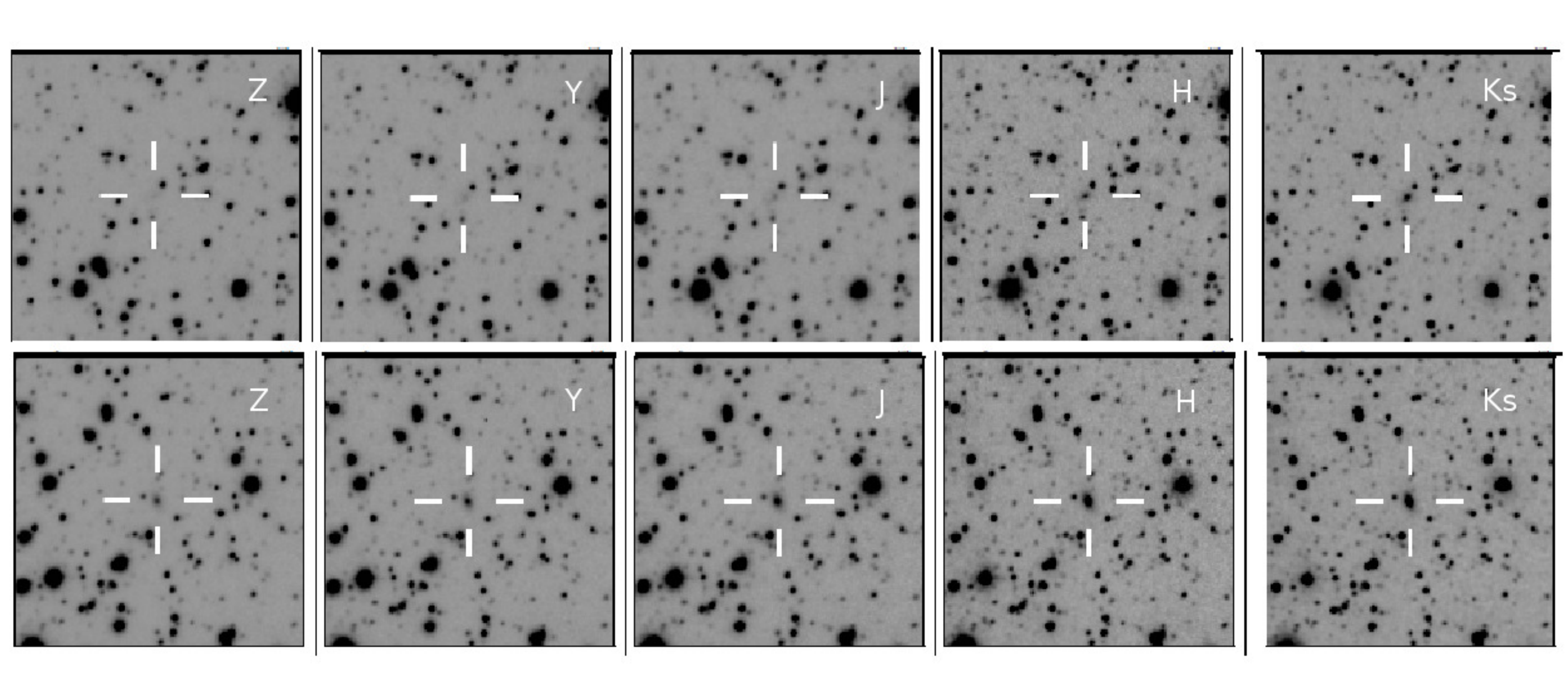}
  \caption{Examples of extragalactic sources detected in the images
  of the VVV survey.}
	\label{gxs_star}
\end{centering}
\end{figure*}

\begin{figure*}
\begin{centering}
          \includegraphics[width=70mm,height=65mm]{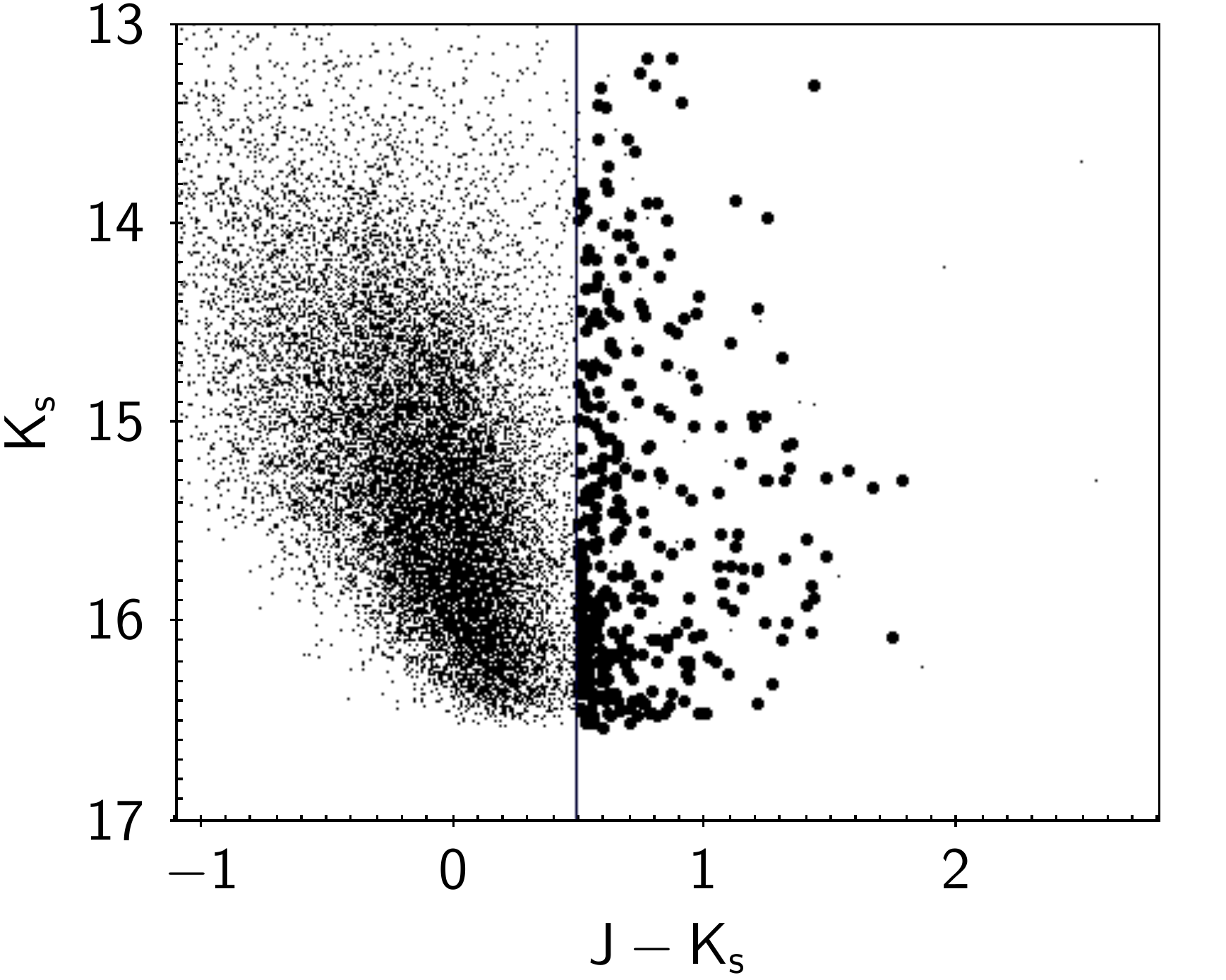}
          \includegraphics[width=70mm,height=65mm]{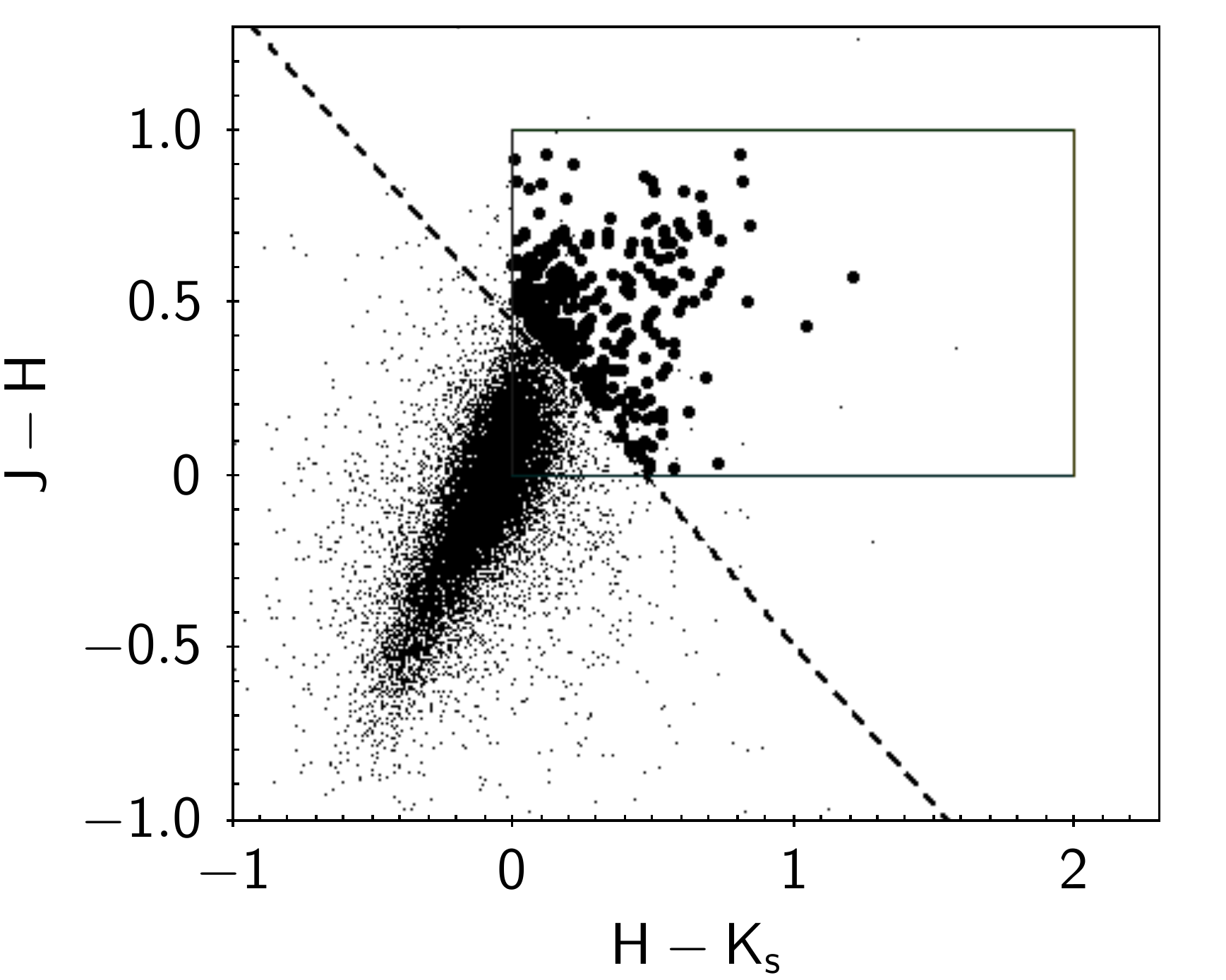}
          \includegraphics[width=70mm,height=65mm]{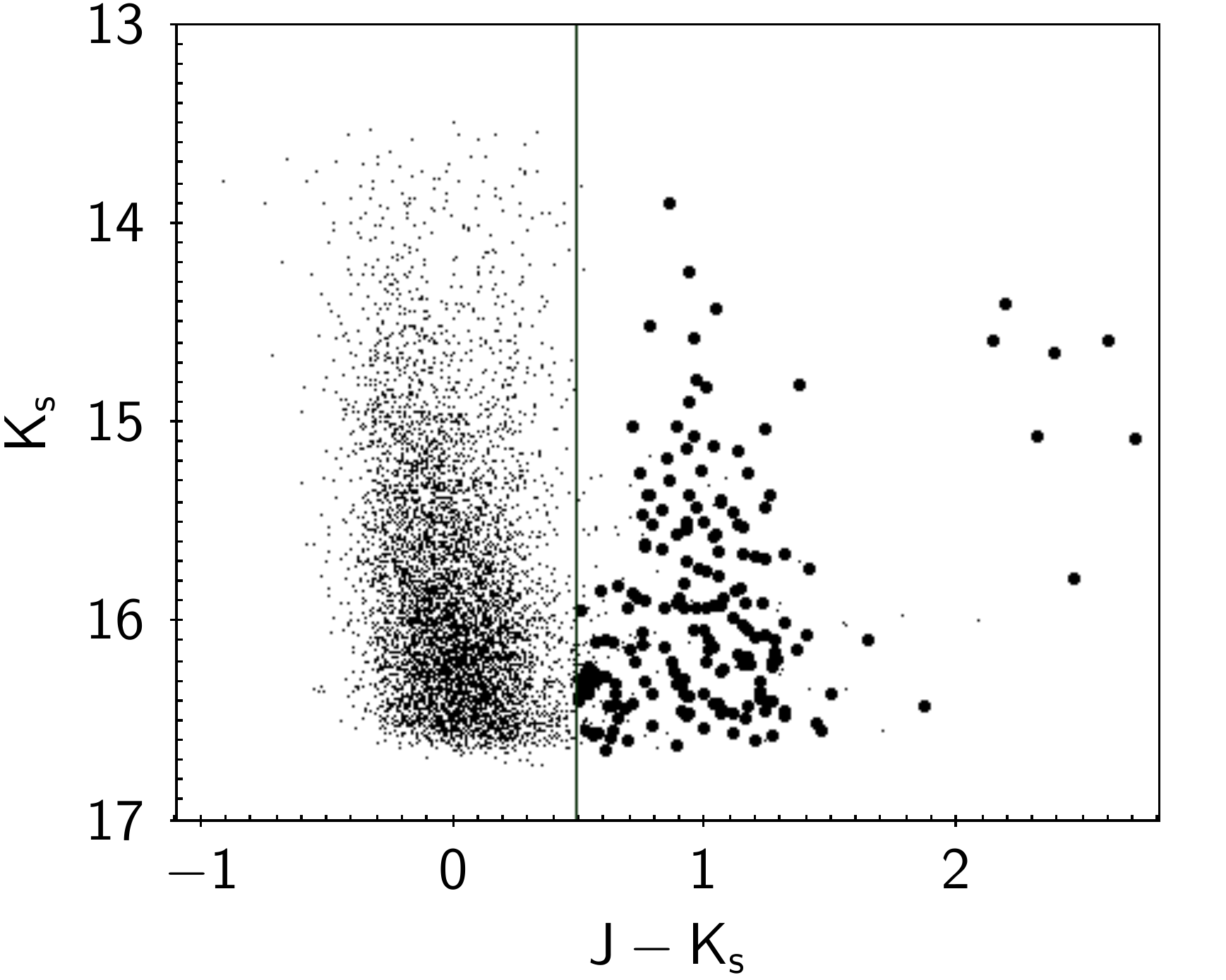}
          \includegraphics[width=70mm,height=65mm]{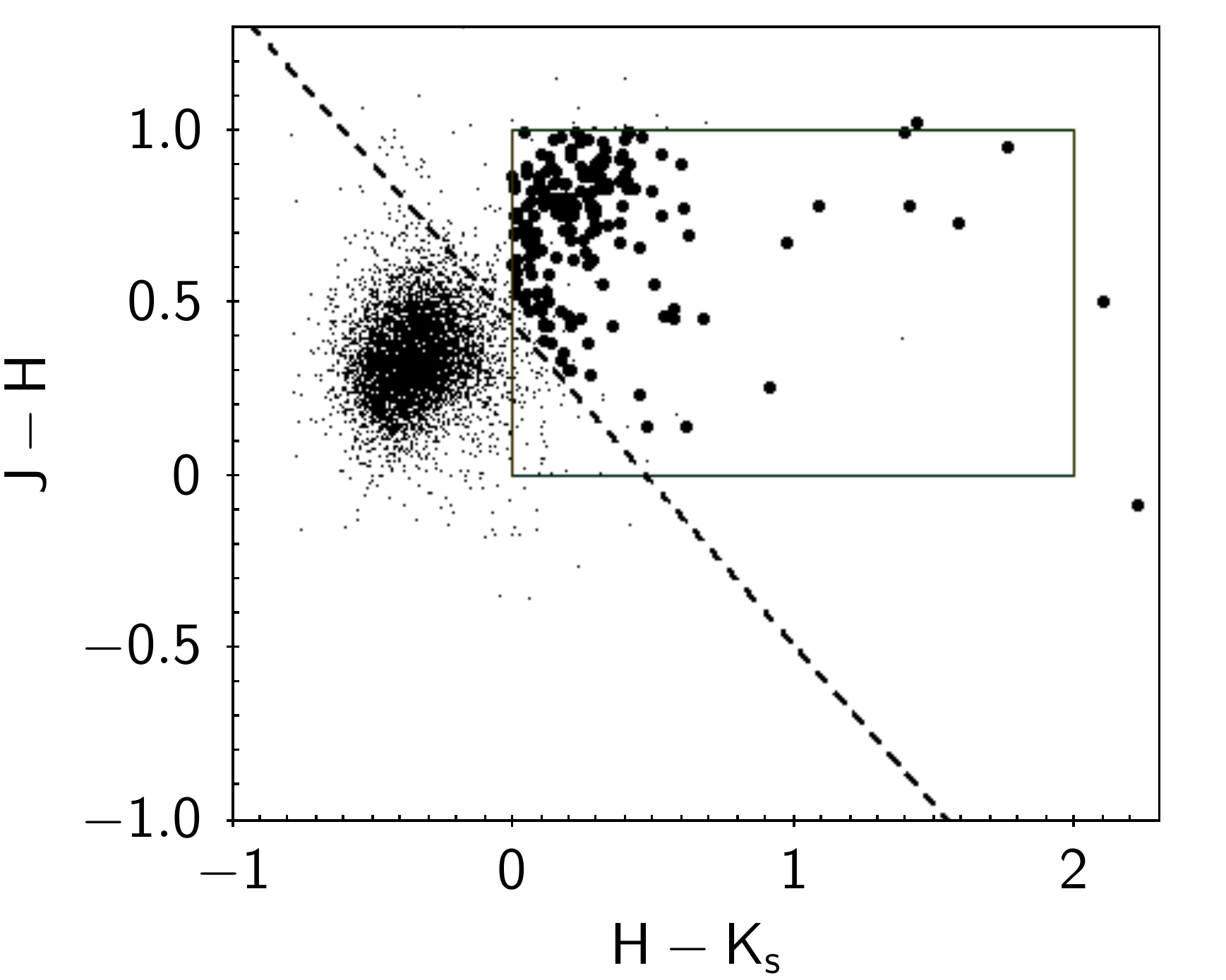}
          \caption{Color-Magnitude and Color-Color Diagrams for extended objects
            detected in J, H and K$_s$ passbands. Upper (lower) panels correspond to d010 (d115) tile. Smaller dots represent extended objects
            and larger dots the extragalactic candidates. 
}
            \label{diagrama1}
\end{centering}
\end{figure*}

\begin{figure*}
\begin{centering}
  \includegraphics[width=70mm,height=65mm]{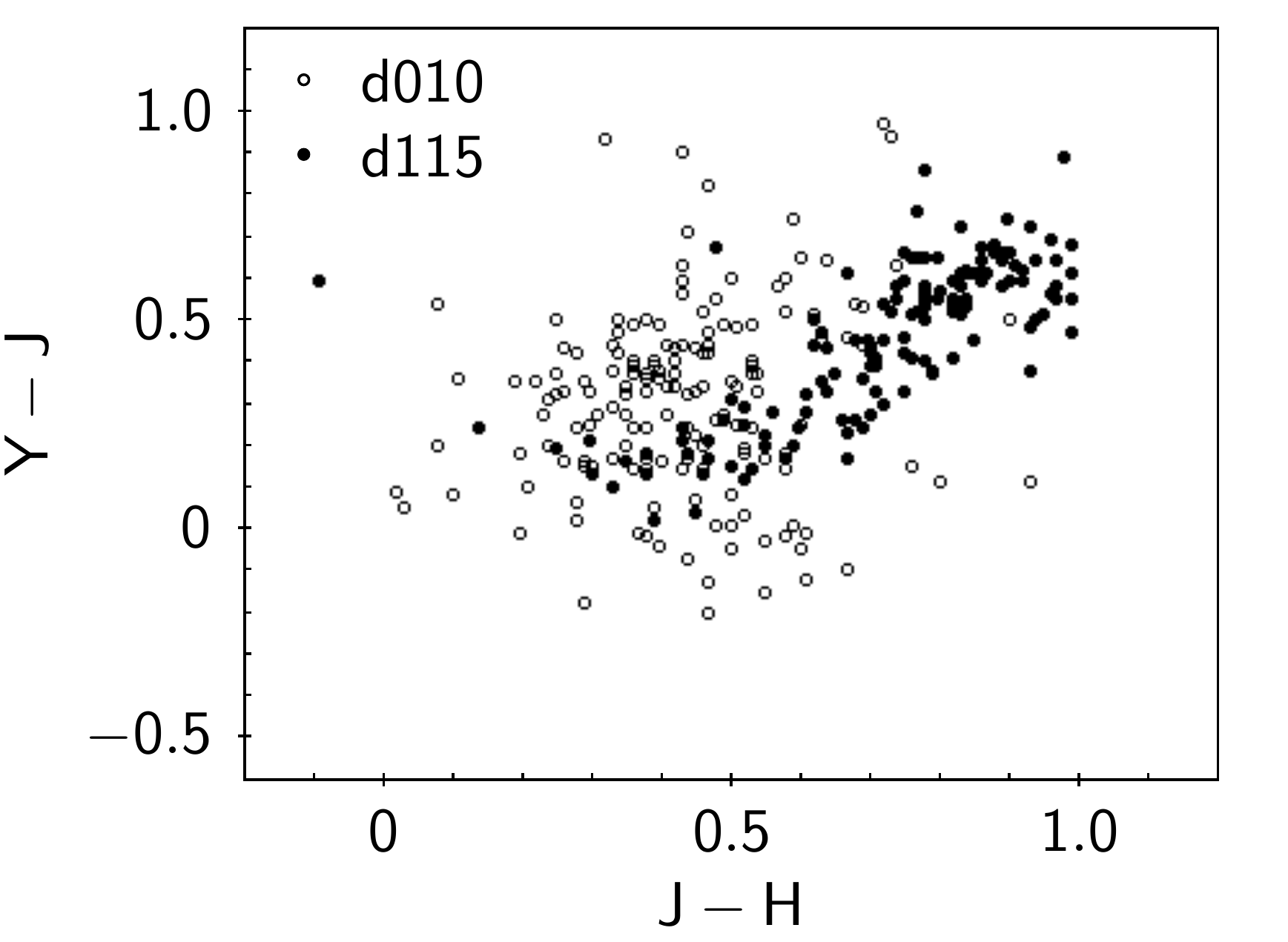}
  \includegraphics[width=70mm,height=65mm]{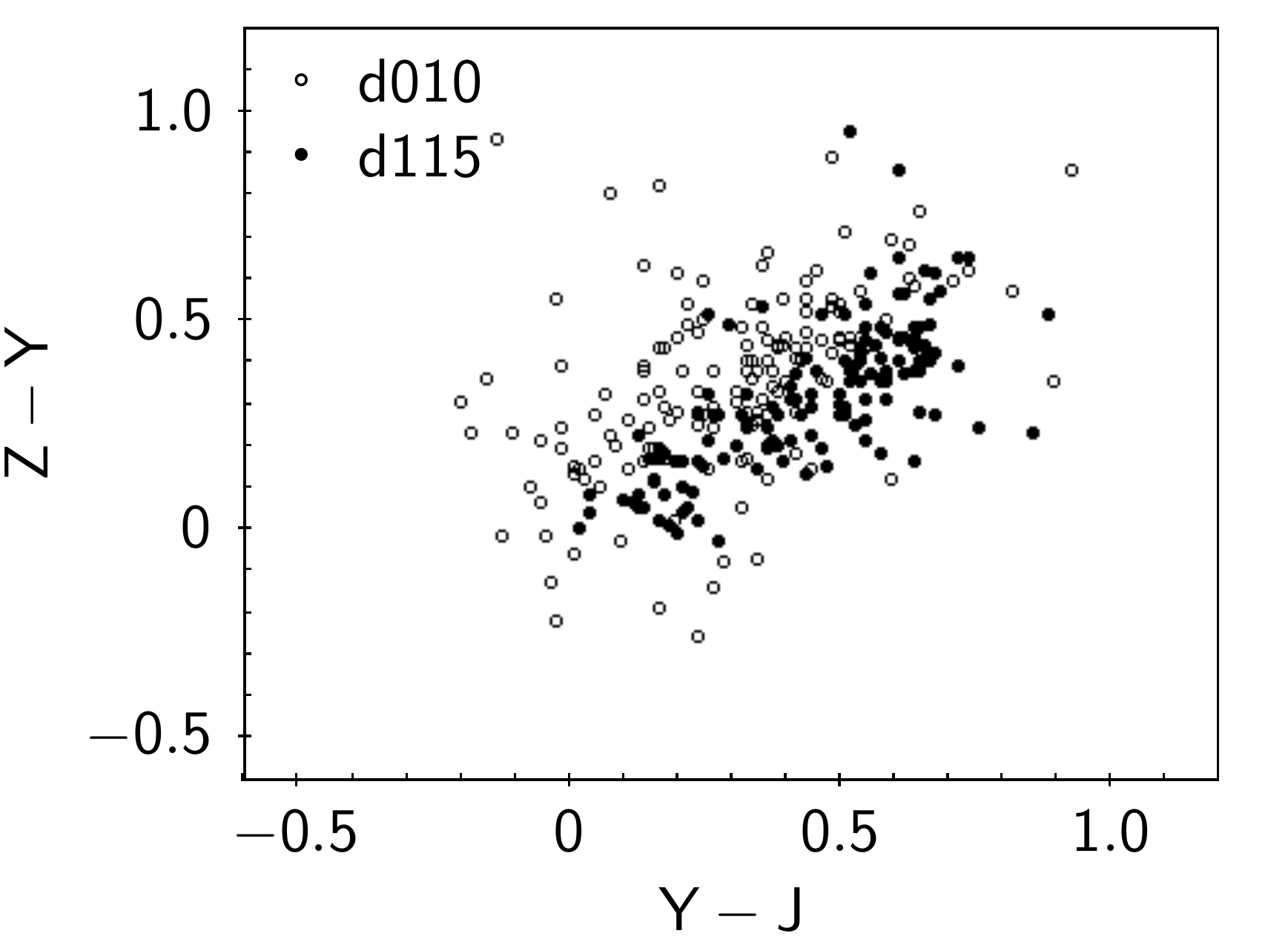}
  \caption{Color-Color Diagrams for extragalactic candidates detected in
    the five passbands.  Open (filled) circles represent the detections in the d010 (d115) tile. }
            \label{color2}
\end{centering}
\end{figure*}

\begin{figure*}
\begin{centering}
            \includegraphics[width=90mm,height=180mm]{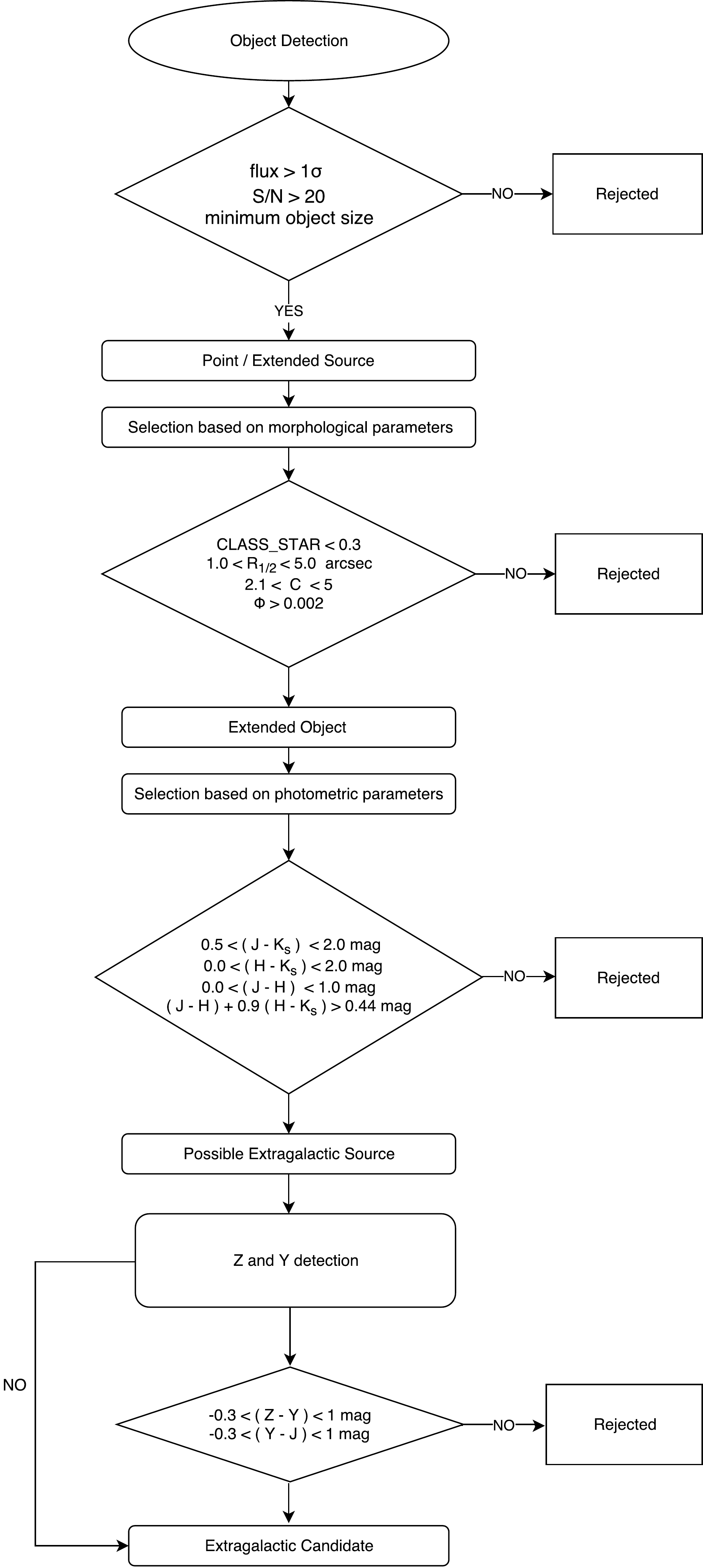}
          \caption{Flowchart showing the complete selection algorithm of
            extragalactic candidates.}
            \label{flowchart}
\end{centering}
\end{figure*}

\begin{figure*}
\begin{centering}
  \includegraphics[width=150mm,height=80mm]{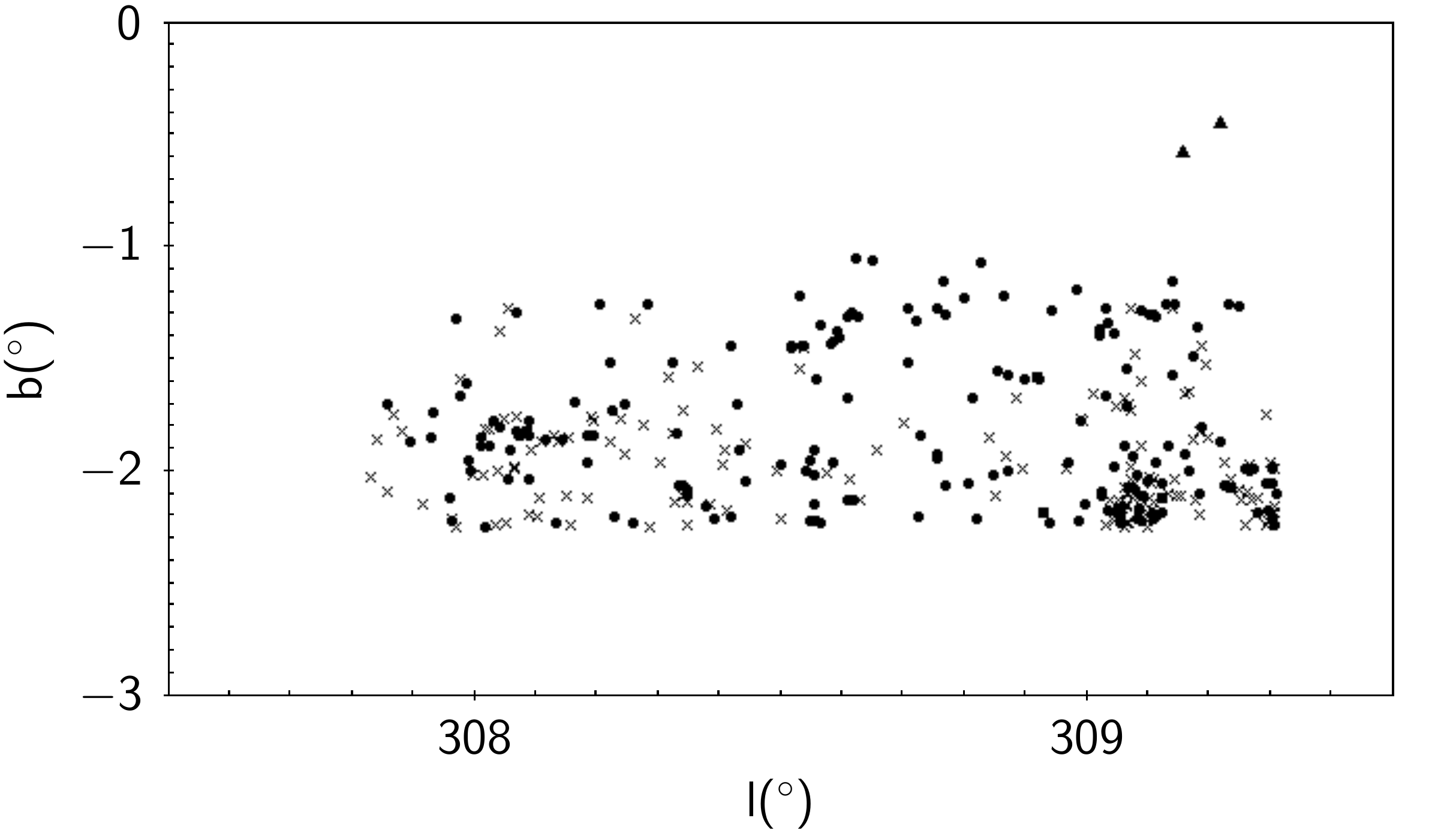}
  \includegraphics[width=150mm,height=80mm]{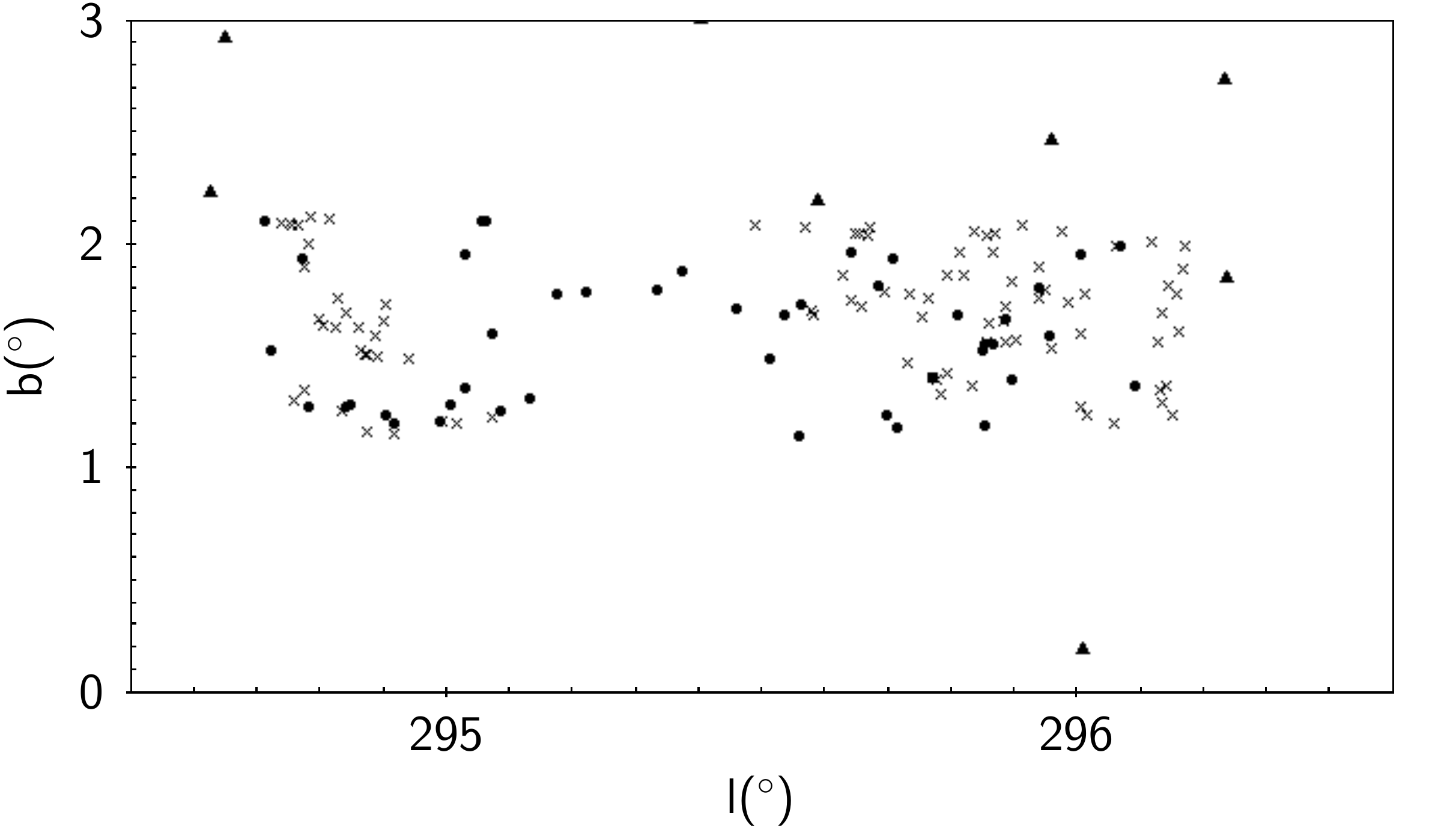}
  \caption{Distribution of extragalactic candidates in Galactic coordinates. For the d010 (upper panel) and d115 (bottom panel) tiles,
    these sources detected in three (filled circles)
    and five (crosses) passbands are shown together with the 2MASS extended
    sources
    represented by filled triangles. }
	\label{lbvvv}
\end{centering}
\end{figure*}

 \begin{figure*}
\begin{centering}
    \includegraphics[width=0.240\textwidth]{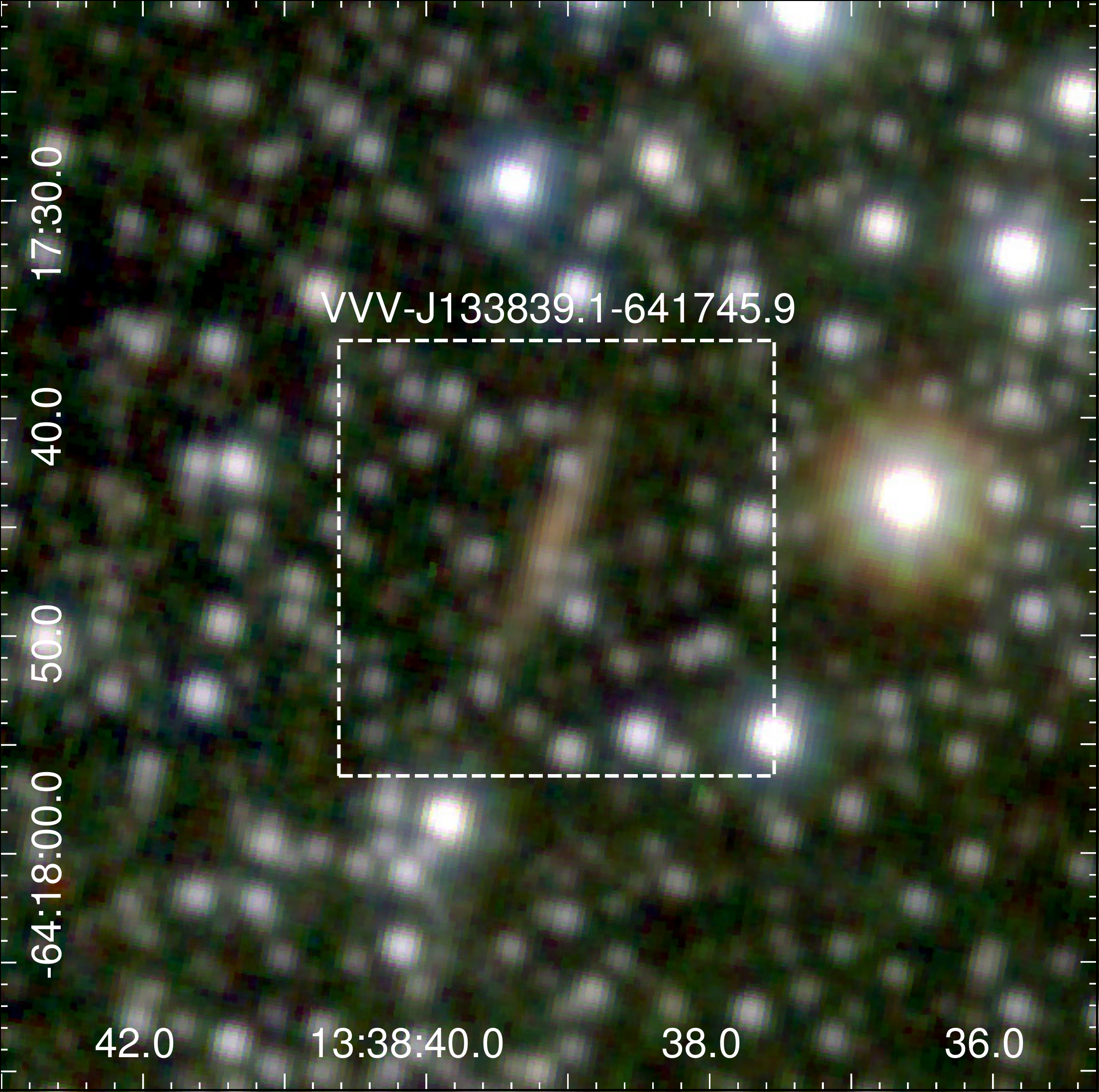}
    \includegraphics[width=0.240\textwidth]{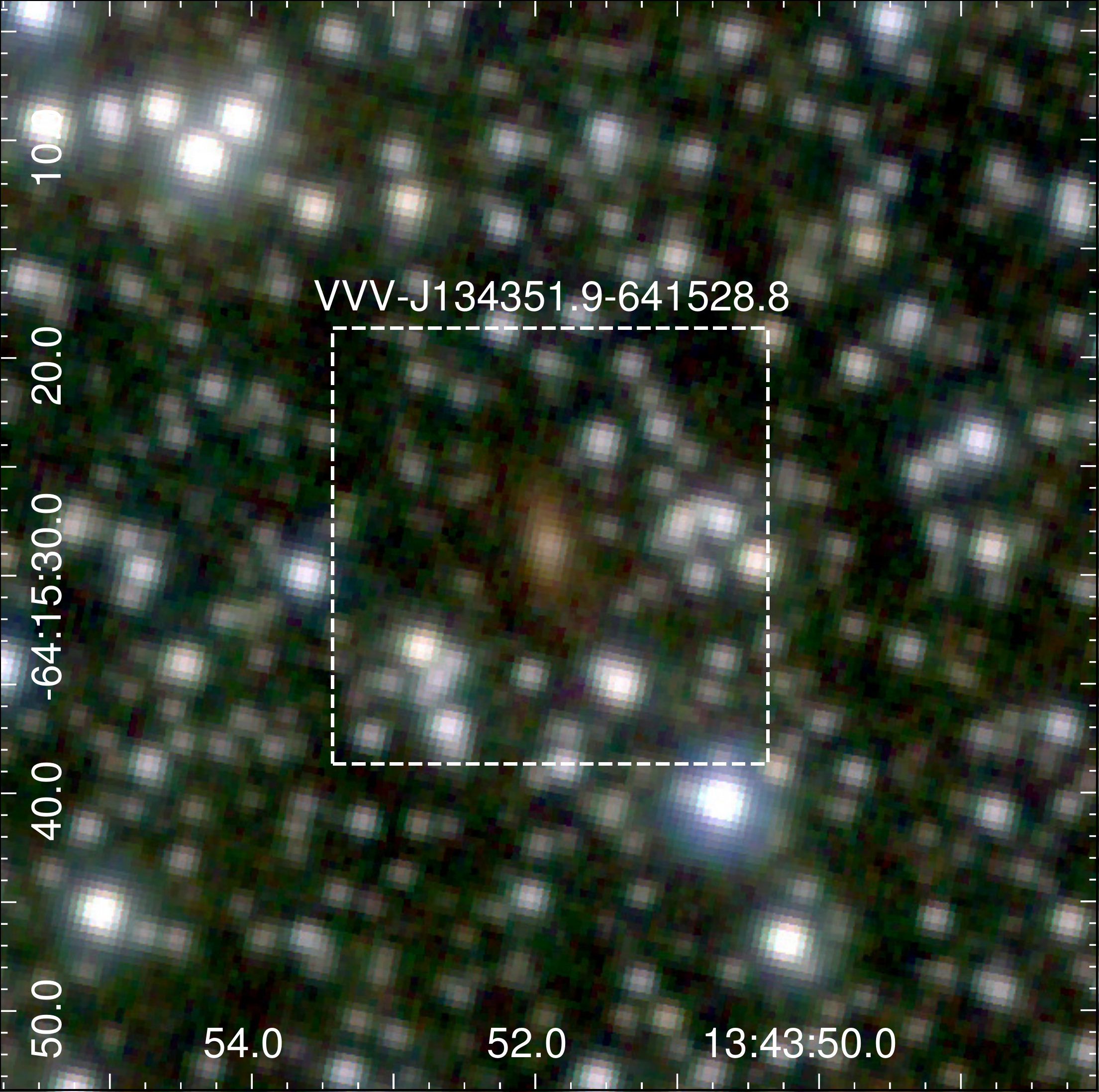}
    \includegraphics[width=0.240\textwidth]{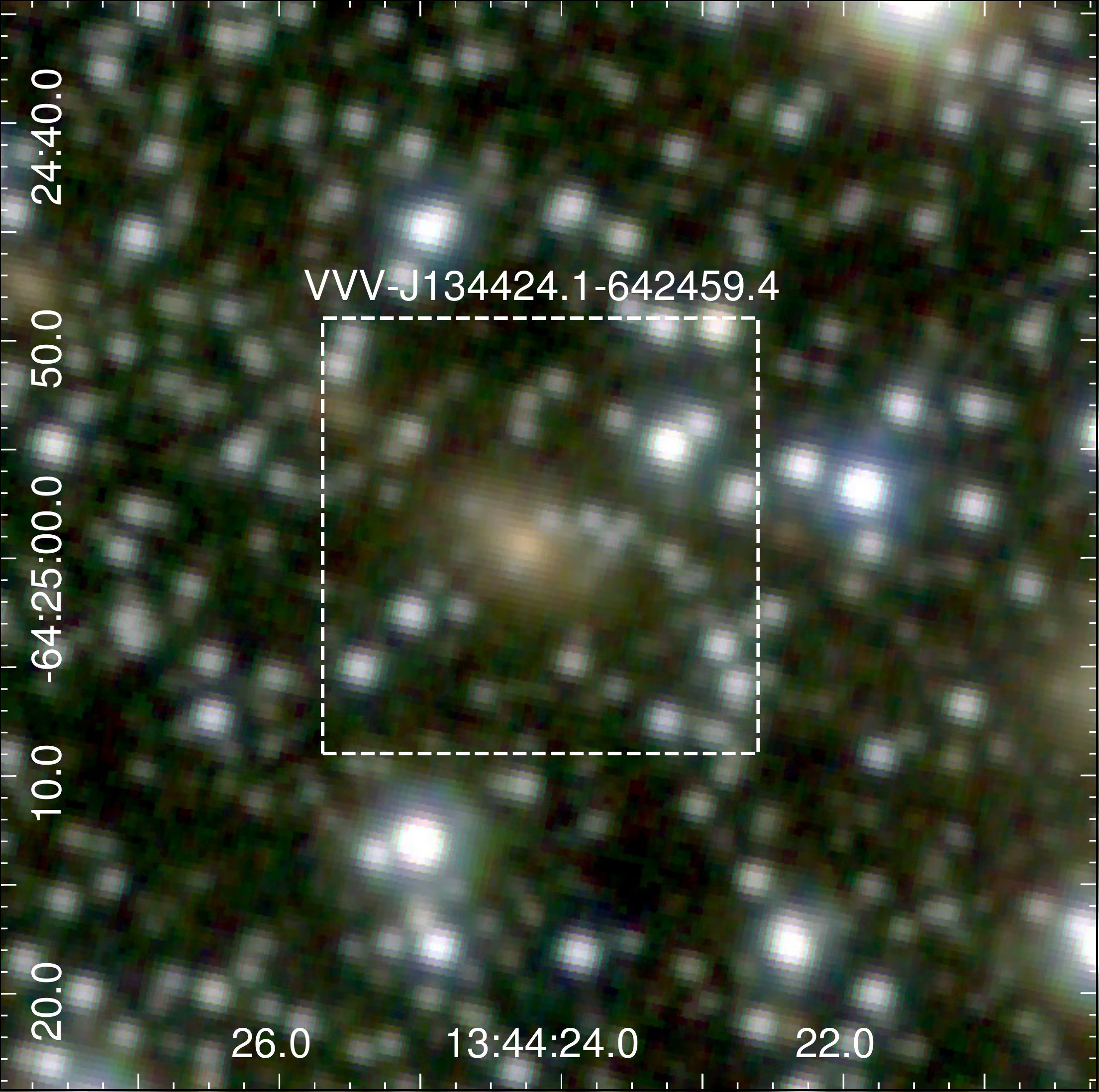}
    \includegraphics[width=0.240\textwidth]{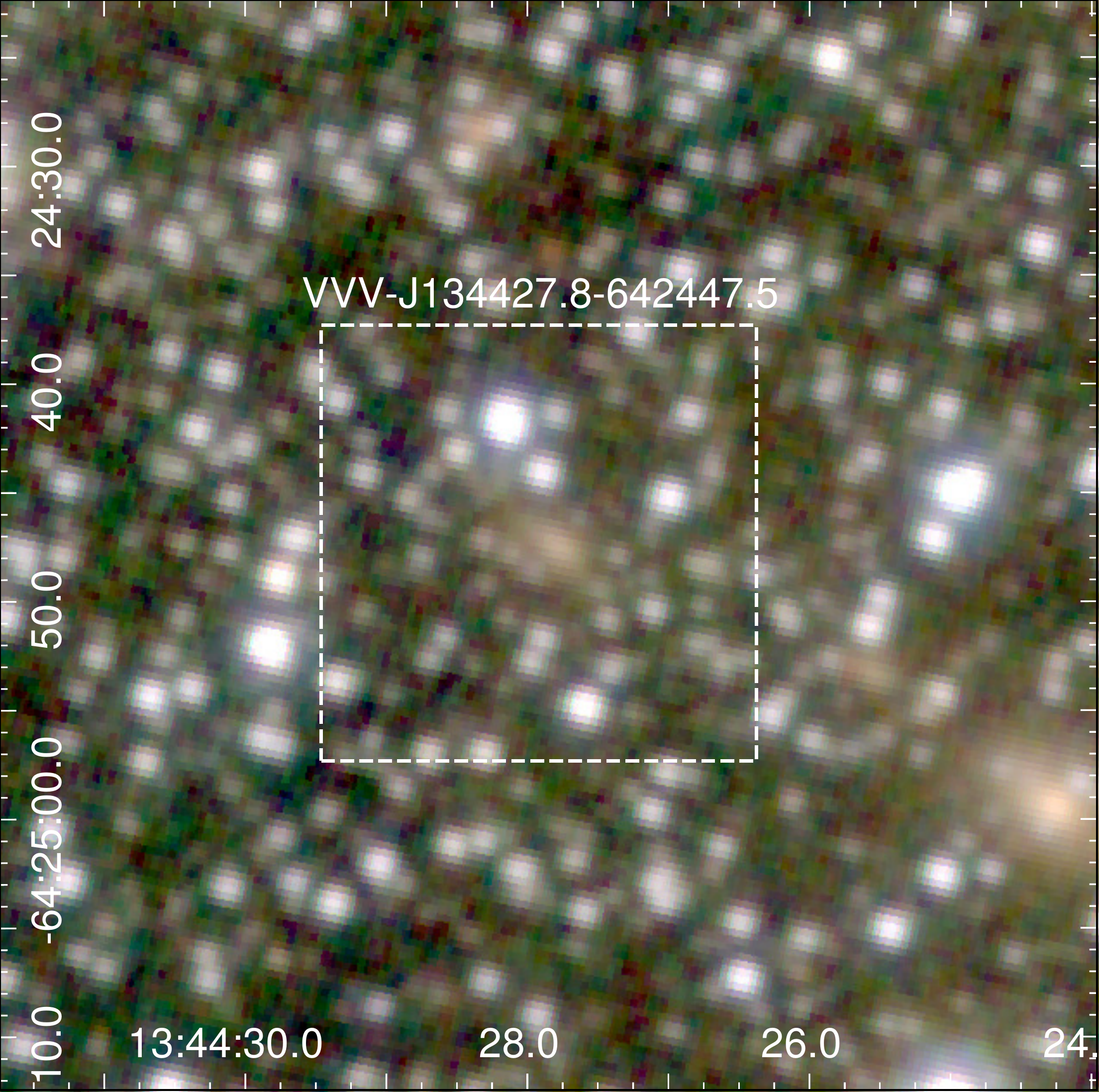}
    \includegraphics[width=0.240\textwidth]{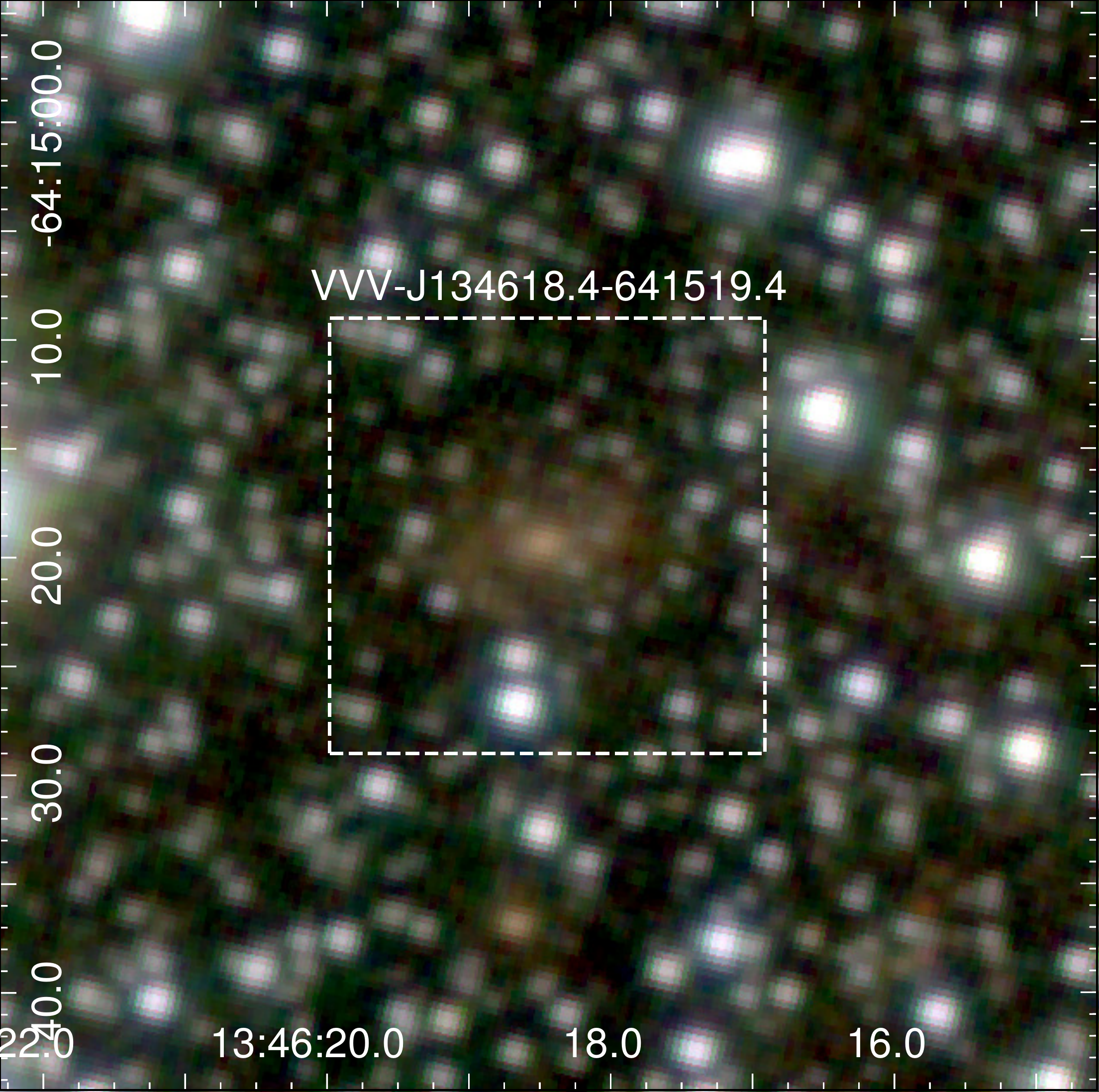}
    \includegraphics[width=0.240\textwidth]{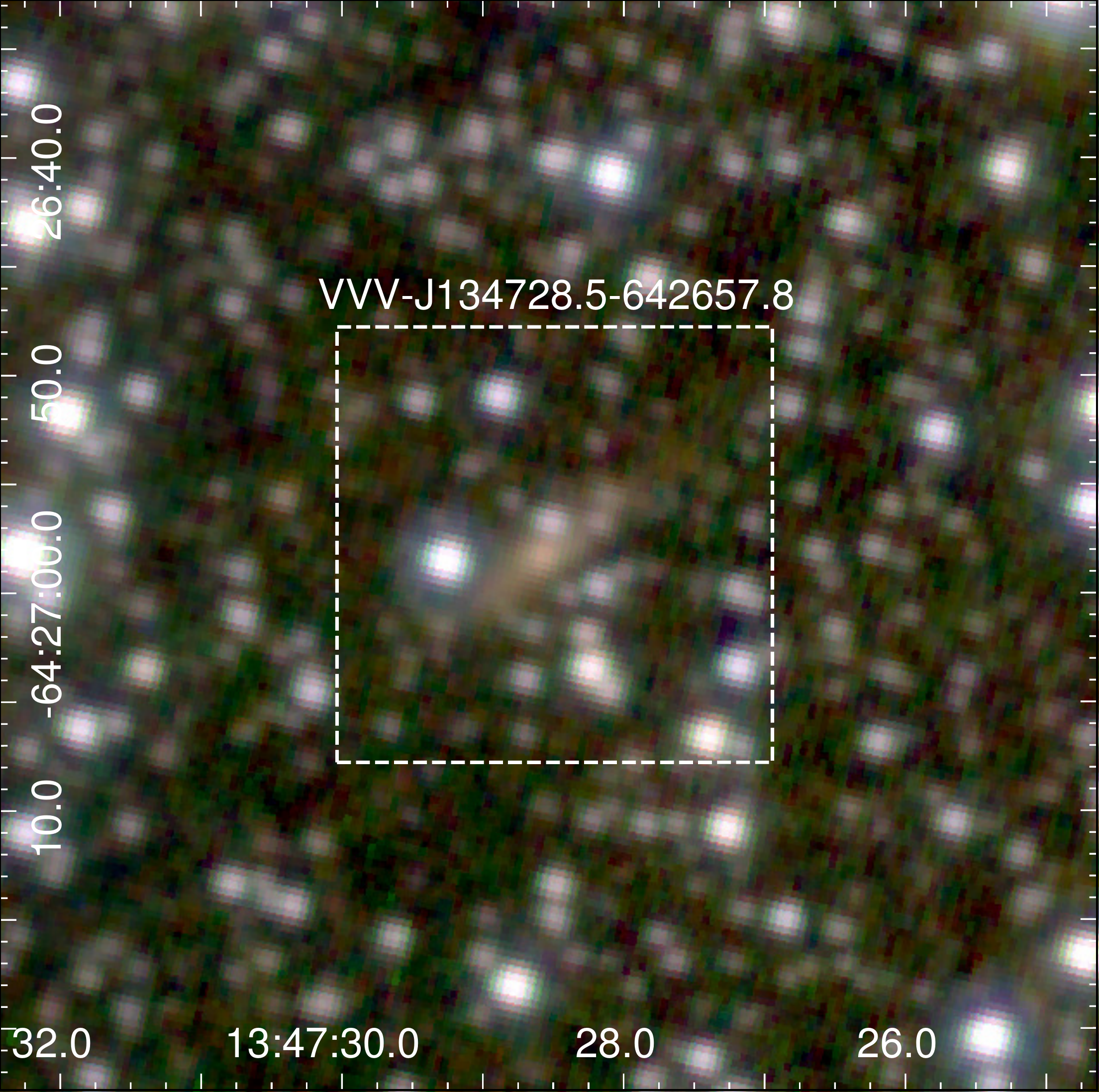}
    \includegraphics[width=0.240\textwidth]{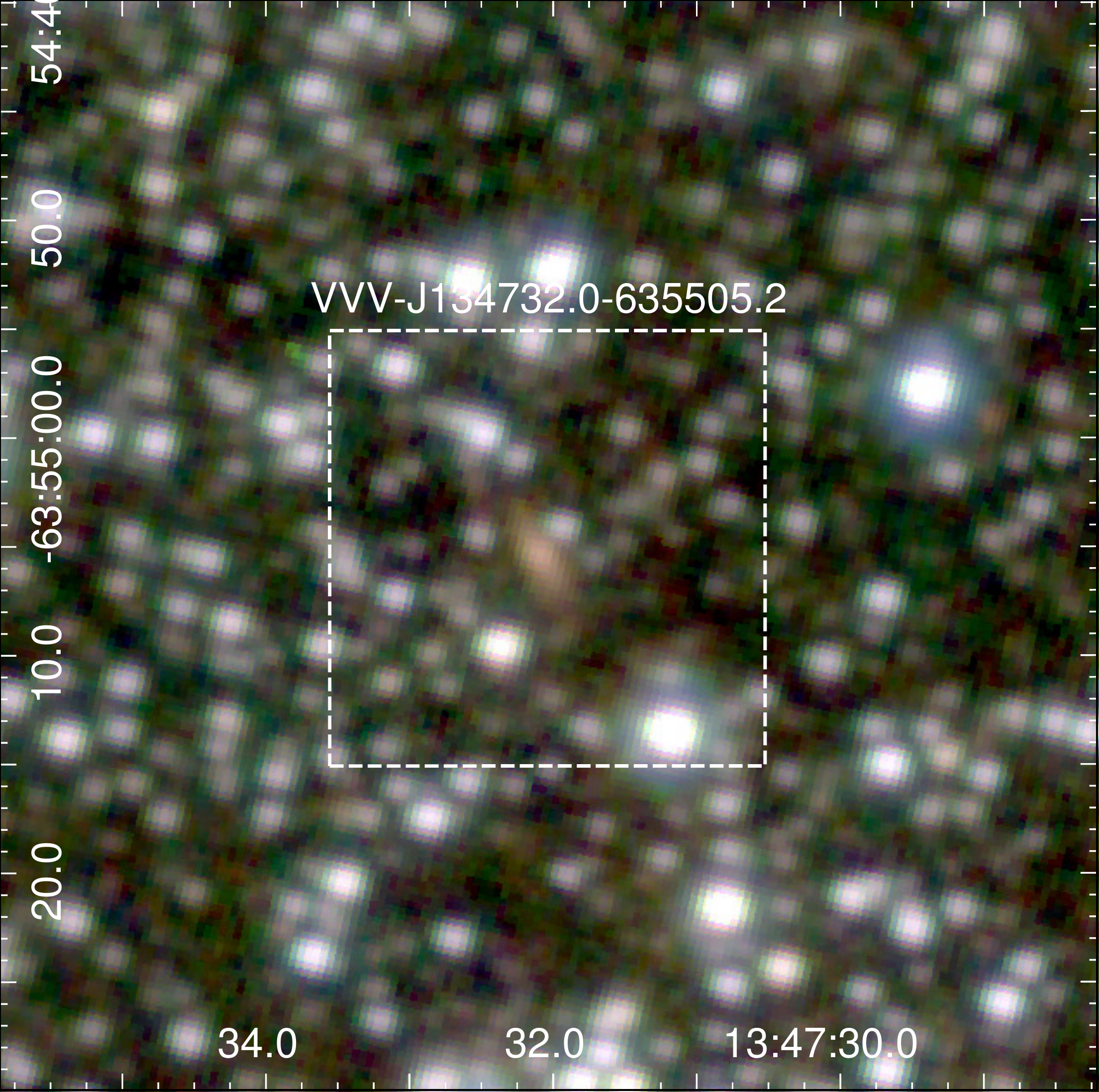}
    \includegraphics[width=0.240\textwidth]{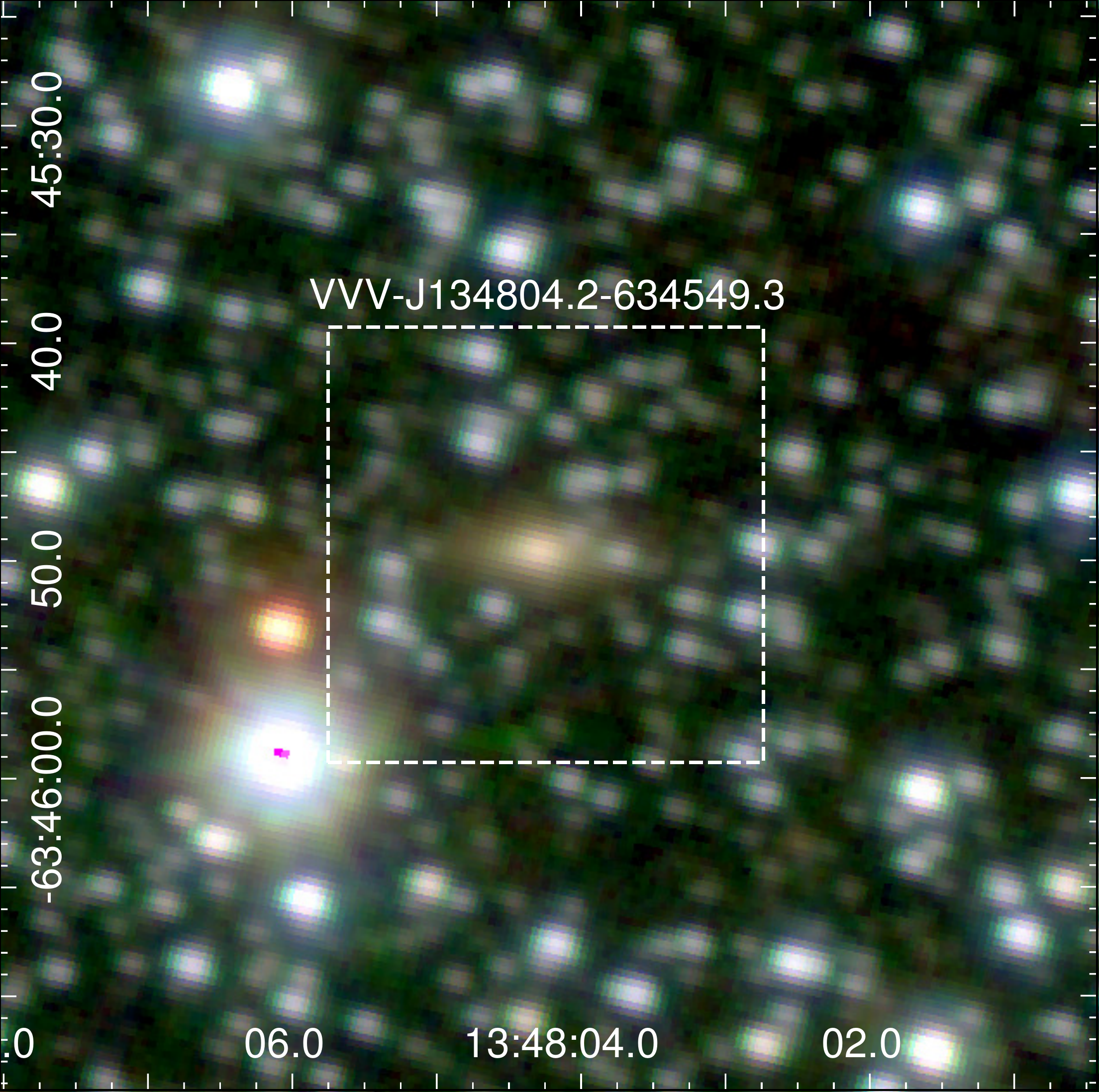}
    \includegraphics[width=0.240\textwidth]{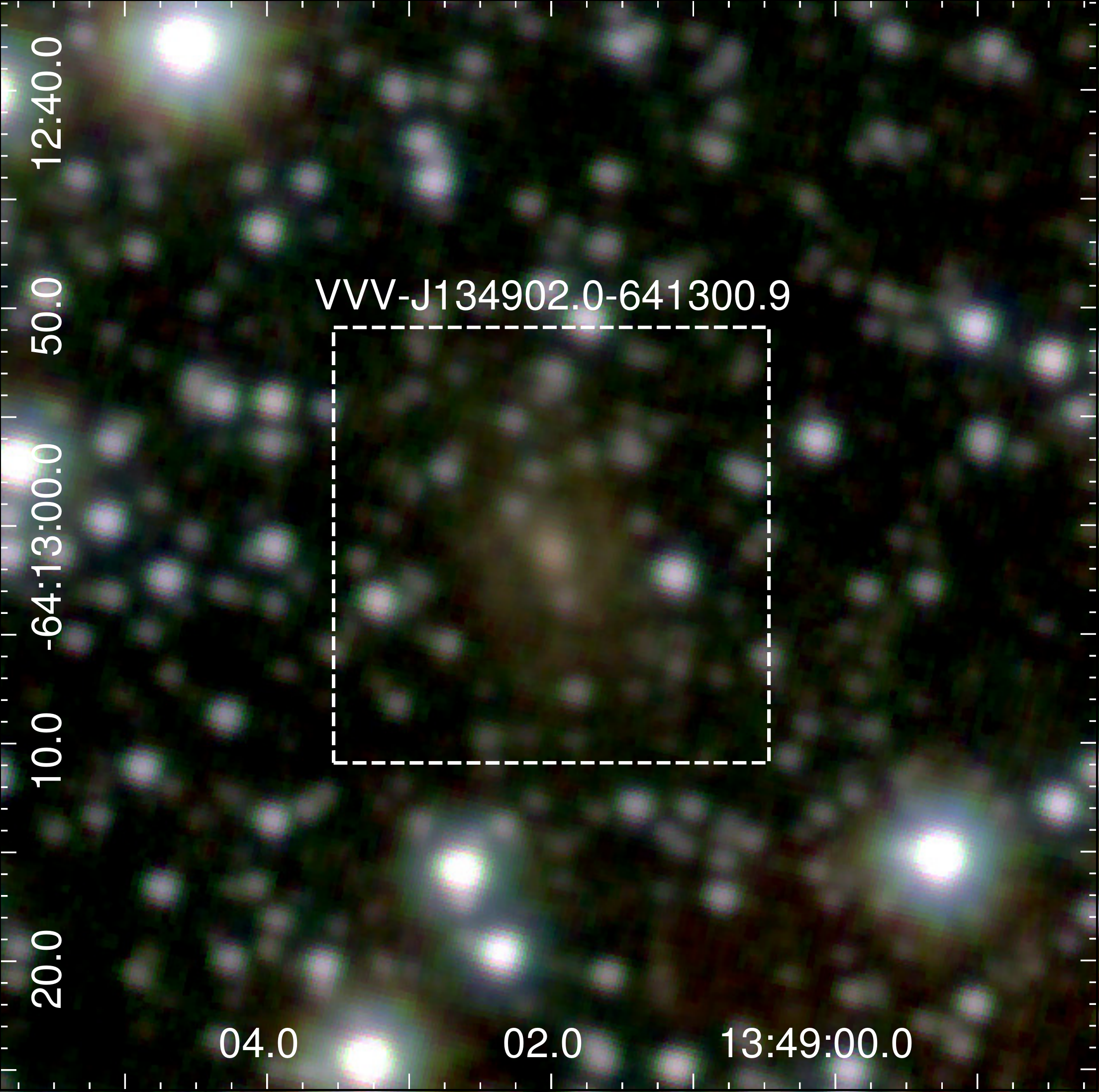}
    \includegraphics[width=0.240\textwidth]{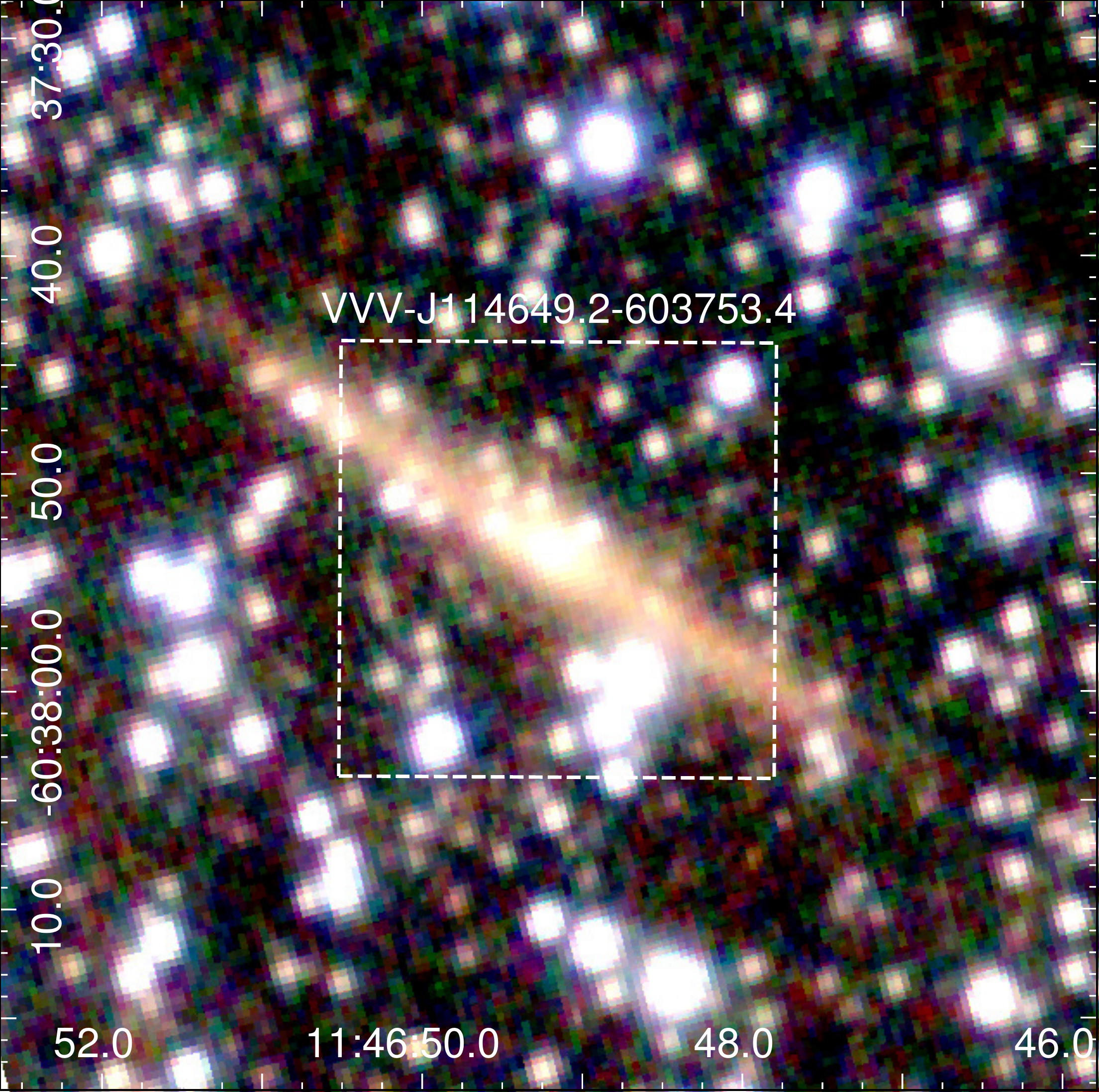}
    \includegraphics[width=0.240\textwidth]{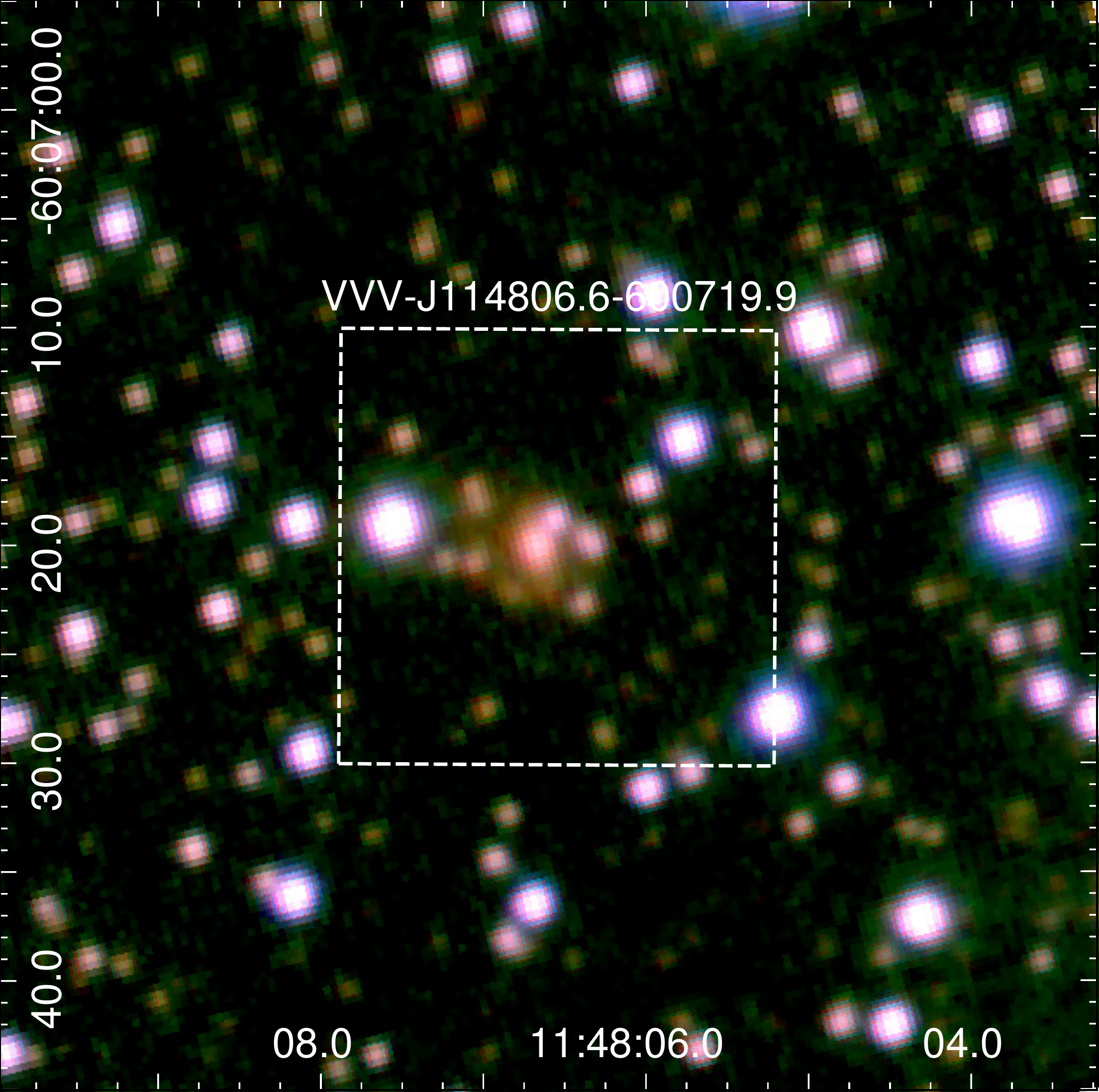}
    \includegraphics[width=0.240\textwidth]{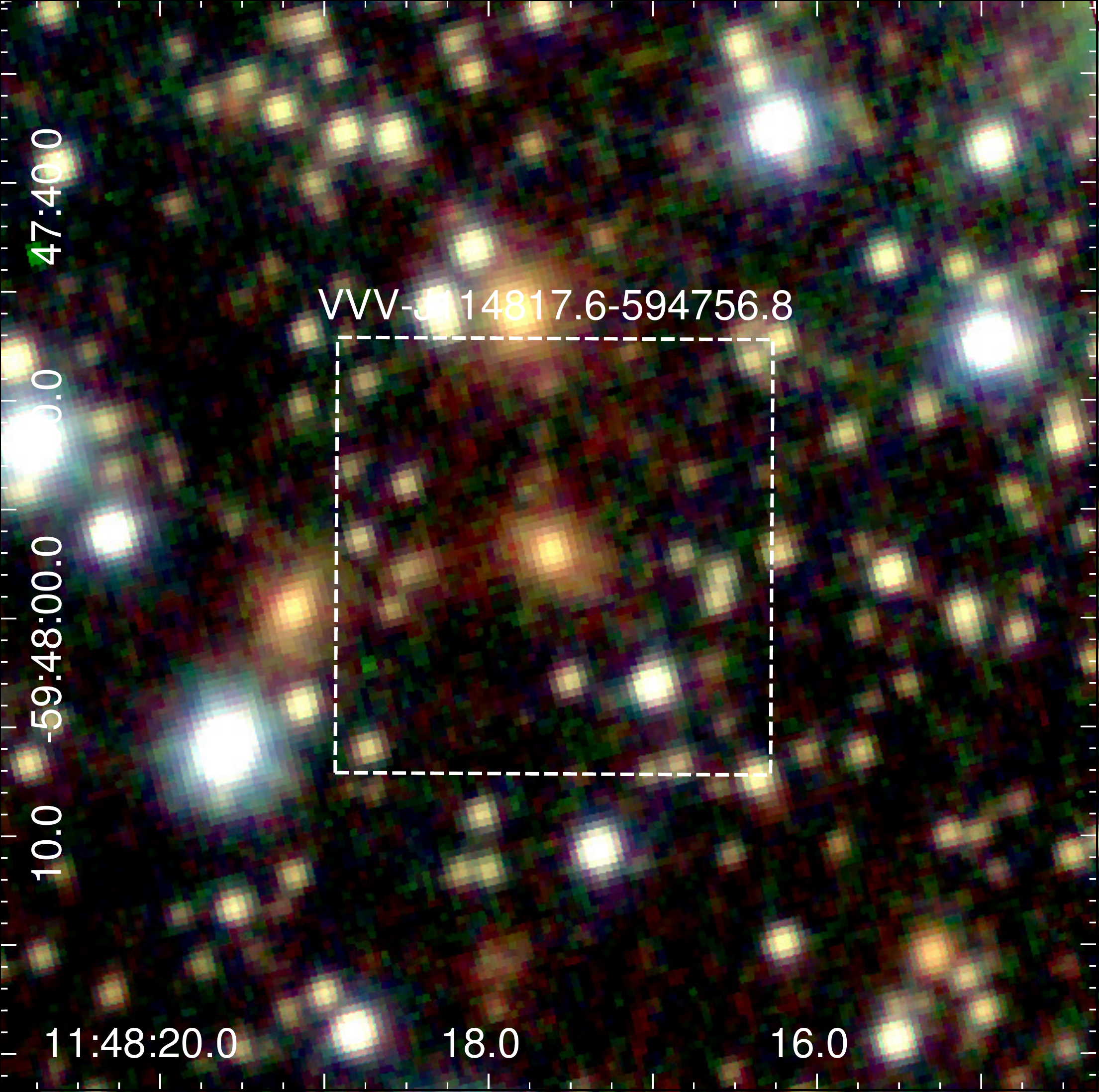}
    \includegraphics[width=0.240\textwidth]{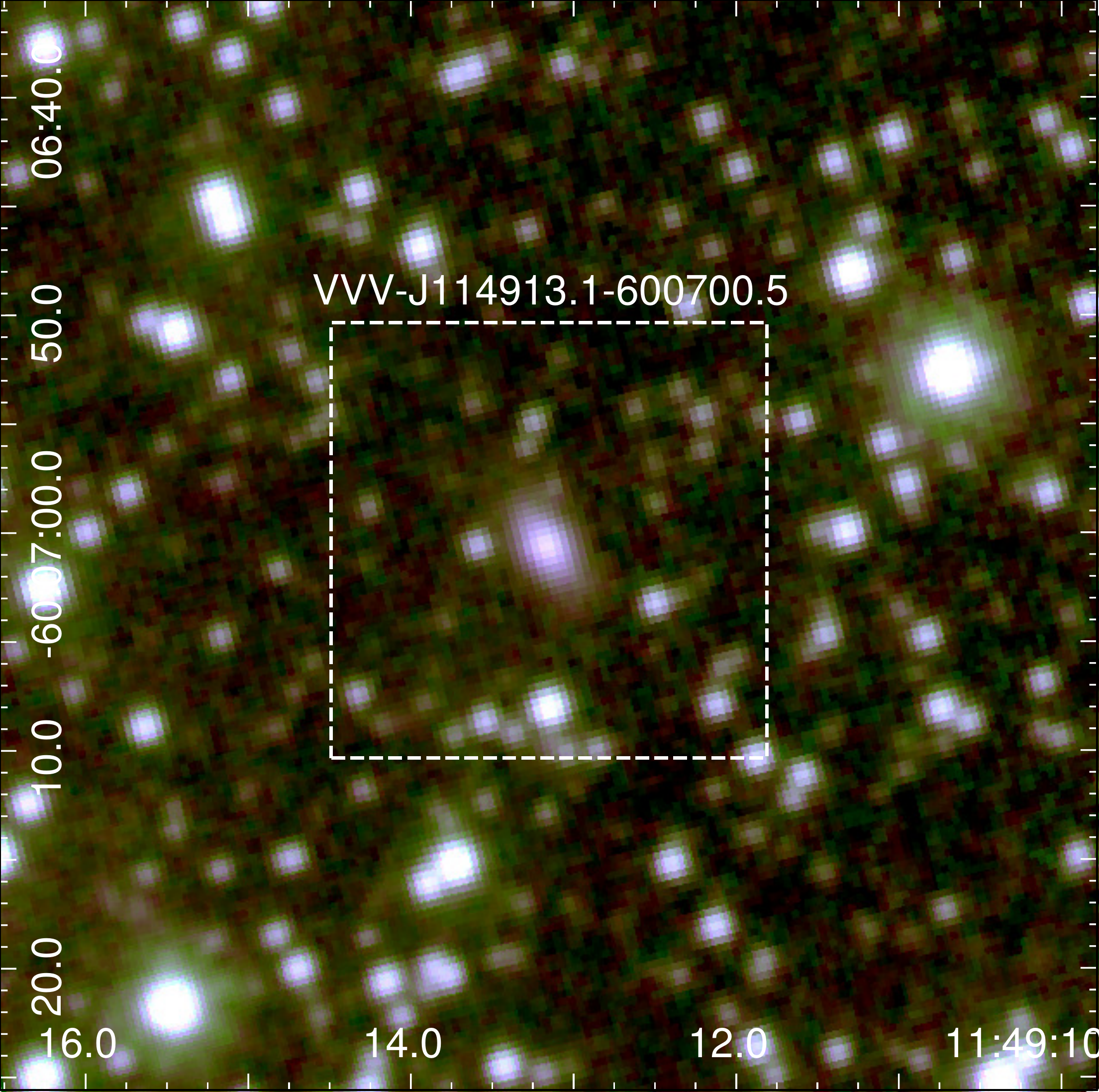}
    \includegraphics[width=0.240\textwidth]{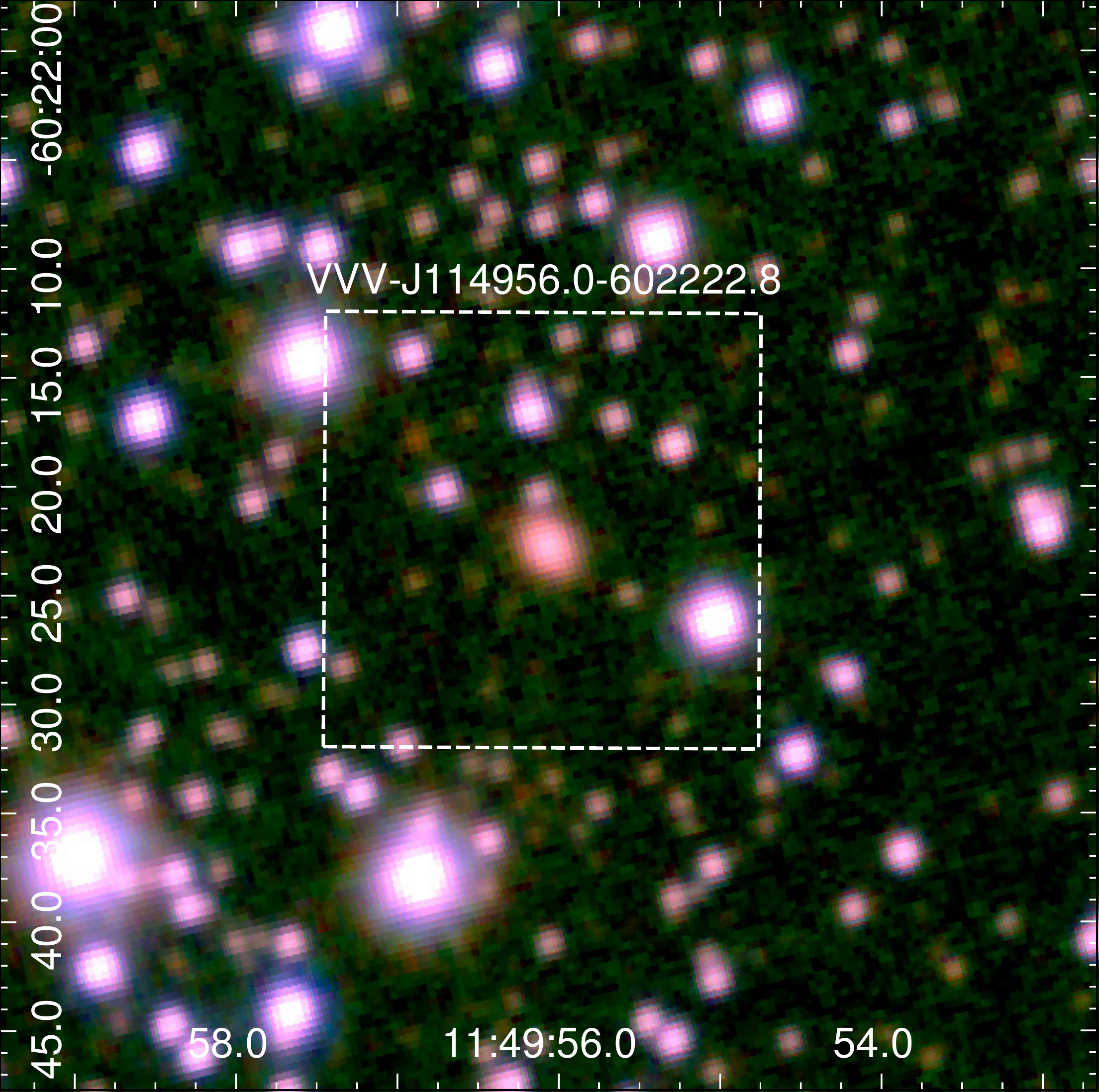}
    \includegraphics[width=0.240\textwidth]{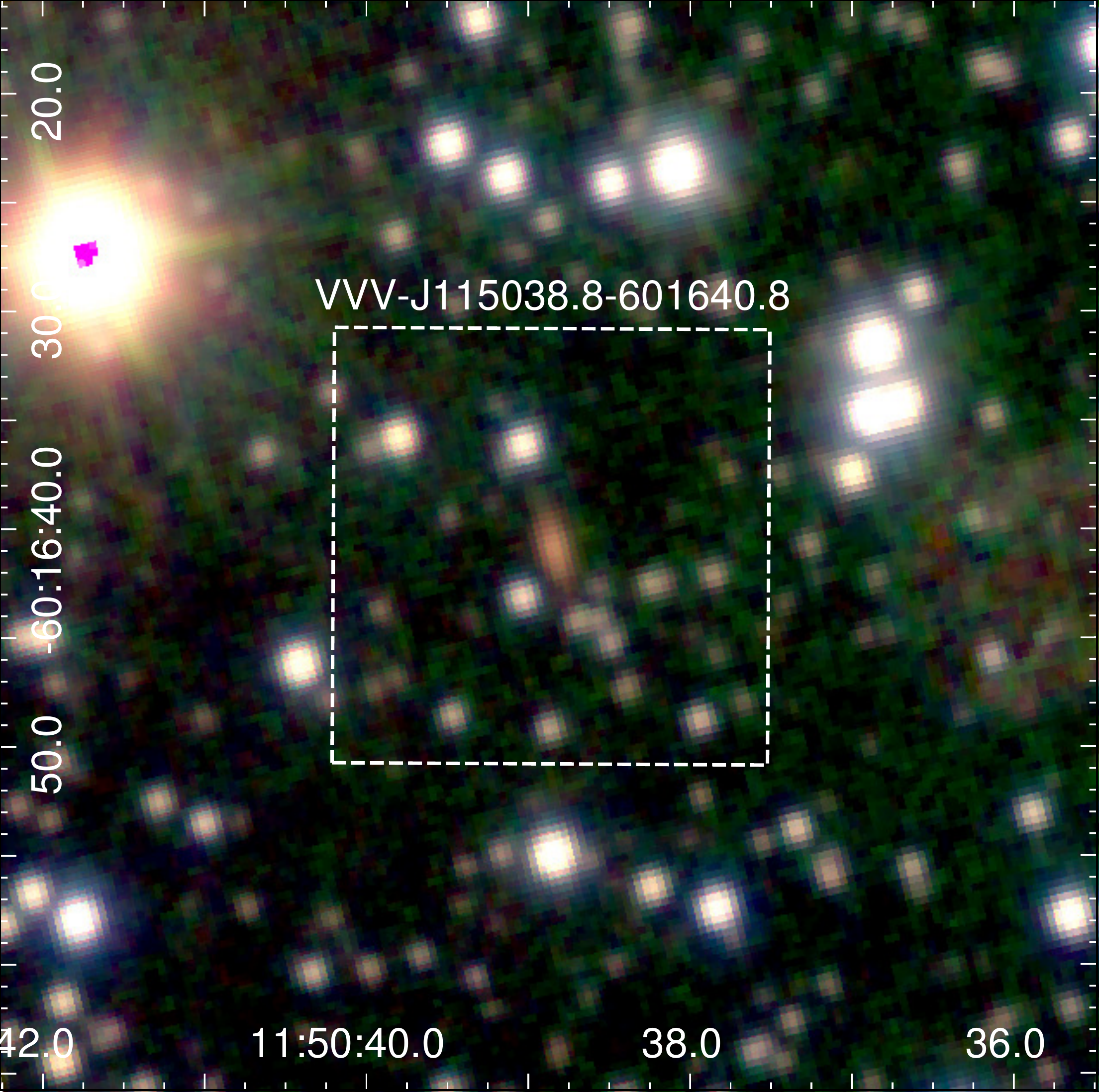}
    \includegraphics[width=0.240\textwidth]{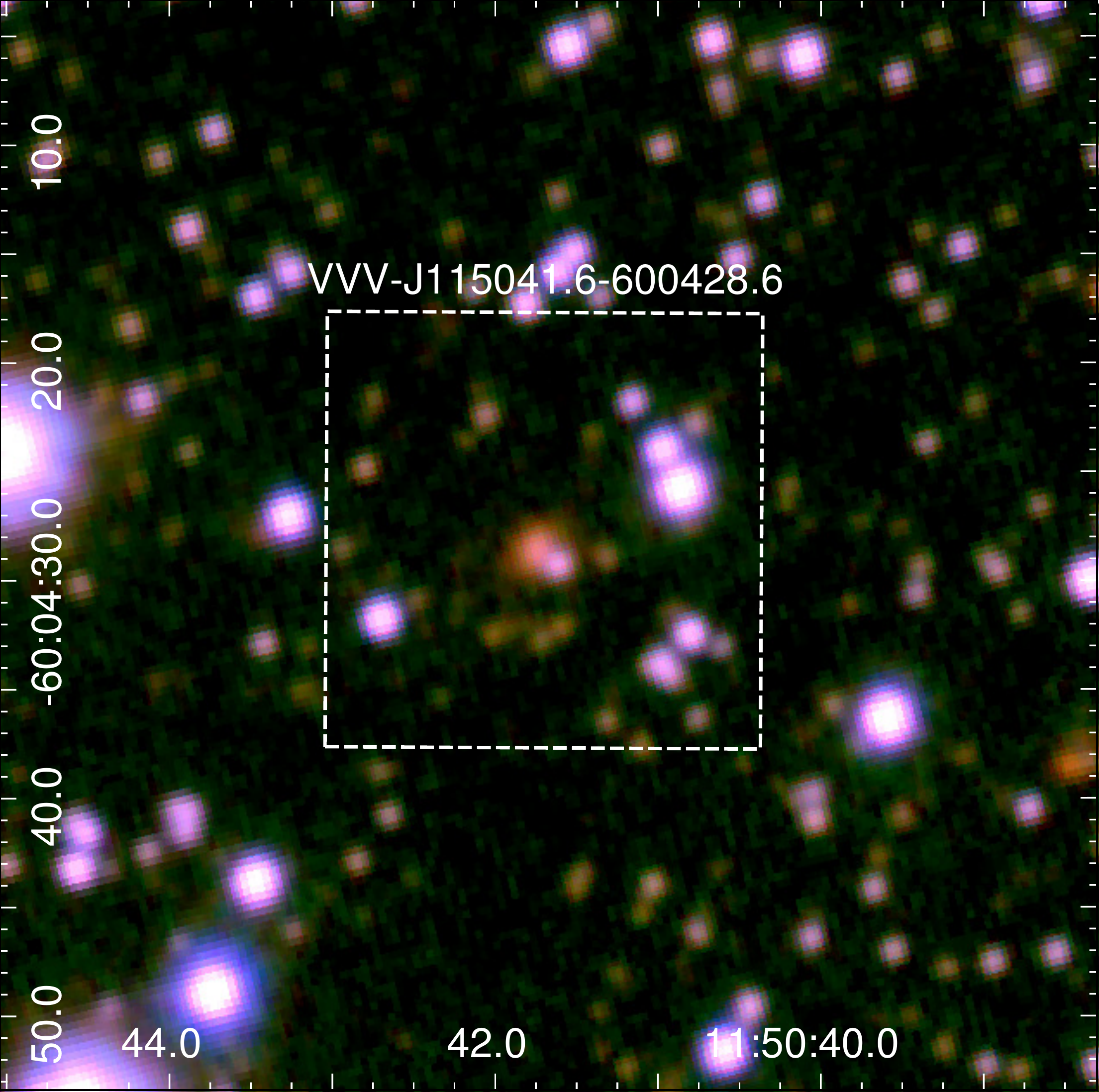}
    \includegraphics[width=0.240\textwidth]{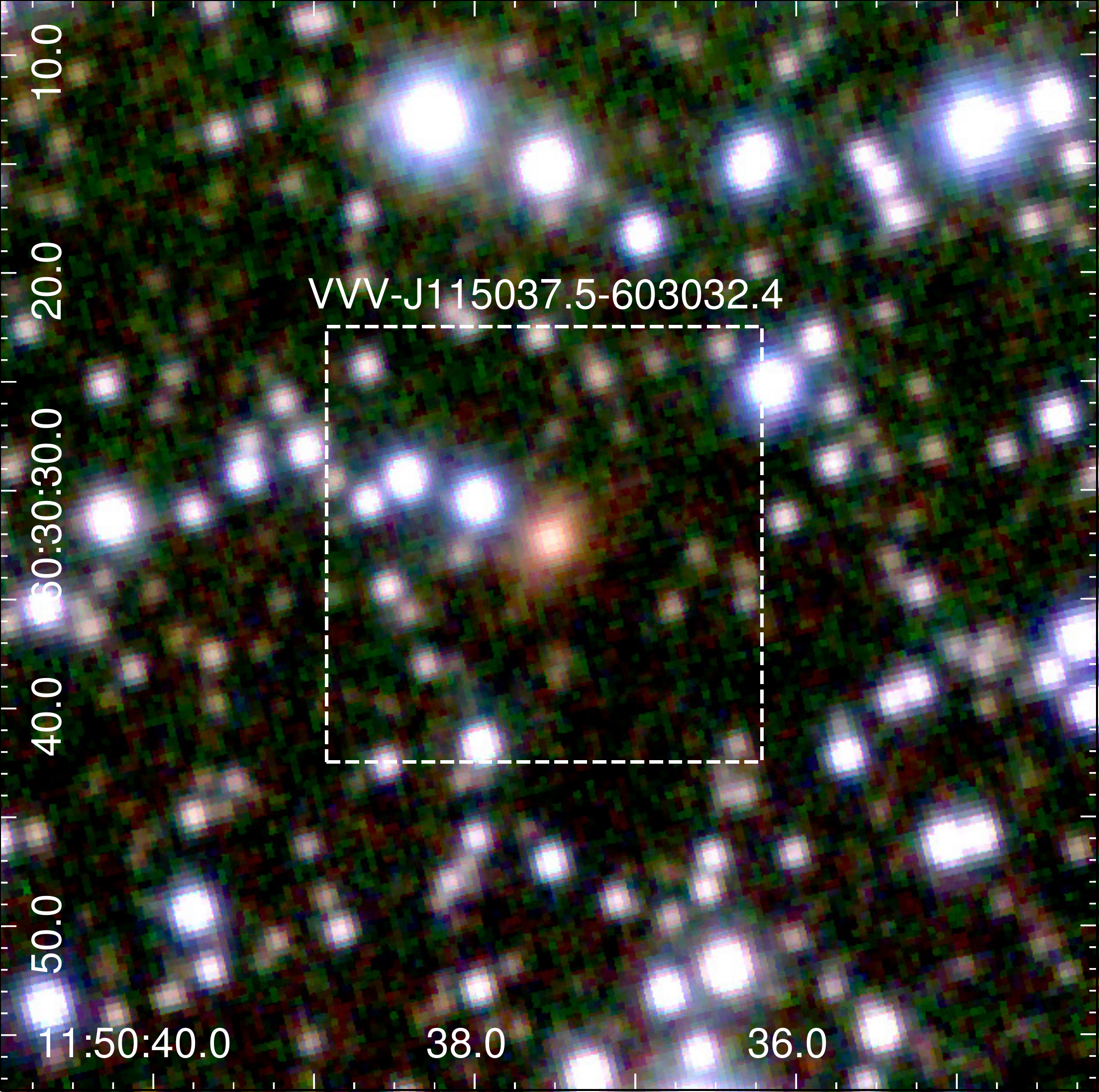}
    \includegraphics[width=0.240\textwidth]{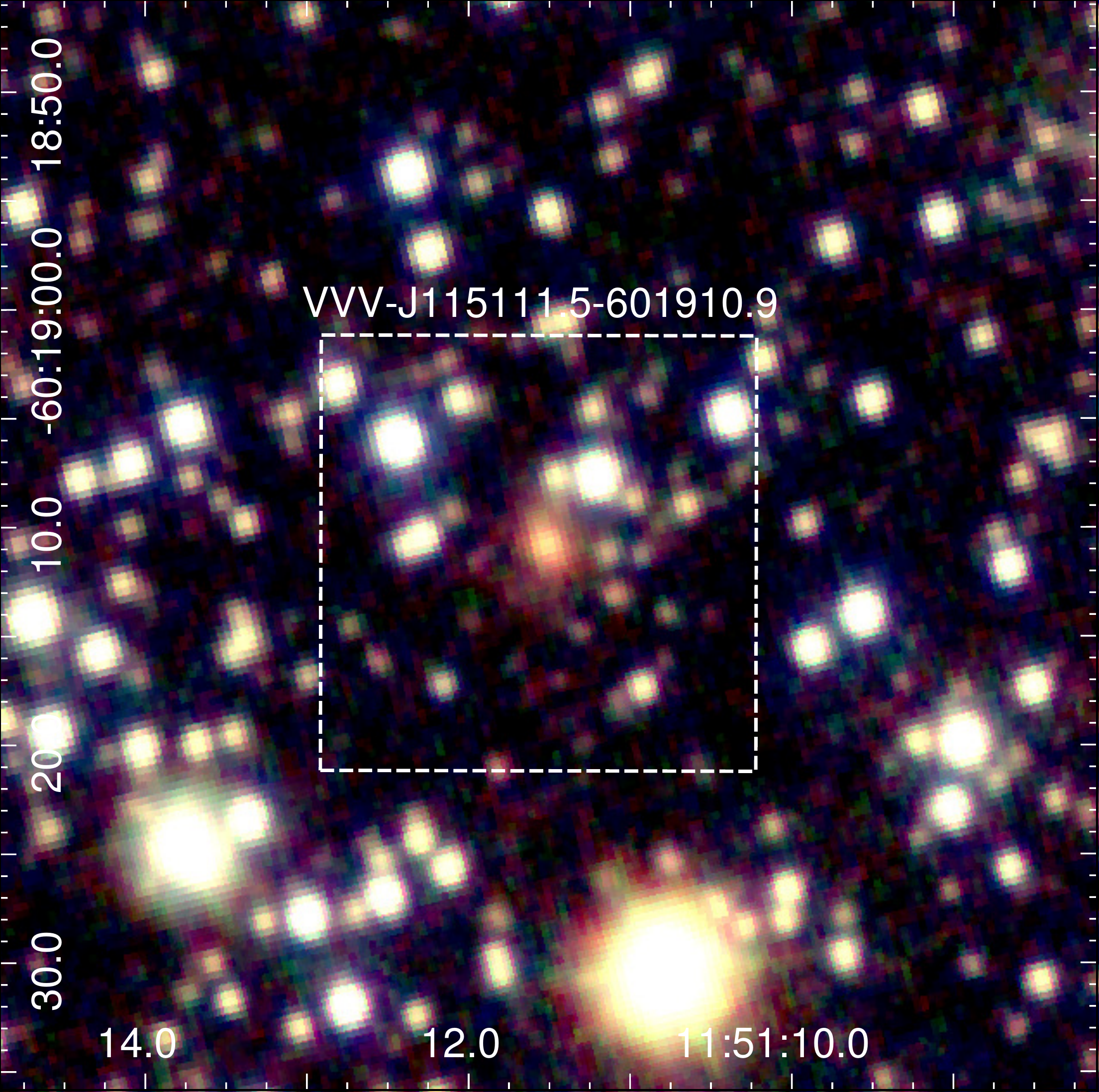}    
    \includegraphics[width=0.240\textwidth]{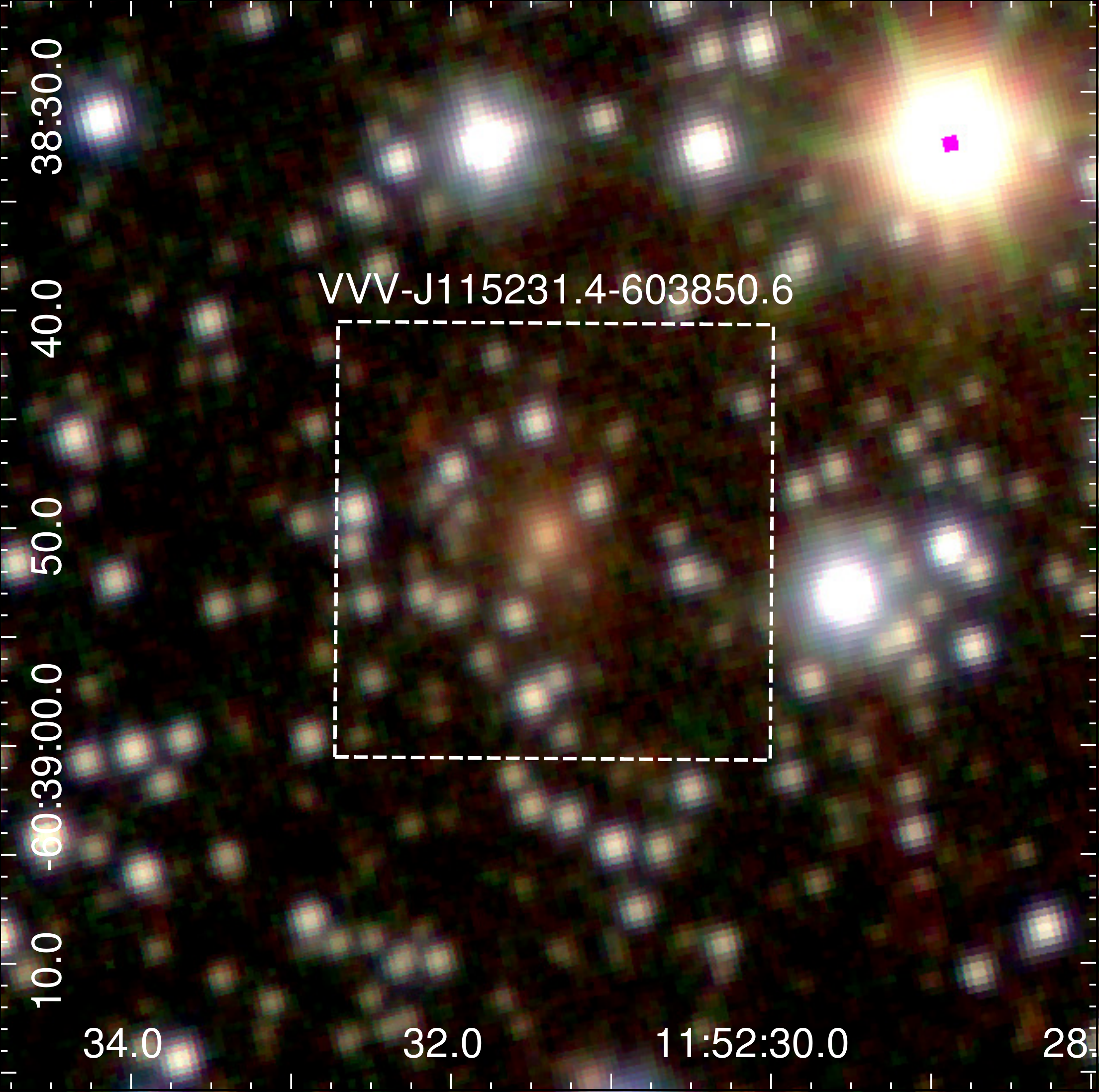}
    \includegraphics[width=0.240\textwidth]{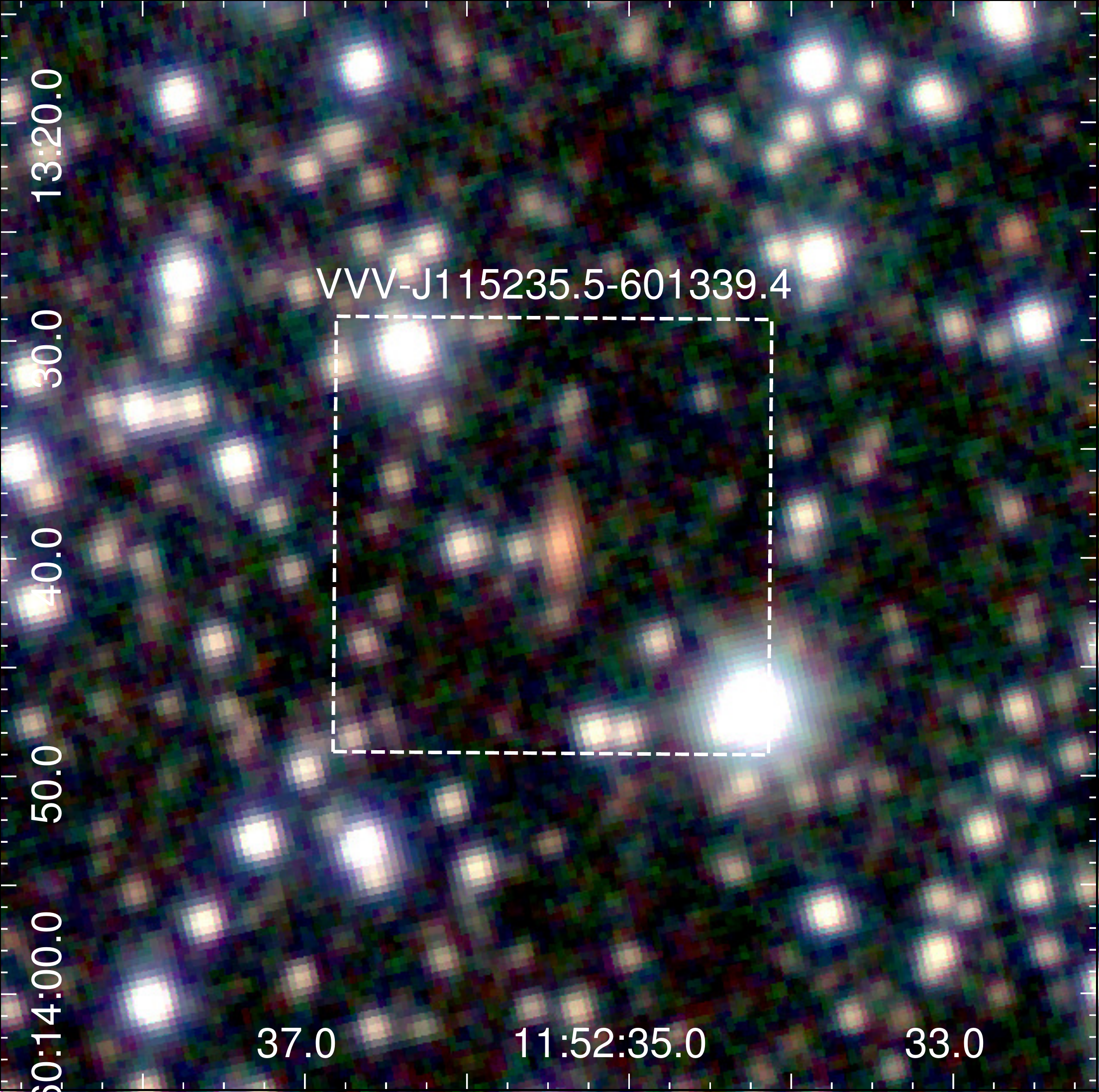}
    \caption{Color composed images of some examples of extragalactic sources.
    North is up and East is to the left.  White 
dashed-line box represents an area of 20$\times$20 arcseconds. }
\label{rgbd115}       
\end{centering}
\end{figure*}

\begin{figure*}
\begin{centering}
         \includegraphics[width=0.45\textwidth]{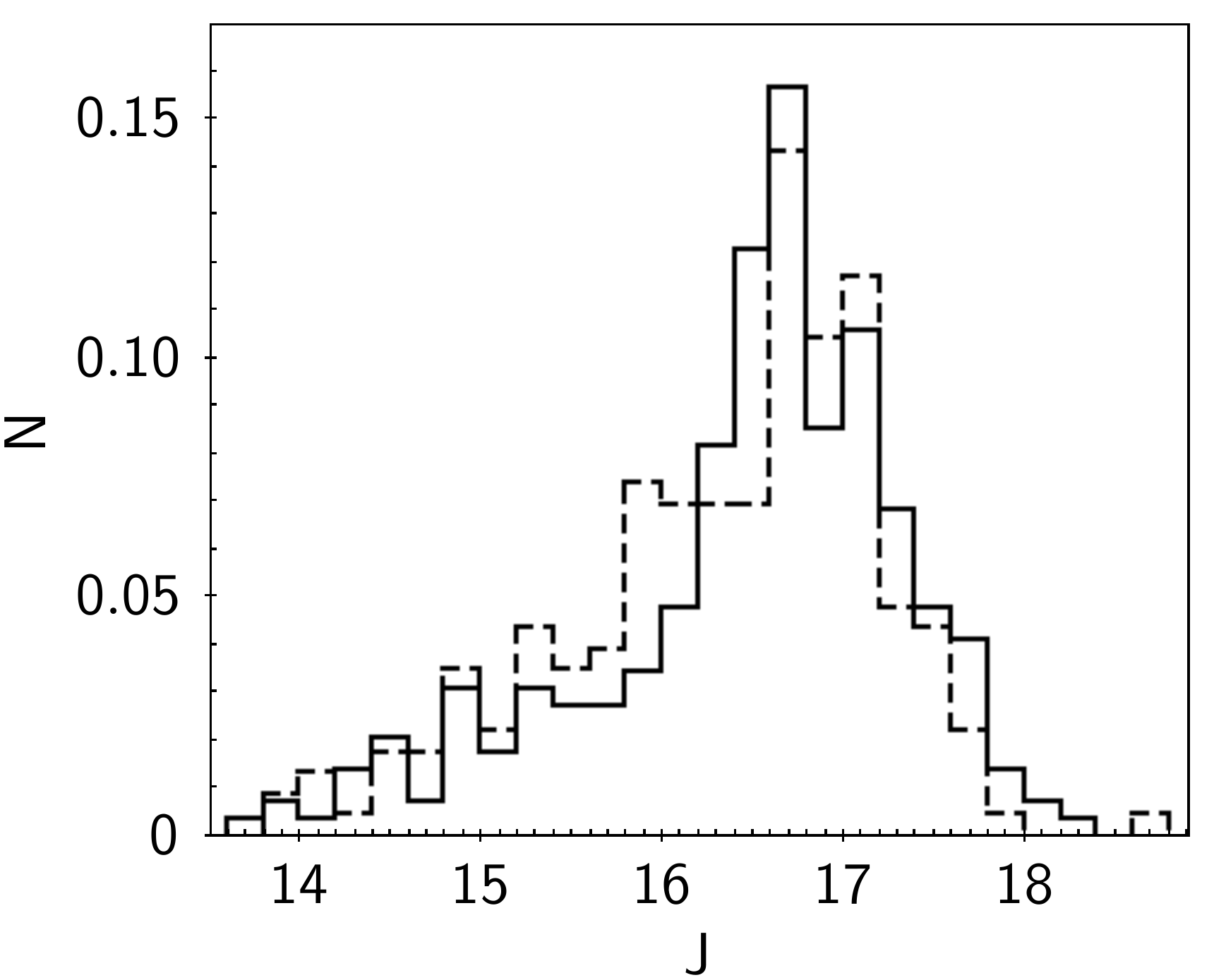}
         \includegraphics[width=0.45\textwidth]{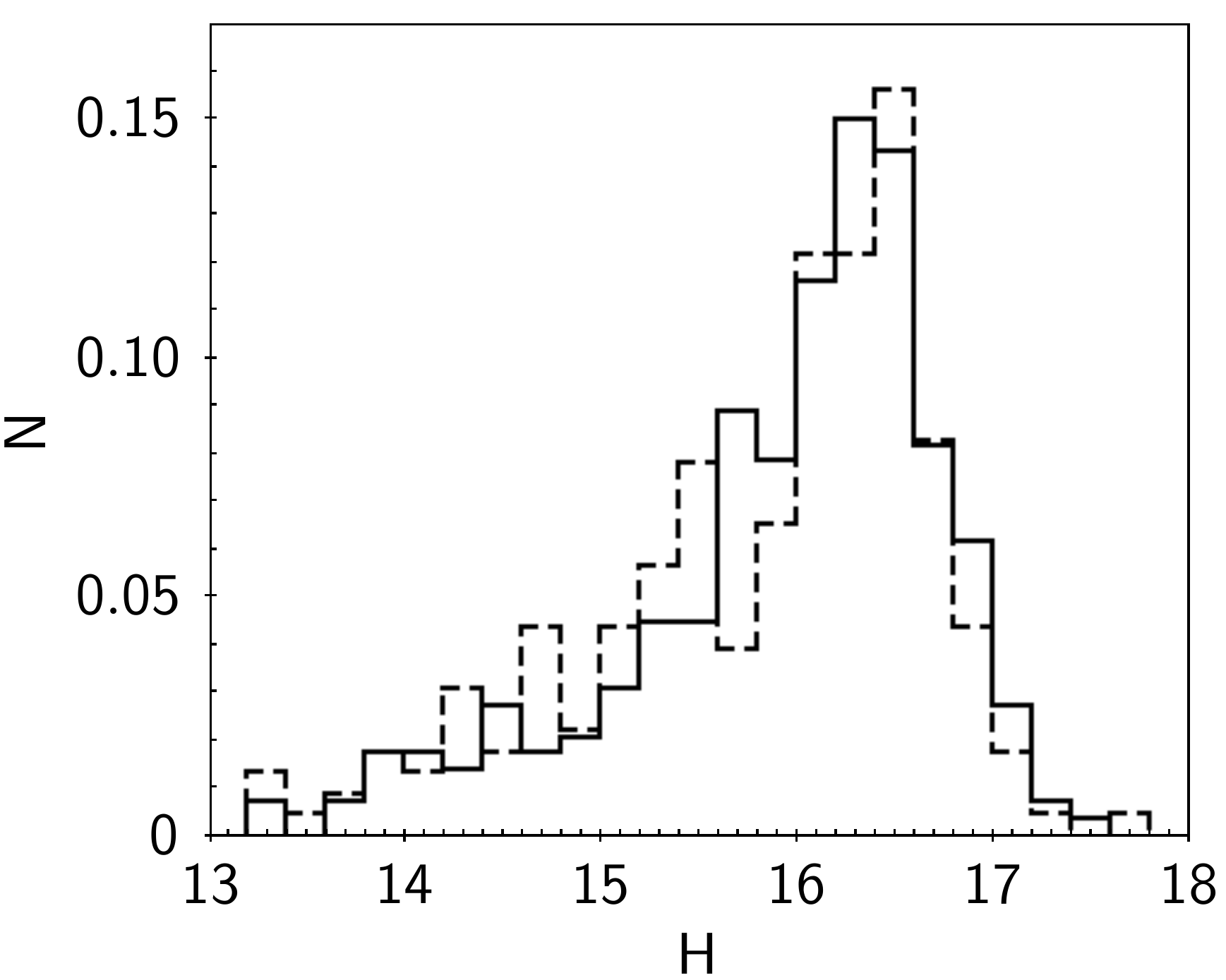}
         \includegraphics[width=0.45\textwidth]{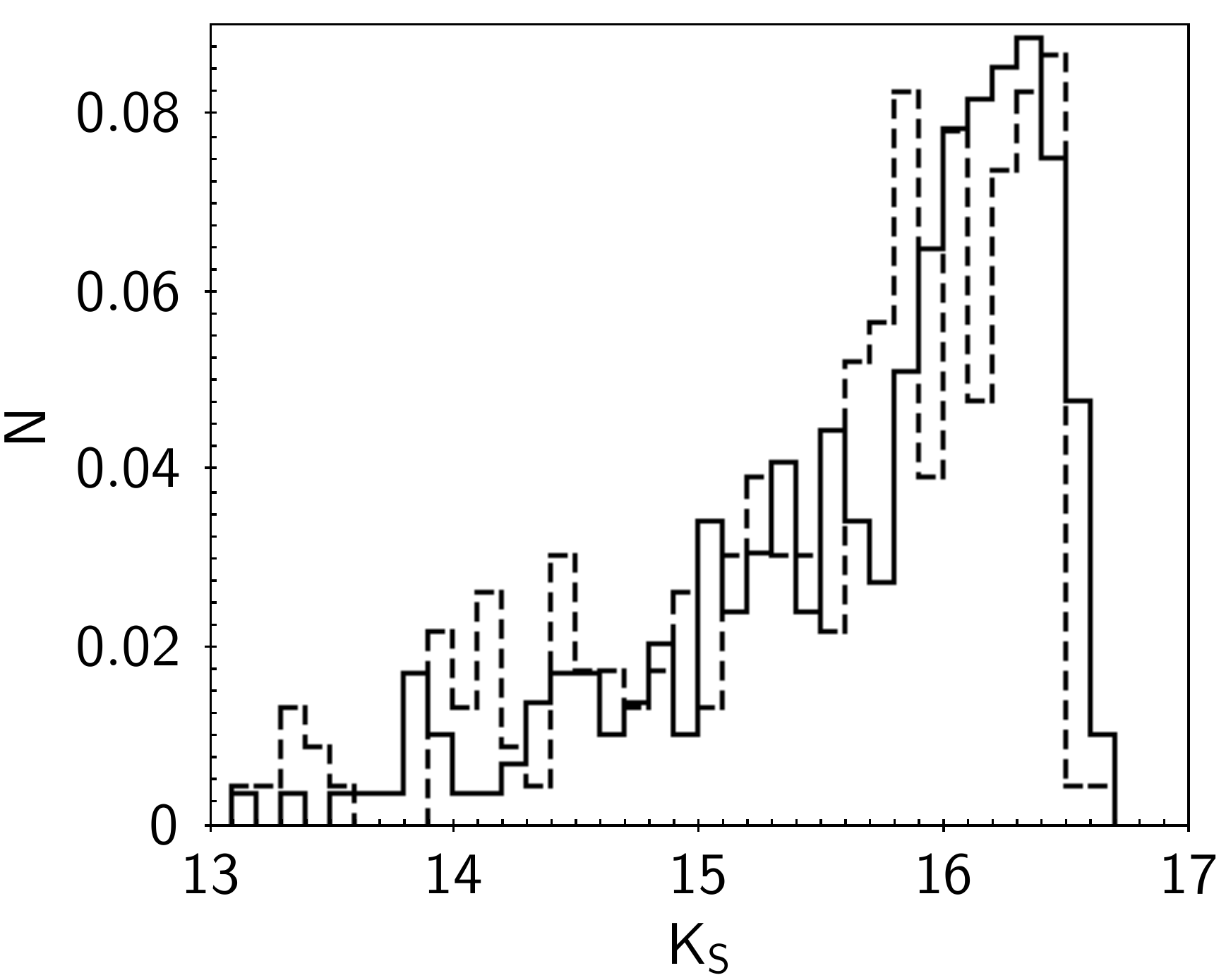}
         \includegraphics[width=0.45\textwidth]{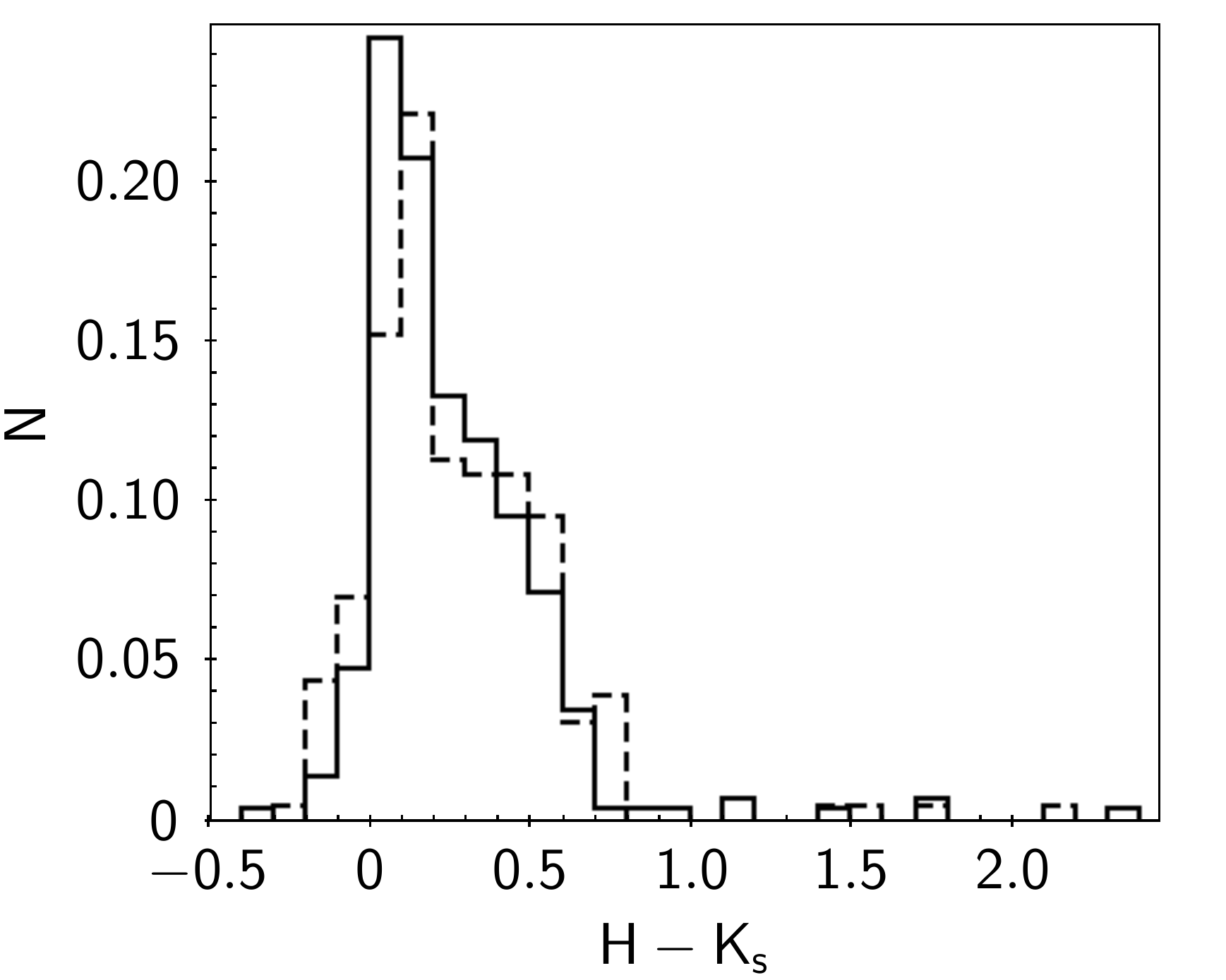}
         \caption{Near-IR magnitude and color distributions for extragalactic candidates. 
              Upper panels show histograms for J and H magnitudes, and lower panels,
              for K$_s$ magnitudes and (H - K$_s$) colors.   The
  distributions for the detections in the five passbands (Z, Y, J, H and K$_s$) are
  represented with solid lines and those detections in only three passbands (J, H and K$_s$) with dashed lines. 
         }
            \label{magdist}
\end{centering}
\end{figure*}

\begin{figure*}
\begin{centering}
         \includegraphics[width=0.45\textwidth]{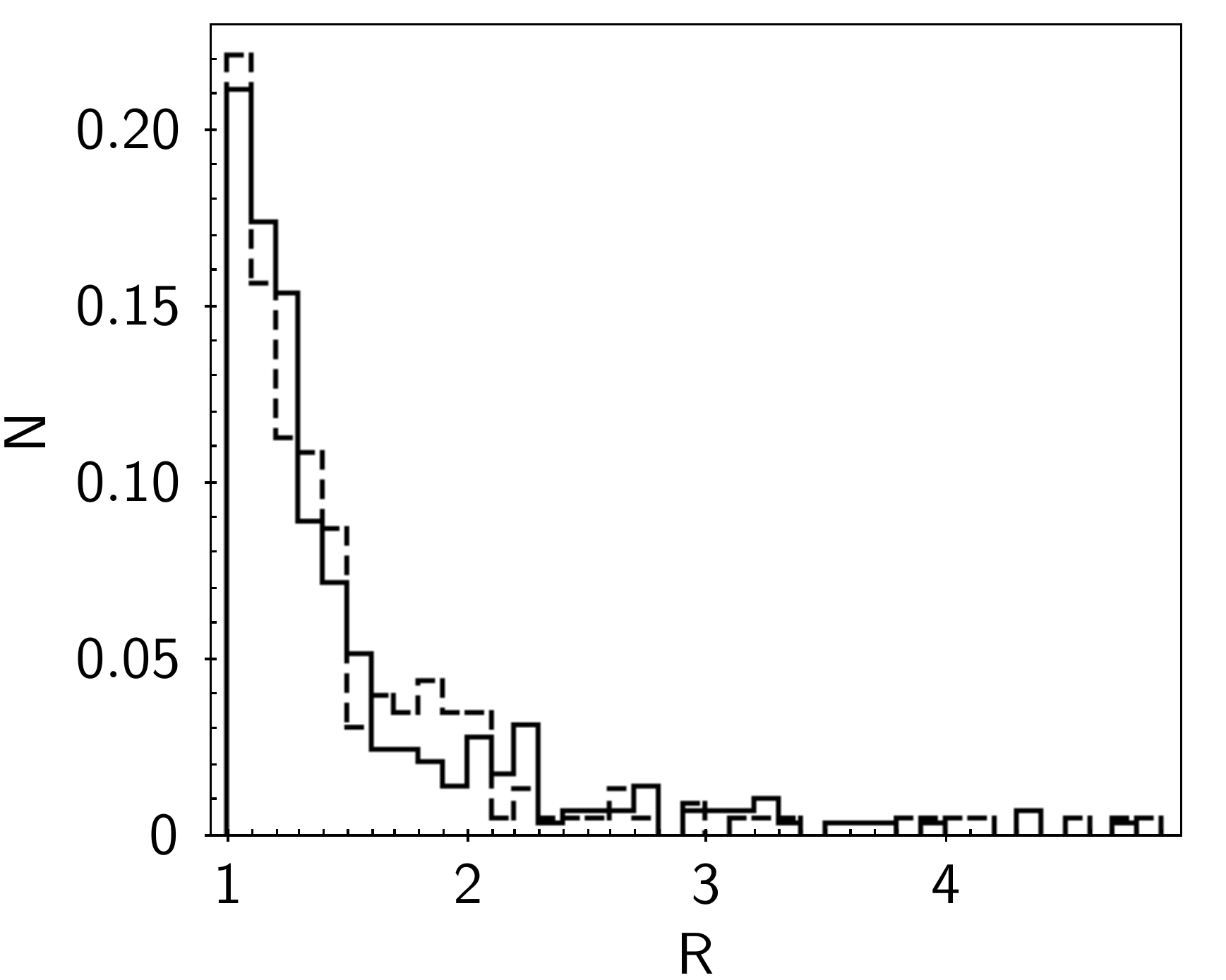}
         \includegraphics[width=0.45\textwidth]{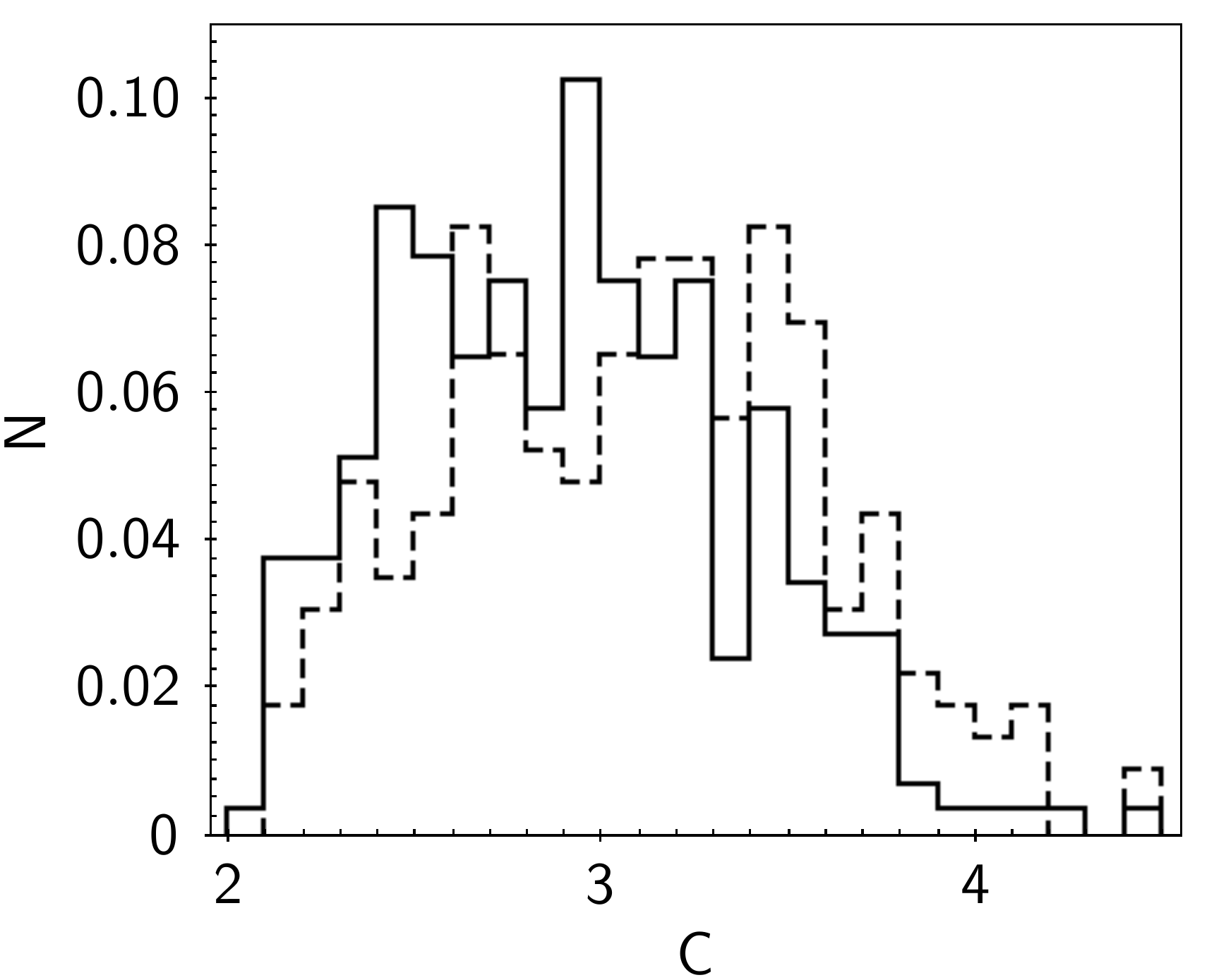}
         \includegraphics[width=0.45\textwidth]{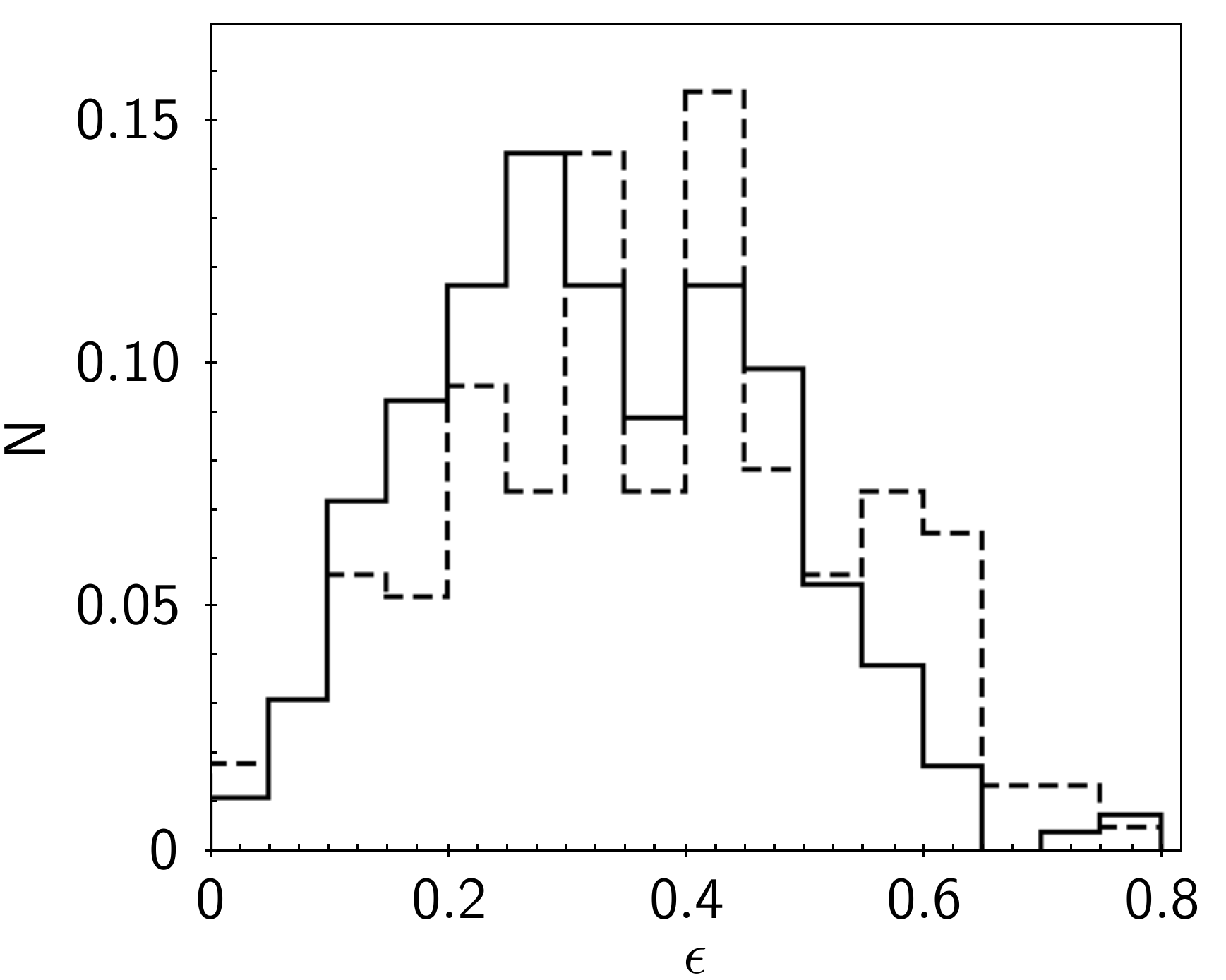}
         \includegraphics[width=0.45\textwidth]{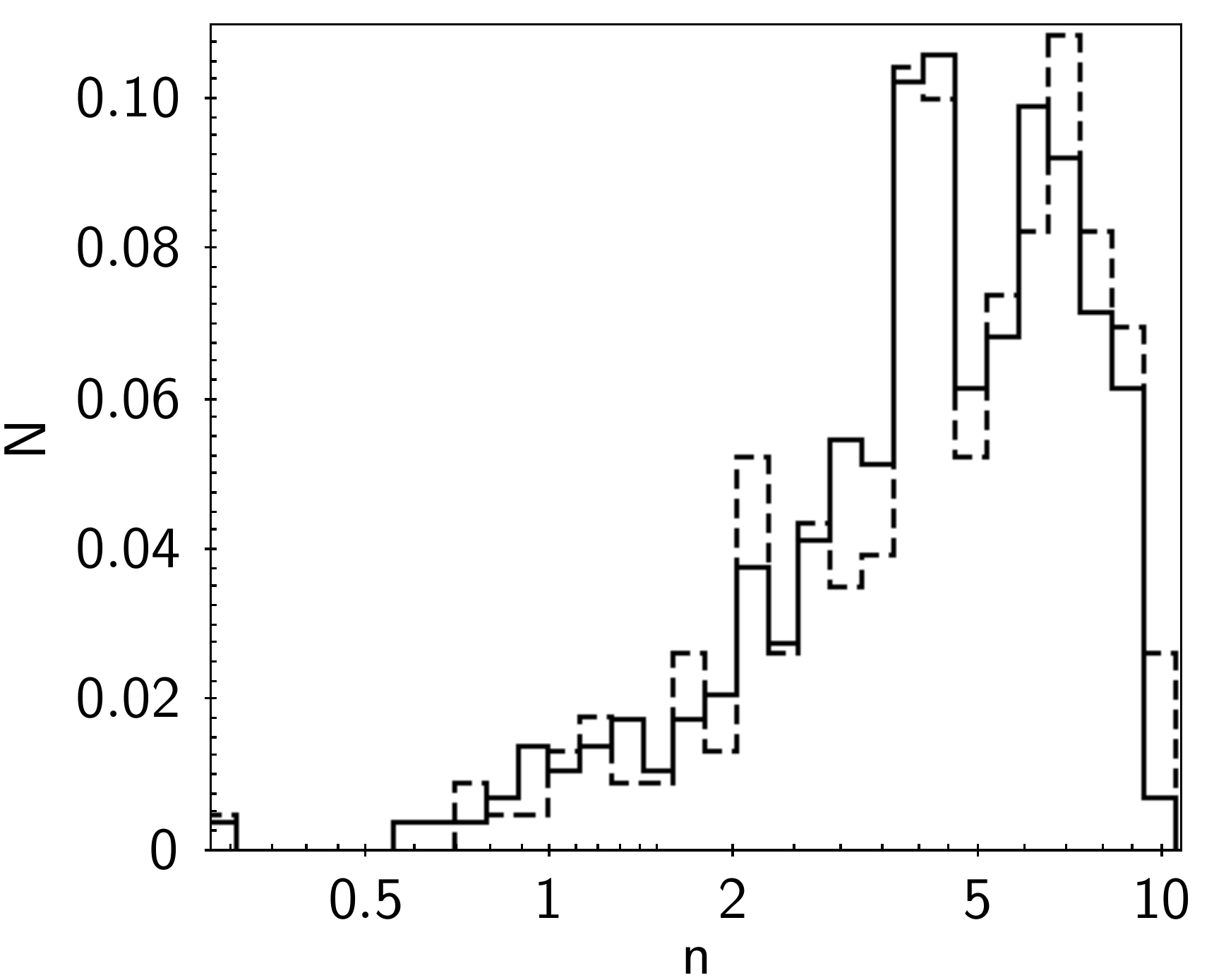}
         \caption{Structural parameter distributions for extragalactic candidates.  Upper panels
           show the histograms for R$_{1/2}$ and C, and lower panels, for
ellipticity and Sersic index. The distributions are represented as in Figure~\ref{magdist}.} 
            \label{all1}
\end{centering}
\end{figure*}


\begin{landscape}
  \begin{table*}

\caption{Photometric catalog of
  extragalactic candidates. }
\label{cat}

  \scalebox{0.75}{
    \begin{tabular}{cllcccccccccclllll}
 & & & & & & & & & & & & & & & & & \\ 
		\hline
		\hline
                 & & & & & & & & & & & & & & & & & \\ 

Id. 	& RA (J2000) & Dec (J2000) & Z & Y & J & H & K$_s$ & Z$_{2\prime\prime}$ & Y$_{2\prime\prime}$ & J$_{2\prime\prime}$ & H$_{2\prime\prime}$ & K$_s$$_{2\prime\prime}$ & R$_{1/2}$ & C & $\epsilon$ & n & Notes	\\
 & & & & & & & & & & & & & & & & & \\ 
		\hline                                                          
 & & & & & & & & & & & & & & & & & \\ 

                VVV-J114419.03-603025.9  & 11:44:19.03  & -60:30:25.9 & 15.86  &  15.85  &  15.76  &  15.30  &  15.02  & 15.96 &  15.94 &  15.77  &  15.19  & 15.06 &  2.46  & 3.54 &  0.14  &  3.97 & \\
                VVV-J114428.39-603158.4  & 11:44:28.39  & -60:31:58.4 &   -    &    -    &  17.17  &  17.00  &  16.44  &   -   &    -   &  17.14  &  16.91  & 16.46 &  1.03  & 2.34 &  0.50  &  7.38 & \\
                VVV-J114431.78-601626.8  & 11:44:31.78  & -60:16:26.8 &   -    &    -    &  16.81  &  16.34  &  16.27  &   -   &    -   &  16.82  &  16.35  & 16.25 &  1.25  & 3.34 &  0.10  &  7.18 & \\
                VVV-J114433.70-602742.8  & 11:44:33.70  & -60:27:42.8 &  16.82 &  16.78  &  16.63  &  16.20  &  16.11  & 16.83 &  16.77 &  16.65  &  16.13  & 16.08 &  1.09  & 2.56 &  0.18  &  4.12 & \\
                VVV-J114450.83-603356.9  & 11:44:50.83  & -60:33:56.9 &  17.18 &  17.02  &  16.75  &  16.22  &  16.06  & 17.16 &   16.97 &  16.80  &  16.13  & 16.05 &  2.20  & 3.65 &  0.43  &  8.36 & \\
                VVV-J114456.52-603249.6  & 11:44:56.52  & -60:32:49.6 &    -   &   -     &  16.56  &  16.02  &  15.86  &   -   &    -   &  16.60  &  15.96  & 15.89 &  1.25  & 2.46 &  0.35  &  2.64 & \\
                VVV-J114457.91-603958.3  & 11:44:57.91  & -60:39:58.3 &  17.80 &  17.76  &  17.60  &  17.31  &  16.43  & 17.77 &  17.76 &  17.57  &  17.32  & 16.40 &  2.51  & 3.09 &  0.42  &  4.90 & \\
                VVV-J114458.73-603252.1  & 11:44:58.73  & -60:32:52.1 &    -   &    -    &  16.57  &  15.83  &  15.45  &   -   &    -   &  16.61  &  15.88  & 15.50 &  1.23  & 2.51 &  0.42  &  1.99 & 1 \\
                VVV-J114516.94-604110.3  & 11:45:16.94  & -60:41:10.3  & 17.39  & 17.05  & 16.71  & 16.16  & 16.09  & 17.32  & 17.05 &  16.73 &  16.12  & 16.12  & 1.02  & 2.43  & 0.33  & 3.74   & \\
                VVV-J114520.81-603555.3  & 11:45:20.81  & -60:35:55.3  &   -   &    -  &   16.86  & 16.3 &   15.77  &   -   &    -  &   16.78 &  16.11 &  15.73 &  1.34  & 2.79 &  0.47  & 4.49 & \\
                \hline
                \tablecomments{Table 1 is published in its entirety in the machine-readable format.
                  A portion is shown here for guidance regarding its form and content.}
                \tablenotetext{}{1. galaxy pair}
                \tablenotetext{}{2. late-type galaxy: elongated shape}
                \tablenotetext{}{3. late-type galaxy: with spiral arms}
                \tablenotetext{}{4. early-type galaxy: bright elliptical galaxy}
                \tablenotetext{}{5. star near the galaxy nucleus}
\end{tabular} 
}  
\end{table*}
\end{landscape}

\begin{table*}
  \centering
  \caption{Median magnitudes, colors and structural parameters of the
    extragalactic candidates. }
      \label{medianas}

    {\small
        \begin{tabular}{c|c|c}
        \hline
        \hline
        & &   \\
Parameter & JHK$_s$ detections & ZYJHK$_s$ detections \\
 & &  \\ 
        \hline                                                          
 & &   \\ 

 Z [mag] &   -- &   17.29 $\pm$ 0.06  \\
 Y [mag] &   -- &   16.99 $\pm$ 0.05  \\
 J [mag] & 16.51 $\pm$ 0.04  & 16.63  $\pm$ 0.05   \\
 H [mag] & 16.05 $\pm$ 0.03  & 16.16  $\pm$ 0.05   \\ 
 K$_s$ [mag] & 15.79 $\pm$ 0.03  & 15.94 $\pm$ 0.04  \\ 
(Z - Y) [mag]    &    --  &  0.340 $\pm$  0.013  \\
(Y - J) [mag]    &    --  &  0.380 $\pm$  0.012  \\
(J - H) [mag]    & 0.530 $\pm$ 0.015  & 0.540 $\pm$ 0.015 \\ 
(J - K$_s$) [mag] & 0.740 $\pm$ 0.024  & 0.730 $\pm$ 0.019  \\ 
(H - K$_s$) [mag] & 0.260 $\pm$ 0.015 & 0.200 $\pm$ 0.015 \\
R$_{1/2}$ [arcsec] & 1.31 $\pm$ 0.03  & 1.28 $\pm$ 0.04  \\
C        & 3.13 $\pm$ 0.03  & 2.91 $\pm$ 0.03  \\ 
$\epsilon$  & 0.370 $\pm$ 0.004  & 0.320 $\pm$ 0.008  \\
n           & 4.60 $\pm$ 0.16  & 4.45 $\pm$ 0.13  \\
        \hline
        \end{tabular}  
    }
\end{table*}

\clearpage


\begin{thebibliography}{}

     \bibitem[Am{\^o}res et al.(2012)]{2012AJ....144..127A} Am{\^o}res, E.~B., Sodr{\'e}, L., Minniti, D., et al.\ 2012, \aj, 144, 127 

\bibitem[Andrews et al.(2014)]{2014PASA...31....4A} Andrews, S.~K., Kelvin, L.~S., Driver, S.~P., \& Robotham, A.~S.~G.\ 2014, \pasa, 31, e004

     \bibitem[Annunziatella et al.(2013)]{2013PASP..125...68A} Annunziatella, M., Mercurio, A., Brescia, M., Cavuoti, S., \& Longo, G.\ 2013, \pasp, 125, 68 

 \bibitem[Arnaboldi et al.(2007)]{2007PASJ...59..419A} Arnaboldi, M., Gerhard, O., Okamura, S., et al.\ 2007, \pasj, 59, 419 

      \bibitem[Baravalle et al.(2017)]{} Baravalle, L.~D., et al.\ 2017, in preparation


   \bibitem[Beard et al.(1990)]{1990MNRAS.247..311B} Beard, S.~M., MacGillivray, H.~T., \& Thanisch, P.~F.\ 1990, \mnras, 247, 311 

   \bibitem[Bertin \& Arnouts(1996)]{1996A&AS..117..393B} Bertin, E., \& Arnouts, S.\ 1996, \aaps, 117, 393 
    
   \bibitem[Bertin(2011)]{2011ASPC..442..435B} Bertin, E.\ 2011, Astronomical Data Analysis Software and Systems XX, 442, 435

     \bibitem[Buta \& McCall(1999)]{1999ApJS..124...33B} Buta, R.~J., \& McCall, M.~L.\ 1999, \apjs, 124, 33 
      
   \bibitem[Catelan et al.(2011)]{2011rrls.conf..145C} Catelan, M., Minniti, D., Lucas, P.~W., et al.\ 2011, RR Lyrae Stars, Metal-Poor Stars, and the Galaxy, 5, 145

     \bibitem[Chiu et al.(2016)]{2016A&C....16...79C} Chiu, I., Desai, S., \& Liu, J.\ 2016, Astronomy and Computing, 16, 79 

    
   \bibitem[Coldwell et  al.(2014)]{2014A&A...569A..49C} Coldwell, G., Alonso, S., Duplancic, F., et al.\ 2014, \aap, 569, A49 
  
\bibitem[Conselice et al.(2000)]{2000ApJ...529..886C} Conselice, C.~J., Bershady, M.~A., \& Jangren, A.\ 2000, \apj, 529, 886 

\bibitem[Dalton et al.(2006)]{2006SPIE.6269E..0XD} Dalton, G.~B., Caldwell, M., Ward, A.~K., et al.\ 2006, \procspie, 6269, 62690X

   \bibitem[Desai et al.(2012)]{2012ApJ...757...83D} Desai, S., Armstrong, R., Mohr, J.~J., et al.\ 2012, \apj, 757, 83 

   \bibitem[Durret et al.(2011)]{2011A&A...535A..65D} Durret, F., Adami, C., Cappi, A., et al.\ 2011, \aap, 535, A65 

   \bibitem[Emerson et al.(2006)]{2006Msngr.126...41E} Emerson, J., McPherson, A., \& Sutherland, W.\ 2006, The Messenger, 126, 41

     \bibitem[Gonz{\'a}lez-Fern{\'a}ndez et al.(2017)]{2017arXiv171108805G} Gonz{\'a}lez-Fern{\'a}ndez, C., Hodgkin, S.~T., Irwin, M.~J., et al.\ 2017, arXiv:1711.08805
         
     \bibitem[Henning et al.(2005)]{2005ASPC..329..199H} Henning, P.~A., Kraan-Korteweg, R.~C., \& Staveley-Smith, L.\ 2005, Nearby Large-Scale Structures and the Zone of Avoidance, 329, Fairall 

   \bibitem[Jarrett et al.(2000)]{2000AJ....119.2498J} Jarrett, T.~H., Chester, T., Cutri, R., et al.\ 2000a, \aj, 119, 2498 
          
   \bibitem[Jarrett et al.(2000)]{2000AJ....120..298J} Jarrett, T.~H., Chester, T., Cutri, R., et al.\ 2000b, \aj, 120, 298 

\bibitem[Kelvin et al.(2012)]{2012MNRAS.421.1007K} Kelvin, L.~S., Driver, S.~P., Robotham, A.~S.~G., et al.\ 2012, \mnras, 421, 1007 

\bibitem[Kilborn et al.(2002)]{2002AJ....124..690K} Kilborn, V.~A., Webster, R.~L., Staveley-Smith, L., et al.\ 2002, \aj, 124, 690 

   \bibitem[Kraan-Korteweg(1999)]{1999ASPC..170..103K} Kraan-Korteweg, R.~C.\ 1999, The Low Surface Brightness Universe, 170, 103 
           
   \bibitem[Kraan-Korteweg(2000)]{2000A&AS..141..123K} Kraan-Korteweg, R.~C.\ 2000, \aaps, 141, 123 

\bibitem[Lawrence et al.(2007)]{2007MNRAS.379.1599L} Lawrence, A., Warren, S.~J., Almaini, O., et al.\ 2007, \mnras, 379, 1599 

   \bibitem[Lewis et al.(2006)]{2006ASPC..351..255L} Lewis, J.~R., Irwin, M.~J., Gonzalez-Solares, E.~A., et al.\ 2006, Astronomical Data Analysis Software and Systems XV, 351, 255 
 
   \bibitem[Mauro et al.(2013)]{2013RMxAA..49..189M} Mauro, F., Moni Bidin, C., Chen{\'e}, A.-N., et al.\ 2013, \rmxaa, 49, 189 

   \bibitem[McIntyre et al.(2015)]{2015AJ....150...28M} McIntyre, T.~P., Henning, P.~A., Minchin, R.~F., Momjian, E., \& Butcher, Z.\ 2015, \aj, 150, 28
     
   \bibitem[Minniti et al.(2010)]{2010NewA...15..433M} Minniti, D., Lucas, P.~W., Emerson, J.~P., et al.\ 2010, \na, 15, 433

   \bibitem[Nilo Castell{\'o}n et al.(2014)]{2014MNRAS.437.2607N} Nilo Castell{\'o}n, J.~L., Alonso, M.~V., Garc{\'{\i}}a Lambas, D., et al.\ 2014, \mnras, 437, 2607 

   \bibitem[Ramatsoku et al.(2016)]{2016MNRAS.460..923R} Ramatsoku, M., Verheijen, M.~A.~W., Kraan-Korteweg, R.~C., et al.\ 2016, \mnras, 460, 923

     \bibitem[Rowe et al.(2015)]{2015A&C....10..121R} Rowe, B.~T.~P., Jarvis, M., Mandelbaum, R., et al.\ 2015, Astronomy and Computing, 10, 121

    \bibitem[Said et al.(2016)]{2016MNRAS.457.2366S} Said, K., Kraan-Korteweg, R.~C., Staveley-Smith, L., et al.\ 2016a, \mnras, 457, 2366 

  \bibitem[Said et al.(2016)]{2016MNRAS.462.3386S} Said, K., Kraan-Korteweg, R.~C., Jarrett, T.~H., Staveley-Smith, L., \& Williams, W.~L.\ 2016b, \mnras, 462, 3386 

   \bibitem[Saito et al.(2012)]{2012A&A...544A.147S} Saito, R.~K., Minniti, D., Dias, B., et al.\ 2012, \aap, 544, A147   
     
   \bibitem[Schlafly \& Finkbeiner(2011)]{2011ApJ...737..103S} Schlafly, E.~F., \& Finkbeiner, D.~P.\ 2011, \apj, 737, 103 

     \bibitem[Sersic(1968)]{1968adga.book.....S} Sersic, J.~L.\ 1968, Cordoba, Argentina: Observatorio Astronomico, 1968
     
   \bibitem[Skrutskie et al.(2006)]{2006AJ....131.1163S} Skrutskie, M.~F., Cutri, R.~M., Stiening, R., et al.\ 2006, \aj, 131, 1163 
   
   \bibitem[Staveley-Smith et al.(2016)]{2016AJ....151...52S} Staveley-Smith, L., Kraan-Korteweg, R.~C., Schr{\"o}der, A.~C., et al.\ 2016, \aj, 151, 52 

   \bibitem[Stetson(1987)]{1987PASP...99..191S} Stetson, P.~B.\ 1987, \pasp, 99, 191
   
   \bibitem[Stetson(1994)]{1994PASP..106..250S} Stetson, P.~B.\ 1994, \pasp, 106, 250

     \bibitem[Tody(1993)]{1993ASPC...52..173T} Tody, D.\ 1993, Astronomical Data Analysis Software and Systems II, 52, 173 

   \bibitem[Tully \& Fisher(1977)]{1977A&A....54..661T} Tully, R.~B., \& Fisher, J.~R.\ 1977, \aap, 54, 661
     
   \bibitem[Varela et al.(2009)]{2009A&A...497..667V} Varela, J., D'Onofrio, M., Marmo, C., et al.\ 2009, \aap, 497, 667

     \bibitem[Williams et al.(2014)]{2014MNRAS.443...41W} Williams, W.~L., Kraan-Korteweg, R.~C., \& Woudt, P.~A.\ 2014, \mnras, 443, 41 

   \bibitem[Woudt \& Kraan-Korteweg(2001)]{2001A&A...380..441W} Woudt, P.~A., \& Kraan-Korteweg, R.~C.\ 2001, \aap, 380, 441 
        

\end{thebibliography}
\end{document}